\documentclass[sigconf]{acmart}
\def\BibTeX{{\rm B\kern-.05em{\sc i\kern-.025em b}\kern-.08emT\kern-.1667em\lower.7ex\hbox{E}\kern-.125emX}}

\usepackage{tabularx}
\let\oldtabularx\tabularx
\renewcommand{\tabularx}{\small\oldtabularx}
\newcolumntype{C}{>{\centering\arraybackslash\hsize=.5\hsize\linewidth=\hsize}X} 
\def\BibTeX{{\rm B\kern-.05em{\sc i\kern-.025em b}\kern-.08em
    T\kern-.1667em\lower.7ex\hbox{E}\kern-.125emX}}
\usepackage{tabularray}
\usepackage{subfig}
\usepackage{wasysym}

\usepackage{nicefrac}
\usepackage{siunitx}
\usepackage{array,framed}
\usepackage{booktabs}
\usepackage{
  color,
  float,
  epsfig,
  wrapfig,
  graphics,
  graphicx,
}

\usepackage{textcomp,amssymb}
\usepackage{setspace}
\usepackage{latexsym,fancyhdr,url}
\usepackage{enumerate}
\usepackage{algpseudocode}
\usepackage{graphics}
\usepackage{xparse} 
\usepackage{xspace}
\usepackage{multirow}
\usepackage{csvsimple}
\usepackage{balance}

\usepackage{xcolor,colortbl}

\usepackage{
  tikz,
  pgfplots,
  pgfplotstable
}
\usepackage{hyperref}

\usetikzlibrary{
  shapes.geometric,
  arrows,
  external,
  pgfplots.groupplots,
  matrix
}

\pgfplotsset{compat=1.9}

\usepackage{mathtools,}
\DeclarePairedDelimiter\abs{\lvert}{\rvert}
\DeclarePairedDelimiter\norm{\lVert}{\rVert}
\DeclareMathAlphabet\mathbfcal{OMS}{cmsy}{b}{n}
\DeclareMathAlphabet{\mathcal}{OMS}{cmsy}{m}{n}

\usepackage{amsmath}

\DeclareGraphicsExtensions{%
    .png,.PNG,%
    .pdf,.PDF,%
    .jpg,.mps,.jpeg,.jbig2,.jb2,.JPG,.JPEG,.JBIG2,.JB2}

\usepackage{xparse}
\newcommand{\bnm}{\begin{newmath}}
\newcommand{\enm}{\end{newmath}}

\newcommand{\bea}{\begin{eqnarray*}}%
\newcommand{\eea}{\end{eqnarray*}}%

\newcommand{\bne}{\begin{newequation}}
\newcommand{\ene}{\end{newequation}}

\newcommand{\bal}{\begin{newalign}}
\newcommand{\eal}{\end{newalign}}

\newenvironment{newalign}{\begin{align}%
\setlength{\abovedisplayskip}{4pt}%
\setlength{\belowdisplayskip}{4pt}%
\setlength{\abovedisplayshortskip}{6pt}%
\setlength{\belowdisplayshortskip}{6pt} }{\end{align}}

\newenvironment{newmath}{\begin{displaymath}%
\setlength{\abovedisplayskip}{4pt}%
\setlength{\belowdisplayskip}{4pt}%
\setlength{\abovedisplayshortskip}{6pt}%
\setlength{\belowdisplayshortskip}{6pt} }{\end{displaymath}}

\newenvironment{newequation}{\begin{equation}%
\setlength{\abovedisplayskip}{4pt}%
\setlength{\belowdisplayskip}{4pt}%
\setlength{\abovedisplayshortskip}{6pt}%
\setlength{\belowdisplayshortskip}{6pt} }{\end{equation}}

\newcounter{ctr}

\newcounter{mytable}
\def\mytable{\begin{centering}\refstepcounter{mytable}}
\def\endmytable{\end{centering}}

\newcounter{myfig}
\def\myfig{\begin{centering}\refstepcounter{myfig}}
\def\endmyfig{\end{centering}}

\newlength{\saveparindent}
\newlength{\saveparskip}
\newcommand{\E}{{\rm I\kern-.3em E}}

\renewcommand{\eqref}[1]{\mbox{Equation~(\ref{#1})}}

\def \part {part}

\renewcommand{\paragraph}[1]{\vspace*{6pt}\noindent\textbf{#1}\;}

\def \blackslug{\hbox{\hskip 1pt \vrule width 4pt height 8pt
    depth 1.5pt \hskip 1pt}}
\def \qed{\quad\blackslug\lower 8.5pt\null\par}

\newcounter{mynote}[section]
\newcommand{\notecolor}{blue}
\newcommand{\thenote}{\thesection.\arabic{mynote}}
\newcommand{\tnote}[1]{\refstepcounter{mynote}{\bf \textcolor{\notecolor}{$\ll$TomR~\thenote: {\sf #1}$\gg$}}}

\newcommand\ignore[1]{}

\newcounter{rcnote}[section]

\newcounter{mrnote}[section]

\newcounter{fknote}[section]

\newcounter{anote}[section]

\DeclareMathSymbol{\mlq}{\mathord}{operators}{``}
\DeclareMathSymbol{\mrq}{\mathord}{operators}{`'}

\newcommand{\rhf}[2]{R_{f, \gamma}}

\DeclareDocumentCommand{\edist}{o o}{
  \ensuremath{
    \IfNoValueTF{#1}{{d}}{{\sf d}(#1,#2)}
  }
}

\newcommand{\olrk}[1]{\ifx\nursymbol#1\else\!\!\mskip4.5mu plus 0.5mu\left(\mskip0.5mu plus0.5mu #1\mskip1.5mu plus0.5mu \right)\fi}

\NewDocumentCommand{\indseq}{ O{1} O{r} }{{#1}\ldots {#2}}

\usepackage{enumitem}
\setlist[itemize]{leftmargin=*}

\usepackage{threeparttable}

\usepackage{listings}
\definecolor{codegreen}{rgb}{0,0.6,0}
\definecolor{codegray}{rgb}{0.5,0.5,0.5}
\definecolor{codepurple}{rgb}{0.58,0,0.82}
\definecolor{backcolour}{rgb}{0.95,0.95,0.92}
\lstdefinestyle{mystyle}{
    float=tp,
    language=C++,
    aboveskip=0pt,belowskip=0pt,
    frame=lines, 
    commentstyle=\color{codegray},
    keywordstyle=\color{blue},
    numberstyle=\tiny\color{codegray},
    basicstyle=\ttfamily\footnotesize,
    xleftmargin=1em,xrightmargin=1em,
    breakatwhitespace=false,         
    breaklines=true,                 
    captionpos=b,                    
    keepspaces=true,                 
    numbers=left,                    
    numbersep=5pt, 
    showspaces=false,                
    showstringspaces=false,
    showtabs=false,                  
    tabsize=1
}
\lstdefinestyle{java}{
    float=tp,
    language=Java,
    aboveskip=0pt,belowskip=-10pt,
    frame=lines, 
    commentstyle=\color{codegray},
    keywordstyle=\color{blue},
    numberstyle=\tiny\color{codegray},
    basicstyle=\ttfamily\footnotesize,
    xleftmargin=1em,xrightmargin=1em,
    breakatwhitespace=false,         
    breaklines=true,                 
    captionpos=b,                    
    keepspaces=true,                 
    numbers=left,                    
    numbersep=5pt, 
    showspaces=false,                
    showstringspaces=false,
    showtabs=false,                  
    tabsize=1
}
\usepackage{filecontents}

\usepackage[vlined, ruled,linesnumbered]{algorithm2e}
\SetKwData{Left}{left}\SetKwData{This}{this}\SetKwData{Up}{up}
\SetKwFunction{Union}{Union}\SetKwFunction{FindCompress}{FindCompress}
\SetKwInput{KwInput}{Input}                
\SetKwInput{KwOutput}{Output}              
\SetCommentSty{textrm}
\SetKwComment{Comment}{~$\triangleright$~}{}
\SetSideCommentRight
\SetFillComment

\newcommand{\myfigureshrinker}{\vspace{-.6cm}}
\newcommand{\framework}{\textsc{FINER}}
\newcommand{\Nexplainers}{six}

\newcommand{\tcolor}{black}

\usepackage{amsthm}

\newtheorem{definition}{Definition}

\begin{document}
\fancyhead{}
\def\thetitle{\framework{}: Enhancing State-of-the-art Classifiers with Feature Attribution to Facilitate Security Analysis}
\title{\thetitle}

\author{\vspace{-.5cm}Yiling He, Jian Lou, Zhan Qin, Kui Ren}
\affiliation{\small{Zhejiang University}}
\email{{yilinghe, jian.lou, qinzhan, kuiren}@zju.edu.cn}

\date{}

\begin{abstract}

Deep learning classifiers achieve state-of-the-art performance in various risk detection applications.
They explore rich semantic representations and are supposed to automatically discover risk behaviors.
However, due to the lack of transparency, the behavioral semantics cannot be conveyed to downstream security experts to reduce their heavy workload in security analysis.
Although feature attribution~(FA) methods can be used to explain deep learning, the underlying classifier is still blind to what behavior is suspicious,
and the generated explanation cannot adapt to downstream tasks, incurring poor explanation fidelity and intelligibility.

In this paper, we propose \framework{}, the first framework for risk detection classifiers to generate high-fidelity and high-intelligibility explanations.
The high-level idea is to gather explanation efforts from model developer, FA designer, and security experts.
To improve fidelity, we fine-tune the classifier with an explanation-guided multi-task learning strategy.
To improve intelligibility, we engage task knowledge to adjust and ensemble FA methods.
Extensive evaluations show that \framework{} improves explanation quality for risk detection. 
Moreover, we demonstrate that \framework{} outperforms a state-of-the-art tool in facilitating malware analysis.

\end{abstract}

\maketitle

\section{Introduction}
\label{sec:intro}

Deep learning~(DL) based classifiers have shown great potential in the risk detection phase.
They automate large-scale detection and achieve considerable accuracy for different risk types including Android/PE malware~\cite{mariconti2016mamadroid, arp2014drebin, obaidat2022jadeite, kim2018multimodal}, code vulnerability~\cite{liu2019deepbalance, lin2020software, cao2021bgnn4vd}, and network intrusion~\cite{elmasry2020evolving, gamage2020deep}.
However, when coming to the security analysis stage, those classifiers fall short as they only produce prediction labels.
This problem is severe since security experts need to actively respond to what the classifiers detect.
Without insight into the reasoning behind each detection, their task becomes immensely challenging. 
Consider malware analysis: in just the first two weeks of 2023, 3 million instances were detected\cite{AV-ATLAS}. Furthermore, analyzing these malware instances is time-intensive, with reverse engineering for a single instance taking roughly 92 minutes\cite{antovaniAFB22}.

To explain DL-based decisions, \textit{feature attribution}~(FA) methods are promising for being compatible with various model architectures~\cite{kim2018interpretability}.
FA methods work on trained classifiers and assign an importance score for each feature of an individual input.
A successful application in image analysis is known as saliency map~\cite{alqaraawi2020evaluating}, where important pixels are visually highlighted on an image for human users to inspect.
Motivated by the success, prior works try to explain risk detection classifiers with the same approach but find a low explanation \textit{fidelity}~\cite{warnecke2020evaluating}.
Although a few security-customized FA methods have been proposed to improve fidelity~\cite{weixnids}, they are task-specific since they add intuitive constraints to a certain domain-general method. 
For example, LEMNA~\cite{guo2018lemna} considers the same black-box setting with LIME~\cite{ribeiro2016should}, but differs from it in handling feature dependency, which is mostly observed for recurrent neural network~(RNN) based applications, e.g., function start detection~\cite{shin2015recognizing}.
Unsurprisingly, as validated by our experiments in Section~\ref{sec:eval}, they need more computational cost while the fidelity improvement is limited / does not hold true in other tasks.
\indent 
Explaining data-driven risk detection classifiers is more challenging.
First, the fidelity problem originates from the diversity of classifiers that use particular data representations~(as shown in Table~\ref{tab:classifiers}).
For example, to represent a binary program, the extracted features can be a long sequence of opcodes, APIs, or hand-crafted statistics~\cite{aafer2013droidapiminer}; when encoded into vector space, the data can be of large sizes and have great variance in shape and distribution. 
Second, an overlooked \textit{intelligibility} problem is caused by the semantic gap between model feature and actionable understanding.
As the example in Figure~\ref{fig:motivating}, the data-driven malware classifier explores low-level language~(i.e., bytecode) that is intrinsically hard to read and individual features~(i.e., opcodes) do not serve as the fundamental unit of maliciousness. 
Under such circumstances, the per-feature explanation style of FA is not insightful to help security experts with security analysis.

In this paper, we address the fidelity and intelligibility issues from the perspective of the explainable risk detection system~(ERDS).
Specifically, ERDS has an end goal to facilitate security analysis with explanations, and it consists of a data-driven classifier and an FA-based explainer.
Our design is based on two intuitions.
First, the classifier should be the major agent of ERDS, and high-fidelity explanations depend on its reasonable decision boundary~\cite{ross2017right}.
For instance, risk detection should be correlated with semantic features instead of artifacts~\cite{arp2022and},  
which is a prerequisite for downstream explanation endeavors~(e.g., customizing FA to handle feature dependency) to be effective.
Second, security experts should be the consumer of ERDS, and intelligibility relies on their desiderata for specific tasks to abstract low-level explanations.
For example, since malicious functionalities are consumable explanations for malware analysts~\cite{downing2021deepreflect}, ERDS with the opcode feature-based classifier should be adjusted to generate function-level explanations.
\indent 
We propose a framework, named \framework{}, to promote data-driven risk detection classifiers into ERDS that generates useful explanations for security analysis.
To improve fidelity, we design an explanation-guided data augmentation strategy for risk samples and leverage it in multi-task learning to fine-tune the classifier.
To improve intelligibility, we define task-aware explanations on the level of \textit{intelligible component}~(IC) and correspondingly adapt a fidelity metric to ensemble different FA methods.
Specifically, \framework{} has an interface to engage with task knowledge, and three modules~(namely \textit{explanation-guided model updating}, \textit{task-aware explanation generation}, and \textit{explanation quality measurement}) to build ERDS with a more interpretable classifier and a more adaptable explainer.
The decoupled architecture also supports the needs of different stakeholders at different stages of building an ERDS.

We apply \framework{} on three state-of-the-art risk detection classifiers targeting tasks including Android malware detection~\cite{mclaughlin2017deep}, Windows malware detection~\cite{downing2021deepreflect}, and vulnerability detection~\cite{li2018vuldeepecker}.
The classifiers are trained on $14$K apps, $48$K binaries, and $32$K gadgets respectively, and the explainers are formed with \Nexplainers{} representative FA methods including white-box methods~\cite{simonyan2014deep,sundararajan2017axiomatic,shrikumar2017learning} and black-box methods~\cite{ribeiro2016should,guo2018lemna,lundberg2017unified}.
The results show that \framework{} significantly improves the explanation fidelity of ERDS across all classifiers and explanation scenarios.
The updating module is effective to improve model interpretability~(from $21.28\%$ to $82.05\%$ depending on the classifier) without accuracy trade-off, and the ensemble module achieves higher explanation fidelity than the baseline~(from $10.12\%$ to $17.00\%$ depending on the scenario). 
We also show that \framework{} outperforms a state-of-the-art tool in malicious functionality localization task.

\noindent \textbf{Contributions.} This paper makes the following contributions.
\begin{itemize}
\item We propose a formalization of the ERDS to establish explanation desiderata for security analysis. We propose to address the fidelity and intelligibility problems by explanation-guided model updating and IC-based explanation ensemble.
\item We implement the framework \framework{} to promote data-driven risk detection classifiers into ERDS with high-fidelity and high-intelligibility explanations.
\framework{} can meet the needs of different stakeholders at each stage of building an ERDS, and we make the dataset and code open-source.
\item We evaluate \framework{} with three critical risk detection tasks and six representative FA methods, showing that explanation fidelity is improved across all ERDS settings, i.e., different combinations of the classifier and explainer.
We also demonstrate that \framework{} outperforms a state-of-the-art tool in localizing malware functions.
\end{itemize}

\begin{table*}[thbp]
\caption{DL-based risk detection classifiers investigated in related works.}
\label{tab:classifiers}
\begin{threeparttable}
\begin{tabularx}{\linewidth}{ccCcc}
\toprule
\textbf{Classifier}   & \textbf{Application}              & \textbf{Feature Space} $\mathcal{F}$\tnote{*}                                   & \textbf{Vector Space} $\mathcal{V}$\tnote{\dag}   & \textbf{Model} $f$             \\ \midrule
Mimicus+     & PDF malware              & Static document statistics~(based on file structure)     & Tabular\,$(135,1)$           & MLP             \\
Drebin+      & Android malware          & Static app statistics~(based on manifest and bytecode)   & Tabular\,$(\geq545000,1)$       & MLP             \\ \midrule
VulDeePecker & Program vulnerability    & Static token sequence~(from source code)                 & Pseudo text\,$(50,200)$      & RNN             \\
DAMD         & Android malware          & Static opcode sequence~(from bytecode)                   & Pseudo text\,$(N+,8)$        & CNN          \\
DR-VGG       & PE malware               & Static ACFG statistics of basic blocks~(from bytecode)   & Pseudo image\,$(20000,18)$   & CNN          \\  
\bottomrule
\end{tabularx} 
    \begin{tablenotes}
    \footnotesize
        \item[*] Feature-space data representation: the ``statistics" is based on several factors chosen by domain experts, according to their experience with risk patterns; since the feature space construction of Mimicus+ and Drebin+ is entirely dependent on hand-crafted patterns, the two classifiers are considered pattern-driven; since VulDeePecker, DAMD, and DR-VGG preserve token, opcode, and basic block, respectively, in feature space without manual selection, the three classifiers are considered as data-driven. 
        \item[\dag] Vector-space data representation: data modality and the vector-space shape $(m,n)$ in our formalization; for DAMD, $m$ is dynamic that depends on the actual number of opcodes; for other classifiers, the vector-space data are of fixed shape, while a special case is that, for Drebin+, $m$ would increase with the increasing of dataset size~\cite{roy2015experimental} and the value reported in the table only applies to an ancient small dataset~\cite{zhou2012dissecting} used in its original paper.
    \end{tablenotes}
\end{threeparttable}
\end{table*}

\section{Explainable Risk Detection System}
\label{sec:erds}

In this section, we introduce \textit{explainable risk detection system}~(ERDS), where the emerging feature attribution~(FA) techniques are used as the \textit{explainer} to interpret the deep learning~(DL) detection of the \textit{classifier}.
We first define the problem of ERDS and propose a novel formalization.
Then, we discuss representative classifiers and explainers from related works and highlight the main parameters in our formalization.
Finally, we provide background for understanding the capability of FA explanations.

\subsection{Problem Definition}

\textcolor{\tcolor}{
\noindent\textbf{Local Post-hoc Explainability.}
Explainable risk detection is a subset of the broader field of eXplainable Artificial Intelligence~(XAI), which aims to make AI systems more interpretable to humans.
As XAI research has progressed in recent years, there is a wide range of techniques designed for explaining different scenarios~\cite{du2019iml_techniques}.
To build the vocabulary, we introduce the taxonomy of XAI methods in terms of what can be explained and how explanations can be generated:
first, an explanation can indicate how the model decisions are affected by individual instances~(\textit{local} methods), or by entire model parameters~(\textit{global} methods); second, the explainability can be achieved by building simple yet inherently interpretable models~(\textit{intrinsic} methods), or by approximating the decision-making of complex models that are already trained~(\textit{post-hoc} methods).
}
\textcolor{\tcolor}{
Recent advances in the local and post-hoc XAI technique, i.e., FA, have led to the development of ERDS~\cite{guo2018lemna, warnecke2020evaluating}.
Different from other techniques, the local post-hoc explainability handles the most common explanation scenario in risk detection that can (1)~provide clues for understanding critical risk samples and (2)~be compatible with trained state-of-the-art classifiers.
}

\vspace{.2cm}

\begin{figure}[t]
    \centering
    \includegraphics[width=.8\linewidth]{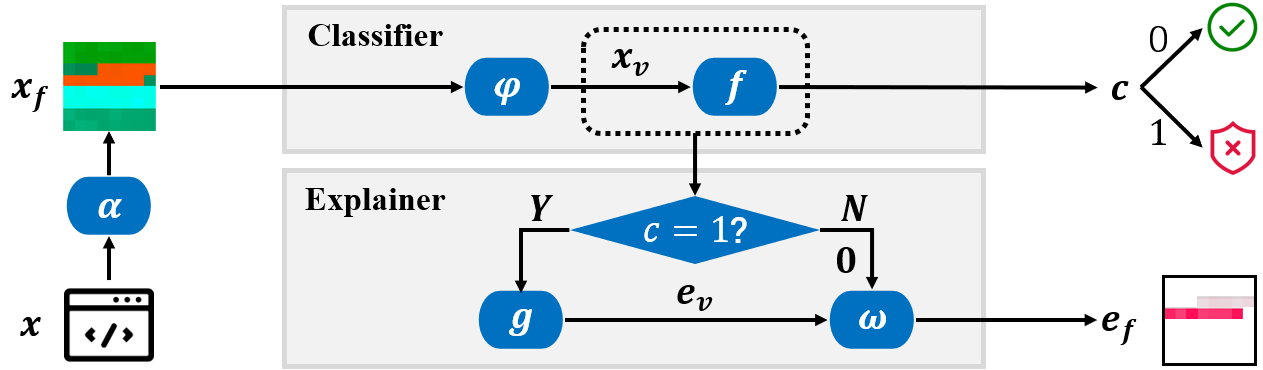}
    \caption{Workflow of the explainable risk detection system~(ERDS). \textcolor{\tcolor}{The formalized notations can be referenced in Table~\ref{tab:notation}}.}
    \label{fig:cls_exp}
\end{figure}

\noindent\textbf{Formalization.}
Under this context, ERDS is generally expanded from a well-trained risk detection classifier by appending a post-hoc explainer, and the ``classification with explanation" workflow for an individual instance can be illustrated as Figure~\ref{fig:cls_exp}.
In the following, we propose the formalization of ERDS and its two components, i.e., the classifier and the explainer.

\begin{definition}[ERDS]
Given a problem-space instance $x \in \mathcal{Z}$, ERDS first performs feature extraction $\alpha: \mathcal{Z} \rightarrow \mathcal{F}$ to embed it into a feature-space representation ${x_f} \in \mathcal{F}$, 
and then generates two outputs respectively from its classifier and explainer:
a predicted class label ${c} \in \mathcal{C} = \{ 0,1 \}$ where the zero value means normal sample and the positive value represents risk sample, and an explanation ${e_f \in \mathcal{F}}$ indicating why the sample can be identified as a risk.
\end{definition}

\begin{definition}[Classifier]
The classification pipeline is defined as 
$c = f(\varphi(x_f))$, 
where $\varphi: \mathcal{F} \rightarrow \mathcal{V} \subseteq{\mathbb{R}^{m \times n}}$ generates a $m$-by-$n$ vector-space representation $x_v \in \mathcal{V}$, such that $\varphi(x_f) = x_v$, and then the trained deep neural network~(DNN) $f: \mathcal{V} \rightarrow \mathcal{C}$ assigns the representation to the binary class label $c$.
\end{definition}

The classifier deals with binary classification tasks~(e.g., whether a computer program is malware or not) and is supervised thanks to the availability of large annotated dataset~\cite{virustotal}.
Unlike image and text domain, the design of feature space $\mathcal{F}$ involves much domain knowledge and varies among different applications.
On the other hand, it is often the case that the classifier would be more accurate/robust if more information is encoded in $\mathcal{F}$~(at the cost of the efficiency of feature extraction)~\cite{he2022msdroid}.

\begin{definition}[Explainer] \label{def:explainer}
For the explanation pipeline, it first calculates a feature importance score $e_v$ with $g: \mathcal{V} \rightarrow \mathcal{V}$ if the classifier outputs $c=1$, or else it would be set to a zero matrix. 
Next, with $\omega: \mathcal{V} \rightarrow \mathcal{F}$, $e_f$ presents the regions of interest~(ROI) on $x_f$ corresponding to values in $e_v$ that are above a certain threshold. 
\end{definition}

The calculation $g(x_v ; \mathcal{M})$ leverages a certain FA algorithm and requires a metadata set $\mathcal{M}$ about the classifier that varies among different algorithms~(see Section~\ref{sec:cls_exp}).
It outputs ${e_v}$ that has the same shape as $x_v$, and each element $(e_v)_{ij}$ quantifies the importance of $(x_v)_{ij}$ with respect to $f(x_v)$.
The ROI is filtered with a constant threshold $\tau_0 > 0$, which we denote as $\omega(e_v;\tau_0) = \{r_f \in x_f : \norm{e_v \odot Ind(x_v, \varphi(r_f))}_{1} \geq \tau_0\}$. 
The indicator function $Ind(x_v, r_v)$ returns a $m$-by-$n$ binary matrix where the $ij$-th element would be positive only if $(x_v)_{ij}$ exists in $r_v$.

\textcolor{\tcolor}{
The ROI implies risky behaviors and thus would only be non-empty for risky samples. 
This type of explanations is beneficial for security analysis since it can help experts to quickly identify the threat and develop corresponding prevention strategies.
On the other hand, the explanation of normality can have a large overlap between the two classes considering the existence of some common utilities.
It will be less distinct for understanding risky behaviors~(but worthy of other investigation as in Section~\ref{sec:discussion}).
}


\subsection{Diverse Classifiers and Explainers} 
\label{sec:cls_exp}

We describe the classifier and the explainer in related literature~\cite{guo2018lemna, warnecke2020evaluating, downing2021deepreflect}, with a focus on the design of $\mathcal{F}$ and $\mathcal{V}$, as well as the implementation of $f(\cdot)$ and $g(\cdot)$.

\noindent\textbf{Risk detection Classifiers.}
We summarize the investigated risk detection classifiers as in Table~\ref{tab:classifiers} and have the following findings.
\begin{itemize}
    \item Applications of Interest: 
    explanations are of high concern for risk detection related with static analysis, especially for malware~(including PDF, Android, and PE malware) detection. It is largely due to the high demanding efforts that are consumed by static reverse engineering and the severe negative consequence that would be brought by malware spread. By providing explanations for these classifiers, ERDS would have the potential to relieve the static analysis process and prevent the unwanted risk spread. 
    \item Different Data Representations: 
    the design of $\mathcal{F}$ and $\mathcal{V}$ varies among the classifiers for encoding different semantic information as different shapes of data. Except for Mimicus+ and Drebin+ that are inherit features from conventional machine learning based methods~\cite{smutz2012malicious, arp2014drebin}, other advanced risk detection classifiers~(i.e., VulDeePecker~\cite{li2018vuldeepecker}, DAMD~\cite{mclaughlin2017deep}, and DR-VGG~\cite{downing2021deepreflect}) adopt raw elements~(i.e., token, opcode, and basic block) from (disassembled)~program code for $\mathcal{F}$. As for $\mathcal{V}$, they encode them as complex text/image data~(called pseudo- text/image to highlight the actual difference in data modalities) instead of tabular data. 
    \item Different Model Architectures: 
    MLP, RNN, and CNN are all model types that are used to implement $f(\cdot)$, with the aim of achieving state-of-the-art performance on a particular dataset. While MLP fits tabular data of hand-crafted features, RNN and CNN can handle the data-driven classification tasks on the pseudo text/image data. Notably, the specific model architectures often cannot be directly adopted from other domains due to the distinct data shapes and distributions in $\mathcal{V}$.  
\end{itemize}

\begin{table*}[tbhp]
\caption{Representative explainers from related works. The involved model metadata~(output layer activation $\mathcal{O}$, middle layer activation $\mathcal{A}$, and baseline inputs $\mathcal{B}$) and computational cost are summarized for the understanding of their different application scenarios.}
\label{tab:explainers}
\begin{threeparttable}
\begin{tabularx}{.74\linewidth}{cCCCcc}
\toprule
\multirow{2}{*}{\textbf{Explainer}} & \multicolumn{3}{c}{\textbf{Metadata $\mathcal{M}$}}      & \multicolumn{2}{c}{\textbf{Computational Cost}}                    \\ \cmidrule(lr){2-4} \cmidrule(lr){5-6}
                           & $\mathcal{O}$ & $\mathcal{A}$ & $\mathcal{B}$ & Propagation  & Supplement   \\ \midrule 
Gradients~\cite{simonyan2014deep} & \checkmark & \checkmark &  & Forward $+$ Backward &  \\
IG~\cite{sundararajan2017axiomatic} & \checkmark & \checkmark & \checkmark & (Forward $+$ Backward) $\times$ (\# Interpolations) &  \\
DeepLIFT~\cite{shrikumar2017learning} & \checkmark & \checkmark & \checkmark & (Forward $+$ Layer-wise Backward)*2 &  \\ \midrule
LIME~\cite{ribeiro2016should} & \checkmark &  & \checkmark\tnote{*} & Forward $\times$ (\# Neighbors) & Training LR \\
LEMNA~\cite{guo2018lemna} & \checkmark &  & \checkmark\tnote{*} & Forward $\times$ (\# Neighbors) & Training MLR with fussed lasso \\
Shapley~\cite{lundberg2017unified} & \checkmark &  & \checkmark\tnote{*} & Forward $\times$ $2^{mn}$ & Calculating Shapley values  \\
\bottomrule
\end{tabularx}
    \begin{tablenotes}
        \footnotesize
        \item[*] Implicitly specified baselines for masking features.
    \end{tablenotes}
\end{threeparttable}
\end{table*}

\noindent\textbf{Feature Attribution Explainers.}
While various FA methods have been proposed to explain different DNN architectures~\cite{selvaraju2017grad, zhou2016learning, dabkowski2017real}, model-agnostic methods are more suitable for ERDS to handle distinct risk detection classifiers~(see examples in Table~\ref{tab:explainers}). 
They can be broadly divided into (1)~gradient-based methods that propagate importance signals backward through all neurons of the network, e.g., Gradients, IG, and DeepLIFT, and (2)~perturbation-based methods that make feature perturbations while analyzing prediction change, e.g., LIME, LEMNA, and Shapley.
As the number of these methods is booming~\cite{saeed2021explainable, belaid2022we}, we introduce three factors to consider when selecting candidate explainers:
\begin{itemize}
    \item Knowledge Assumption:
    FA methods require different knowledge for the metadata set, and thus some methods would be limited when the explainer cannot have access to certain elements in $\mathcal{M}$.
    For example, gradient-based methods cannot be applied to black-box scenarios as they require full access to the DNN pipeline~\cite{jeyakumar2020can}.
    That is, in addition to output layer activation $\mathcal{O}$, they need middle layer activation set $\mathcal{A}$ of the model to accomplish the gradient calculation. 
    Many methods would also involve a baseline set $\mathcal{B}$ to provide counterfactual intuition~\cite{lundstrom2022rigorous}. For gradient-based methods, $\mathcal{B}$ is explicitly specified as a single model input $b_v$ that goes through the same propagation process as the original input $x_v$.
    For the perturbation-based method, it is implicitly a background dataset used for feature masking.
    However, baselines must be carefully chosen with domain knowledge~\cite{mamalakis2022carefully} and sometimes are sampled from the training dataset of the model.
    \item Computational Costs: the cost for explanation cannot be too heavy compared to the classification pipeline. From this perspective, gradient-based methods are usually superior to perturbation-based methods, especially for most risk detection classifiers where the feature size is large. 
    This is because gradient-based methods take a single backward pass for each input data to their approximation rule~\cite{ancona2017towards}, while perturbation-based methods sample a neighborhood around the instance, perform forward passing for all neighbors, and fit another model for feature influence estimation. 
    Typically, for LIME and LEMNA, training one linear regression model would have $O((mn)^3)$ in time complexity, and LEMNA would be extremely hard to converge with large features; for Shapley, the number of coalitions is related to feature size~\cite{yeh2022threading}, resulting in $O(2^{mn})$ complexity in forward passing.
    \item Instance-level Performance: since FA methods rely on general assumptions about the classifier's data and model, their performance is unstable among instances. 
    In other words, they will all fail to handle some corner cases~\cite{liao2021human}, for example, Gradients is found to have a saturation problem when the activation of an input is capped at zero.
    It is uncertain which method will be the most appropriate until they are evaluated on a specific risk, particularly given the fact that most FA methods are domain-general and primarily cares about data modalities like image and text. 
    LEMNA is a pioneering method proposed for security applications, and it customizes LIME to handle feature dependency and non-linear decision boundary.
    Intuitively, when working with pseudo-text data and RNN architecture, LEMNA would have a better performance than LIME.
\end{itemize}
\subsection{Understanding Risk Explanations}

\noindent\textbf{Semantic Capacity.}
\textcolor{\tcolor}{The semantic capacity of risk explanations depends on its associations to problem-space samples.} 
FA has an intrinsic limitation in that they merely explore the model and do not learn from other resources in the problem space, and thus FA explanations cannot provide risk semantics that are not encoded in the feature space.
For instance, if the malware classifier works on tabular features that encode the statistical summary of program patterns~(crafted by domain experts), then the information lost during feature extraction~\textcolor{\tcolor}{(e.g., pattern selection for Drebin~\cite{arp2014drebin} and hashing trick for Ember~\cite{anderson2018ember})} cannot be captured in explanations. 
For this type of classifiers, the semantic capacity of FA explanations is relatively low since they only reflect known patterns~(Figure~\ref{fig:mimicus-drebin-exp}), which would be too shallow for complex tasks such as malware reverse engineering. 
\textcolor{\tcolor}{To conclude, the higher the level of abstraction in feature extraction $\alpha$ is, the lower the semantic capacity of explanations $e_f$ will be.}


\noindent\textbf{Non-trivial Evaluation.}
Unlike the classification task, evaluating the explanation performance is non-trivial due to the unavailability of labels.
It is often impractical to label risk explanations on a large scale since the features are prohibitively long and require much expert knowledge to inspect.
For some black-box explainers that use surrogate models~(e.g., LR and MLR for LIME and LEMNA), evaluation can be translated into comparing the output probabilities of the surrogate model and the original model, but this approach do not generalize to other FA methods.
Therefore, a more general feature deduction-based approach is widely adopted to measure local explanation \textit{fidelity}, which is named descriptive accuracy~(DA)~\cite{warnecke2020evaluating}.
It uses the model prediction of an altered sample where the $k$ most important features are nullified in the feature space.
Let $\tau_{k}$ denote the threshold that equals to the $k$-th largest value in $e_v$, then with our formalization, the DA metric at $k$ is
\begin{equation}
\begin{aligned}
    & DA_k(x_v, e_v, f) = f(x_v^{\prime})_{[c=1]}, \\
    & \operatorname*{s.t.} x_v^{\prime}=x_v \odot (\mathbf{1}-Ind(x_v, r_v)),~~ \text{where } r_v=\omega(e_v;\tau_{k}),
\end{aligned}
\label{equ:DA}
\end{equation}
and the larger the value drops from $f(x_v)$, the better the explanation is thought to be faithful to the model~(i.e., accurate). 







\section{Scope and Motivation}


\subsection{Our Research Scope} \label{sec:scope}

\noindent \textbf{Goal.}
We aim to promote ERDS to \textit{assist security experts} in security analysis. 
The targeted ERDS differ in explanation object~(i.e., various classifiers) and application domain~(i.e., various risk types).
The desired assistance is to reduce human workload by pointing out ROI in risk samples. 
For example, the ERDS for inspecting malware bytecode should be able to localize malicious functionalities, so that analysts can make less effort in reverse engineering.  
To this end, ERDS should \textit{accurately} explain why a sample is identified as a risk by the classifier and \textit{comprehensively} convey it to security experts.

\noindent \textbf{Focus.}
We focus on ERDS that has a high semantic capacity for risk explanations.
Thereby, the classifier to be explained are data-driven as algorithmic explanations can be mapped back into raw program elements, such as the last three classifiers in Table~\ref{tab:classifiers}.

\noindent Compared with explaining pattern-driven classifiers~\textcolor{\tcolor}{(see discussion in Appendix~\ref{app:pattern})}, we work on a more \textit{challenging} problem due to the complexity of the feature space and the model design.
However, this allows ERDS to benefit more \textit{practical} scenarios, e.g., automatic detection of malicious components~(instead of pointing out predefined properties) in malware.

\noindent \textbf{Assumption.}
We assume that problem-space data samples~(e.g., code chunks, Android applications, PE binaries) can be correctly embedded into feature space~\cite{kong2013discriminant}.
That is, problems such as code obfuscation and packing can be solved with existing tools~\cite{cheng2018towards} so that the feature extraction $\alpha$ functions well.
We also assume that inputs and models are benign for the explainer, while deliberate attacks can be avoided with existing defense methods~\cite{zhang2020interpretable, zantedeschi2017efficient}.

\subsection{Why Current ERDS is Insufficient} 

\begin{figure}[tb]
\centering
\includegraphics[width=\linewidth]{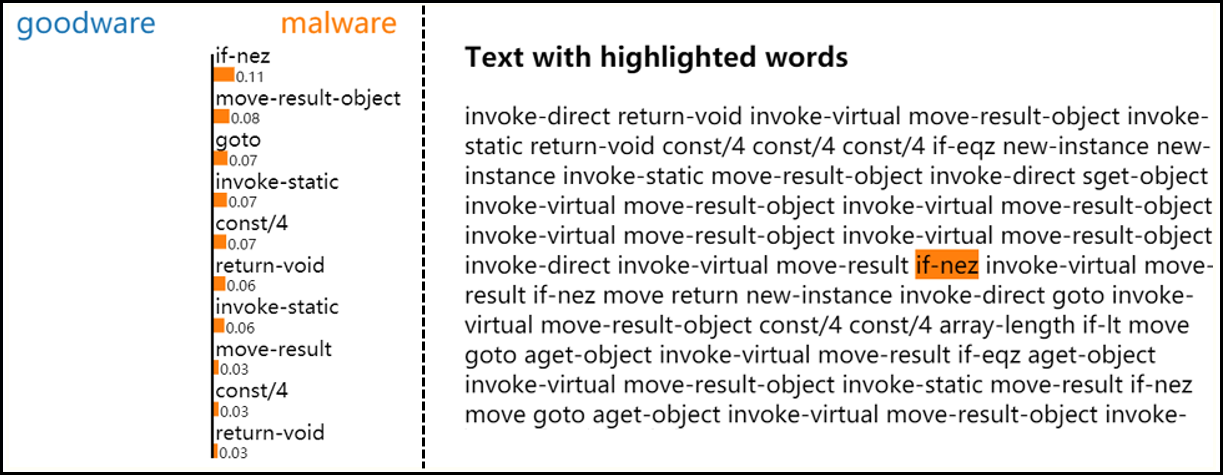} 
\caption{A showcase of the failure of current ERDS: explaining DAMD with the text explainer in LIME toolbox~\cite{lime}. The selected sample is predicted as malware with $100\%$ confidence by the classifier. For the visualized explanation, the left panel shows the top $10$ opcodes that make the sample malicious, and the right panel shows the feature-space opcode sequence with the explanation highlighted. Since the original sequence has a length of 28\,245, we display part of the results on the right panel, yet it is representative enough as the explanation is sparse throughout the sequence.}
\label{fig:motivating}
\end{figure}

Existing works build ERDS by stacking an off-the-shelf classifier with a specific explainer at hand.
Through this approach, the generated explanations usually fail to be accurate or comprehensive, making them useless for security analysis.
For example, as the explanation illustrated in Figure~\ref{fig:motivating}, the sparsely dispersed opcodes hardly provide analysts with insights about malware behaviors.
The failure can be attributed to the following two causes.

\noindent \textbf{Ill-suited Explainer for Classifier.}
Due to the post-hoc property, the explainer makes some general assumptions about the classifier, leading to unavoidable approximation error for ERDS. 
Firstly, domain-general FA methods are obviously not suitable for explaining risk detection classifiers.
As these methods are designed intuitively for image and text domain, the clear differences from ERDS in dataset distribution, data modality, and model architecture can lead to poor explanation performance. 
Secondly, although a few security-customized methods have been proposed, they require much higher computational costs and still do not fit ERDS well.
For one thing, these methods make efforts to increase the complexity of the approximation model~\cite{xnids}.
For another, they rely on observations about security applications to intuitively adapt the approximation assumption, which usually does not summarize risk detection classifiers.
For example, the statement that RNN and MLP are more widely adopted might apply to function start detection and tabular feature-based malware detection~\cite{guo2018lemna} but does not hold true for advanced risk detection.
To be fair, since there exist distinct types of security applications, making a general approximation would be extremely hard.


\noindent \textbf{Explanation on Low-level Features.}
Current ERDS follows FA to provide feature-space explanations, and little user perspective is involved to adapt the explanation level.
However, there exists a gap between model features and user understanding, especially for data-driven risk detection classifiers that work on a low abstraction level. 
As for the example in Figure~\ref{fig:motivating}, it is the opcode-level where discrete opcodes require high skills to read and cannot serve as a unit of malicious behavior.
However, existing FA applications in security are not fully aware of this issue, as much attention is paid to facilitating internal use cases like model debugging~\cite{severi2021explanation, Pirch21TagVet}. 
For rare external use cases that aim to facilitate human analysis, they study pattern-based classifiers where the feature-space elements are originally simple for understanding~(e.g., manually abstracted PDF file features for Mimicus+) yet with limited semantic capacity. 
As a matter of fact, the proper explanation level is task-specific, requiring ERDS to engage with domain knowledge.

To conclude, in order to generate useful explanations for security analysis, building ERDS should be a task-aware process with respect to its internal components and external users.
The key challenges lie in
(1)~handling the mismatch between classifier and explainer; (2)~handling the intelligibility gap between classifier and human.






\begin{figure}[tb]
    \centering
    \includegraphics[width=\linewidth]{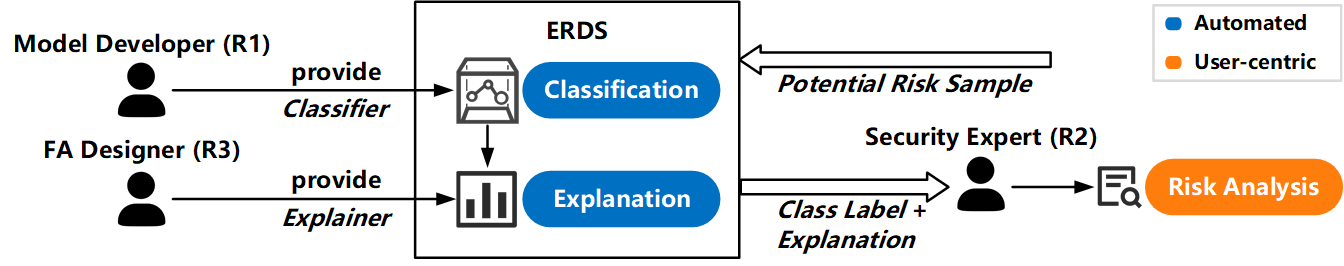}
    \caption{Different Stakeholders involved in building ERDS.}
    \label{fig:stakeholders}
\end{figure}

\begin{figure}[tb]
    \centering
    \includegraphics[width=.9\linewidth]{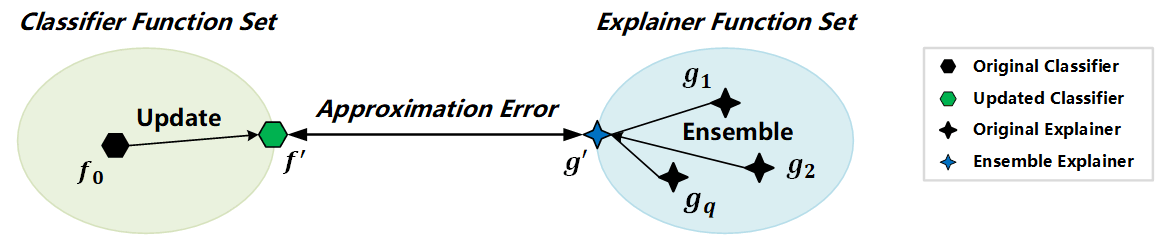}
    \caption{Insights for minimizing the approximation error.}
    \label{fig:insight}
\end{figure}

\begin{figure*}[ht]
    \centering
    \includegraphics[width=.85\textwidth]{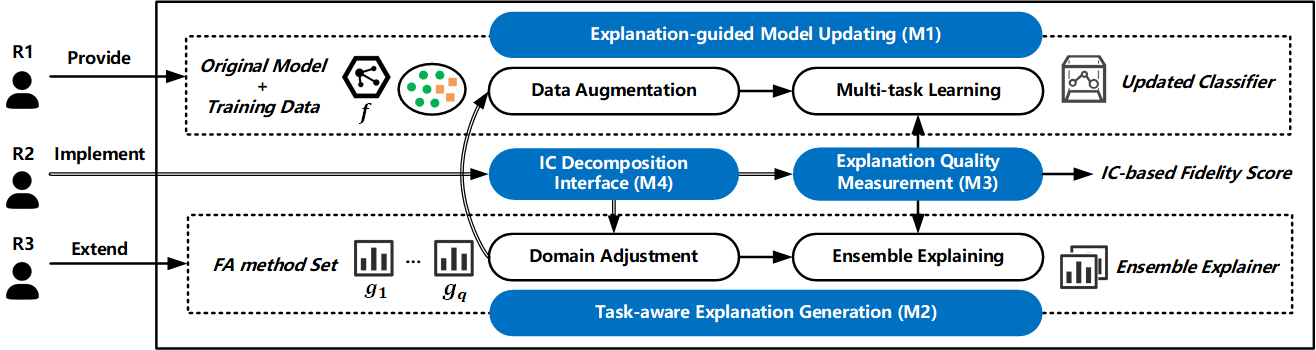}
    \caption{The overall architecture of \framework{}.}
    \label{fig:framework}
\end{figure*}

\section{\framework{} Framework}
\label{sec:methodology}

\subsection{Insights Behind Our Design}
\label{sec:insight}

To address the above challenges, our solutions are (1)~fidelity patch: updating the targeted classifier and ensembling available explainers; 
(2)~intelligibility redesign: engaging with users and generating abstracted explanations. 
The high-level idea is to improve the explainability of ERDS with joint efforts from different stakeholders.
In the following, we first identify the stakeholders with their abilities and expectations.
Then, we respectively introduce the insights behind the two solutions. 

\noindent \textbf{Identifying Stakeholders.}
As shown in Figure~\ref{fig:stakeholders}, the three main stakeholders differ in their roles in ERDS. We describe them as: 
\begin{itemize}
    \item Model Developer~(R1) who develops the classifier to accomplish a risk detection product. They have knowledge about training data and certain DL skills. For them, making the model easier to be explained would make the product more trustworthy. 
    \item Security Expert~(R2) who is the end user of the system. They are typically non-data scientists but with high domain skills. They want to leverage the automated explanation to reduce their manual work in security analysis, e.g., malware reverse engineering. 
    \item FA Designer~(R3) who design FA algorithms to encourage the development of XAI~\cite{gunning2019xai}. They are often data scientists motivated by creating a ``right to explanation"~\cite{casey2019rethinking}, but care little about the application domain of the classification and the explanation. 
\end{itemize}
Note that different roles can be played by the same identity.
Telling them apart actually helps with establishing clear desiderata~\cite{bhatt2020explainable}.
For example, under a typical scenario where R1 and R2 belong to the same security company, supporting black-box setting would no longer be desired for R3.
We highlight that our design path is different from related works: we investigate how to maximize the contributions that can be made by different stakeholders instead of leaving all the explanation burden to R3. 


\noindent \textbf{Minimizing Approximation Error.}
The essence of improving explanation accuracy is to minimize the approximation error between the exact function $f(\cdot)$ and the approximate function $g(\cdot)$.
As illustrated in Figure~\ref{fig:insight}, we find two design points in ERDS:
\begin{itemize}
    \item Fine-tuning the classifier for post-hoc explanation task. As model training can be regarded as looking for the best in a function set, the definition of the best should be revised when enforcing post-hoc model explainability. Therefore, we would like to update the model parameters of $f(\cdot)$ with multi-task learning. 
    Practitioners who have full access to the classifier and training data, e.g., R2, should be responsible to do this step. 
    \item Ensembling an FA set for local explanations. Given the diversity of FA algorithms and the uncertainty of local performance, it would be helpful to balance different attributions dynamically for different inputs. With this aim, we want to leverage a FA set $\mathcal{G} = \{g_i; i\in \mathbb{N}\}$ and weight the outputs locally. 
    With this design, explanation performance can be improved whenever new FA methods are extended into the system by R3.
\end{itemize}

\noindent \textbf{Raising Abstraction Level.}
To make a risk explanation comprehensive, we move the explanation target of ERDS to a high abstraction level that attends R2's understanding.
For example, explaining malware with malicious functions instead of discrete opcode features.
We define such an explanation as domain-space explanation $e_d \in \mathcal{D}$, which indicates the ROI on \textit{intelligible components}~(ICs).
To generate and evaluate $e_d$, our intuitions are 
\begin{itemize}
    \item Adding an IC decomposition interface. The definition of the IC varies among security analysis tasks.
    Thus, we want to leverage the skills of R2 to achieve an appropriate abstraction level.
    Considering their abilities, we hide the algorithmic details and design the interface that accepts an IC decomposition function $h:\mathcal{Z} \rightarrow \mathcal{D}$. 
    This function should generate the domain-space risk representation $x_d = h(x)$, an IC set of which the size can be variable depending on specific instances~(e.g., the unfixed number of functions in malware).
    \item Adjusting data operations with IC. 
    To move the explanation target of ERDS from $\mathcal{F}$ to $\mathcal{D}$, we switch all feature-space data operations into the domain space. 
    Typically, both the feature perturbation used by $g(\cdot)$ and the feature deduction used by DA should nullify domain-space regions $r_d \in \mathcal{D}$. 
    \textcolor{\tcolor}{We also constrain that the nullifying should be meaningful with IC in whatever space, for which we introduce the function $mask(x,r;\mathcal{B})$. 
    It would replace the specified region $r$ on $x$, i.e., $Ind(x,r)$, with regions sampled from non-risk baseline instances $\mathcal{B}$~(e.g., benign software).} 
\end{itemize}

\subsection{Overview} \label{sec:overview}

Based on the key insights, we propose a framework called \framework{} to enhance the explanation performance of ERDS.

\noindent \textbf{Architecture.}
Figure~\ref{fig:framework} shows the architecture of \framework{}.
It includes two major modules~(\textit{M1: explanation-guided model updating} and \textit{M2: task-aware explanation generation}) to minimize approximation error and two auxiliary modules~(\textit{M3: explanation quality measurement} and \textit{M4: IC decomposition Interface}) to raise abstraction level. 
For classification, M1 accepts the DNN model and training data of the original classifier to produce an updated classifier that has high model interpretability.
For explanation, M2 uses an extensible FA algorithm set, incorporates application-specific knowledge about IC, and combines multiple results to get an ensemble explainer that generates optimal explanations.
Being invoked by the major modules, M3 and M4 constitute IC-based fidelity metrics, which are used in M1 to measure the explanation task for updating and in M2 to assign explanation weights for ensembling.


\noindent 
\textcolor{\tcolor}{\textbf{Roadmap.}
The remainder of this section proceeds as follows.
Section~\ref{sec:M1}~(M1) presents the introduction and optimization of the explanation task in fine-tuning; Section~\ref{sec:M2}~(M2) introduces the adjustment of data operations and the local weights for different explainers; 
Section~\ref{sec:M3}~(M3) provides the explanation evaluation approaches in the context of ERDS with the new framework. 
Section~\ref{sec:M4} addresses deployment problems such as user instructions for the interface~(M4) and the applicable scope and functionalities of the framework.
}

\subsection{Explanation-guided Model Updating}
\label{sec:M1}

This module fine-tunes $f(\cdot)$ with multi-task learning, where the model jointly learns how to classify and how to explain. 
\textcolor{\tcolor}{
For the explanation task, the label-based regularization~\cite{ross2017right} would be impractical considering the challenge of labeling risk explanations, e.g., identifying malicious functions for the whole training dataset.
To overcome this problem, we adopt the idea from self-supervised learning~\cite{chen2020simple}, which novelly introduces explanations into data augmentation and then regularizes model predictions of the augmented samples. 
Specifically, given a training batch of data $\mathbf{X}$ and labels $\mathbf{Y}$, three new sets of data are constructed from positive risk samples with a surrogate explainer $g_{train}(\cdot)$.
To update model parameters $\theta$, the prediction of the original sample set and the three constructed sample sets are treated as different tasks with different aims.}

\noindent \textbf{Data Augmentation.}
With the current model state fixed, the ROI for a batch would be $\mathbf{R} = \omega(g_{train}(\mathbf{X}); \tau_{k})$.
\textcolor{\tcolor}{
Since instances that are predicted as risks by the model have non-zero ROI, we utilize them to create new tasks. 
Those positive instances can be divided into true positives~(TPs) and false positives~(FPs).
For TPs, the ROI is associated with true risky behaviors, while for FPs, it indicates why the model regards them as outliers from the training data.
Therefore, the three new data sets are constructed as follows.
\begin{itemize}
    \item Sanitized risk set: ROI is nullified in TPs that
    \begin{equation}
        \mathbf{X_{san}} = \{mask(\mathbf{X}_i, \mathbf{R}_i) ; f_\theta(\mathbf{X}_i)=1, \mathbf{Y}_i=1\}.
    \end{equation}
    \item Variant risk set: non-ROI in TPs is exchanged that
    \begin{equation}
        \mathbf{X_{var}} = \{mask(\mathbf{X}_i, \mathbf{X}_i - \mathbf{R}_i) ; f_\theta(\mathbf{X}_i)=1, \mathbf{Y}_i=1\}.
    \end{equation}
    \item Counter example set: ROI or non-ROI is nullified in FPs that 
    \begin{equation}
        \begin{aligned}
            \mathbf{X_{cou}} & = \{mask(\mathbf{X}_i, \mathbf{R}_i) ; f_\theta(\mathbf{X}_i)=1, \mathbf{Y}_i=0\} \\
            & \cup \{mask(\mathbf{X}_i, \mathbf{X}_i - \mathbf{R}_i) ; f_\theta(\mathbf{X}_i)=1, \mathbf{Y}_i=0\}.
        \end{aligned}
    \end{equation}
\end{itemize}
For the function $mask(\cdot)$, the baselines $\mathcal{B}$ are omitted~(as well as in the rest of the paper) for brevity, and they would be true negatives among the data batch in the context of fine-tuning.}
For the surrogate explainer, $g_{train}(\cdot)$ can be implemented with any FA method since all metadata about the classifier is available; we use Gradients that naturally constrain the decision boundary and has the least computational cost.
\textcolor{\tcolor}{We explain the intuition of the three augmentation approaches with malware classification: the generation of $\mathbf{X_{san}}$ can be regarded as malicious code removal, the generation of $\mathbf{X_{var}}$ can be considered as piggybacking malicious payload into goodware, and the generation of $\mathbf{X_{cou}}$ mimics the process of implementing different benign utilities.}

\noindent \textbf{Multi-task Learning.}
Given the original sample set and the three new sample sets constructed by explanation-guided data augmentation, our aim is to ensure high accuracy on the original classification task and to achieve high fidelity on additional explanation tasks.
Specifically, for the explanation tasks, we want to make the prediction probability of the positive class drop for sanitized risk set while holding for the other two, compared with their origins~(TPs for $\mathbf{X_{san}}$ and $\mathbf{X_{var}}$, and FPs for $\mathbf{X_{cou}}$).
Therefore, multi-task learning would be formulated as the following optimization problem
\begin{equation}
\begin{aligned}
    &\operatorname*{argmin}_{\theta^*} \mathcal{L}_0(\mathbf{X}, f_\theta, \mathbf{Y}) 
    + \lambda_1\mathcal{L}_1(\mathbf{X_{san}}, \mathbf{X}, f_\theta; \mathbf{Y=1}) \\
    & \quad + \lambda_2\mathcal{L}_2(\mathbf{X_{var}}, \mathbf{X}, f_\theta; \mathbf{Y=1}) 
    + \lambda_3\mathcal{L}_2(\mathbf{X_{cou}}, \mathbf{X}, f_\theta; \mathbf{Y=0}), \\
    & \operatorname*{s.t.} \theta^* \subseteq \theta,
\end{aligned}
\label{equ:loss0}
\end{equation}
where the first item $\mathcal{L}_0$ is the loss function of the original classifier that deals with supervised binary classification~(typically cross entropy loss); the last three items measure the relative ``drop" and ``hold" with $\mathcal{L}_1$ and $\mathcal{L}_2$, and they are respectively balanced by weight coefficients $\lambda_1$, $\lambda_2$, and $\lambda_3$; the constraint means the model parameters are partially unfrozen as $\theta^*$.

Nevertheless, the optimization objective is difficult to minimize due to the measurement of relative values.
For one thing, the two items compared are both related to $\theta^*$.
For another, the results can have high variance among different instances considering the bias brought by both predictions and explanations.
Therefore, as the origins of a constructed sample set belong to the same class, we simplify the problem by classifying sanitized samples as the opposite class while others as their original classes.
The last three items in Equation~\ref{equ:loss0} are replaced with:
\begin{equation}
    \lambda_1 \mathcal{L}_0(\mathbf{X_{san}}, f_\theta, \mathbf{0}) + \lambda_2 \mathcal{L}_0(\mathbf{X_{var}}, f_\theta, \mathbf{1}) + \lambda_3 \mathcal{L}_0(\mathbf{X_{cou}}, f_\theta, \mathbf{0}).
\label{equ:loss1}
\end{equation}
Then, we can use any standard gradient-based optimization method to solve it~\cite{kingma2014adam}. 
To understand the three modified items better, the intuition is that the first two constrain the model's decision-making process with more boundary values and the last one broadens the model's horizon with unexplored normal spaces.

\begin{algorithm}[t]
\DontPrintSemicolon
\KwData{input $x$, classifier metadata $\mathcal{M}$, explainer set $\left\langle \mathcal{G}, \omega \right\rangle$; \\
\noindent \,\,\,\, transformation functions $(h, \alpha, \varphi)$, workload expectation $k$}
\KwResult{domain-space explanation $e_d$.}
Get $\mathbf{I}$, $\mathcal{I}$ with $get\_ic\_indicator($x$, h, \alpha, \varphi)$  \Comment*[r]{IC Indicator}
Init $\mathbf{w}$ to $[0]_{\vert\mathcal{G}\vert \times 1}$ \Comment*[r]{Explainer Weight}
Init explanations $\mathbf{E}$ to $[0]_{\vert\mathcal{G}\vert \times \vert\mathcal{I}\vert}$, current index $i$ to $0$ \;
\For{each FA method $g \in \mathcal{G}$}{ 
    \uIf(\Comment*[f]{Domain Adjustment} ){$g$ is black-box}{ 
        Get $g_\mathbf{I}$ by modifying data operations in $g$ with $\mathbf{I}$ \;
        Get IC attributions $e_{v^\prime} \in \mathbb{R}^{\vert\mathcal{I}\vert}$ with $g_\mathbf{I}$\;
    }
    \uElse{
        Get importance score $e_v$ with $g$ \;
        Get IC attributions $e_{v^\prime} \in \mathbb{R}^{\vert\mathcal{I}\vert}$ with $e_v$ and $\mathbf{I}$ \;
    }
    Assign $normalize(e_{v^\prime})$ to $\mathbf{E}[i]$    \Comment*[r]{Ensemble Explaining}
    Get score $s$ by inputting $(\mathbf{x}_v, e_{v^\prime}, f, k, \mathbf{I})$ to M3\;
    Assign $s$ to $\mathbf{w}[i]$; Add $1$ to i \;
}
Get $e_{v^\prime} \in \mathbb{R}^{\vert\mathcal{I}\vert}$ with $\mathbf{E}^\intercal \cdot normalize(\mathbf{w})$ \;
Get $e_d$ with $\omega(e_{v^\prime}; \tau_{k})$ defined on $\mathcal{I}$\;
\Return $e_d$\; 
\caption{Task-aware explanation generation}
\label{algorithm_1}
\end{algorithm}

\subsection{Task-aware Explanation Generation}
\label{sec:M2}

\textcolor{\tcolor}{
The explanation generation process with \framework{} is described in Algorithm~\ref{algorithm_1}, which can be divided into two steps.
First, domain adjustment customizes individual FA methods in $\mathcal{G}$ to generate IC-based attributions~(Lines~5-10).
Then, the ensemble explaining leverages the IC-based local fidelity scores to weight different attributions~(Lines~11-15).
}

\textcolor{\tcolor}{
Compared with the original explanation pipeline in Definition~\ref{def:explainer}, the newly involved data include the transformation functions and the workload expectation. 
First, the transformation functions $(h, \alpha, \varphi)$ are used to generate the local \textit{IC indicator} $(\mathbf{I}, \mathcal{I})$ where $\mathbf{I} \in \mathcal{V}$ and $\mathcal{I} \in \mathcal{D}$~(see the generation process in Appendix~\ref{app:interface}), which supports domain-space data operations on vector-space inputs. 
Specifically, the $ij$-th element of the vector-space indicator object $\mathbf{I}$ maps to the IC index in the domain-space indicator object $\mathcal{I}$ that $(x_v)_{ij}$ belongs to; then when $x\in\mathcal{V}$ and $r\in\mathcal{D}$, the implementation of $Ind(x,r)$ would be checking whether $(\mathbf{I})_{ij}$ exists in $r$ denoted with IC indexes, which would be leveraged by data operations such as the $mask(\cdot)$ function.
To highlight the cross-domain data operations, we will add the subscript $\mathbf{I}$ to related functions in this section.
Second, the workload expectation $k \in \mathbb{N}^+$ determines the number of ICs that should be nullified for fidelity measurement, and then the local \textit{explainer weight} $\mathbf{w} \in \mathbb{R}^{\vert \mathcal{G} \vert}$ can be generated for ensembling.
}



\noindent \textbf{Domain adjustment.}
To generate the domain-space attribution vector that has a size of $\abs{\mathcal{I}}$~\textcolor{\tcolor}{(Lines~5-10)}, the adjustment for FA algorithms are divided into two cases in terms of the data operation used in their approximation approaches.
As shown in Table~\ref{tab:fa-decompose}, gradient-based methods explicitly work in $\mathcal{V}$ where gradients should be calculated for inputs that can propagate in the model neurons, but for perturbation-based methods, vector-space representations are not strictly required by the approximation rules.
That is, for LIME and LEMNA, the regression models can accept vectors of arbitrary length and the output size complies with $x_n \in neighbor(x)$; for Shapley, the cooperative game theory uses arithmetic operations that work on any set of coalition predictions and generates attributions with a length of $\abs{\mathbf{X}_c}$ where $\mathbf{X}_c=\{ coalition(x_i);i \leq \vert x \vert \}$.
Therefore, for perturbation-based methods, we switch the neighborhood sampling function and the coalition enumeration function into $\mathcal{D}$\textcolor{\tcolor}{~(Line 7), i.e., $neighbor_{\mathbf{I}}(\cdot)$ and $coalition_{\mathbf{I}}(\cdot)$}, by (1)~performing random sampling and subset permutation with $\mathcal{I}$ to get the IC indexes that should be nullified and (2)~sending those indexes to \textcolor{\tcolor}{$mask_{\mathbf{I}}(\cdot)$} as the second parameter. For gradient-based methods, we generate the $i$-th IC attribution with $\norm{e_v \odot Ind(x_v, \mathbf{I}_{[i]})}_{1}$~\textcolor{\tcolor}{(Line 10)}. 
Note that the step of querying $\mathcal{O}$~(with $x_n$ and $\mathbf{X}_c$) still involves vector-space representations, and that is when we should translate these operations into vector space with the indicator matrix. 

\begin{table}[t]
\vspace{.45cm}
    \caption{Data operations inside different FA methods.}
    \label{tab:fa-decompose}
    \centering
    \begin{tabularx}{\linewidth}{Ccc} 
    \toprule
    Explainer & Data & Rule \\ \midrule
    Gradients & $\mathbf{x}_v, \mathcal{A}, \mathcal{O}$ & Partial gradients \\
    IG & $\mathbf{x}_v \sim b_v, \mathcal{A}, \mathcal{O}$ & Accumulated gradients \\
    DeepLIFT & $\mathbf{x}_v, b_v, \mathcal{A}, \mathcal{O}$ & Discrete gradients \\ \midrule
    LIME & $neighbor(x), \mathcal{O}$ & Model coefficients \\ 
    LEMNA & $neighbor(x), \mathcal{O}$ & Model coefficients \\
    Shapley & $\{ coalition(x_i);i \leq \vert x \vert \}, \mathcal{O}$ & Averaged differences 
    \\ \bottomrule
    \end{tabularx}
\vspace{-.2cm}
\end{table}

\noindent \textbf{Ensemble Explaining.}
Given multiple IC attributions $e_{v^\prime} \in \mathcal{S} \subset \mathbb{R}^{\vert \mathcal{I} \vert}$, we generate their weights $\mathbf{w}$~\textcolor{\tcolor}{(Lines~11-13)} by evaluating explanation fidelity with the module M3.
To balance the effect of different IC attributions $e_{v^\prime}$, we need a metric that indicates higher IC-based fidelity with larger values, for which we define the model prediction drop~(MPD) at $k$ as 
\begin{equation}
\label{equ:mpd}
    \begin{aligned}
        & MPD_{k}(x_v, e_{v^\prime}, f) = \left(f(x_v) - f(x_v^\prime)\right)_{[c=1]}, \\
        & \operatorname*{s.t.} x_v^\prime = \textcolor{\tcolor}{mask_{\mathbf{I}}(x_v, r_d)} \, where \, r_d = \omega(e_{v^\prime}; \tau_{k}).
    \end{aligned}  
\end{equation}
Besides the high-is-better property, MPD is different from DA in (1)~refined meaning: describing the relative drop by establishing the connection to original predictions~(that are different among instances), and (2)~modified nullifying: translating IC deduction into vector space instead of setting separate elements to zero~(as in Equation~\ref{equ:DA}). 
Additionally, there are two normalizing steps that act on the IC attribution $e_{v^\prime}$ and the explainer weight $\mathbf{w}$, respectively: the first reduces the variance of attribution distributions among different explainers~\textcolor{\tcolor}{(Line~11)}; the second adjusts the weight vector to sum up to one after clipping negative values~\textcolor{\tcolor}{(Line~14)}.

\subsection{Explanation Quality Measurement}
\label{sec:M3}

\begin{figure*}[tb]
\myfigureshrinker
\centering
    \subfloat[DAMD]{%
        \includegraphics[width=0.26\textwidth]{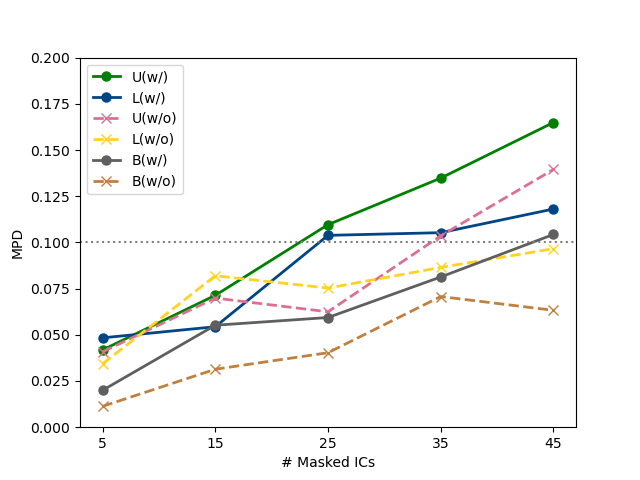}%
        \label{fig:damd}%
        }
    \subfloat[DR-VGG]{%
        \includegraphics[width=0.26\textwidth]{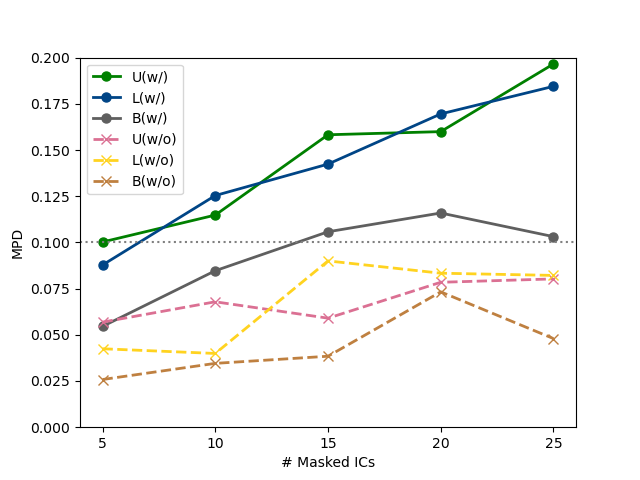}%
        \label{fig:dr}%
        }
    \subfloat[VulDeePecker]{%
        \includegraphics[width=0.26\textwidth]{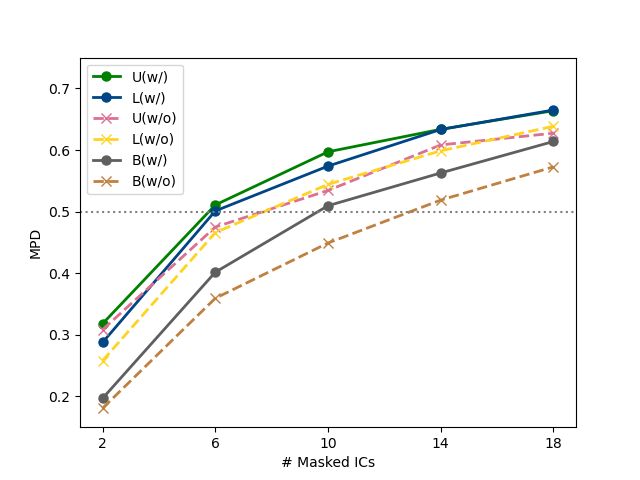} 
        \label{fig:vuldeepecker}%
        }%
    \caption{IC deduction test for ERDS with~(w/) and without~(w/o) \framework{}: model prediction drop~(MPD) measured at different number of masked ICs. The higher MPD means higher explanation fidelity of the system.}
    \label{fig:sys_fidelity}
\vspace{.3cm}
\end{figure*}

We have introduced the internal usage of M3 in the above sections.
The explanation fidelity measurements involved in the two major modules are summarized as: 
(for M1)~the loss item $\mathcal{L}_1$ in Equation~\ref{equ:loss1} that measures how well the \textit{classifier} can predict sanitized samples as the flipped class;
(for M2)~the MPD in Equation~\ref{equ:mpd} that quantifies how closely an \textit{explainer} can approximate the model prediction locally. 
In this section, we introduce two scenarios where M3 can be used as a standalone module.
\begin{itemize}
    \item \textit{Model interpretability} evaluation: to select the model function for a certain risk detection classifier trained with different configurations. It performs IC deduction on a test dataset $X$ with a deduction percentile $p \in \mathbb{R}$~(where the threshold is denoted as $\tau_{p}$) and a candidate FA algorithm set $\mathcal{G}$. 
    To calculate the model interpretability score, samples are altered with explanations from all possible explainers, and the metric named average model prediction~(AMP) is defined as $\frac{1}{\abs{\mathcal{G}}\abs{X}}\sum_{i=1}^{\abs{\mathcal{G}}} \sum_{j=1}^{\abs{X}} f(x_j^\prime; x_j, g_i, \tau_{p})$.
    \item \textit{Global fidelity} evaluation: to prioritize better explainers in $\mathcal{G}$ when the task-specific workload $k \in \mathbb{N}^+$~(e.g., the number of functions in the explanation that is acceptable for analyst to read) is known. For the test dataset $X$, it firstly filters out the samples with an IC size no greater than $k$, resulting in $X_k=\{x;\abs{\mathcal{I}_x}>k, x\in X\}$.
    Then, the score for each (ensemble) explainer is the average MPD on remained samples, i.e., $\frac{1}{\abs{X_k}}\sum MPD \circ g(X_k; \tau_{k})$.
\end{itemize}
Different from feature deduction, IC deduction should take the variant IC size into concern. 
Thus, the percentile-based approach is used for model interpretability evaluation where the downstream explanation task is unknown and all test samples are considered as explanation targets;
the traditional number-based approach is used for global fidelity evaluation with the aim of reducing the human workload on samples that are tough for analysis.
Typically, users should sample the percentiles or numbers within an acceptable workload range for their evaluation trials~(see examples in Figure~\ref{fig:sys_fidelity} and Figure~\ref{fig:fidelity}), to find the best model function or explainer set. 






\subsection{Framework Deployment} \label{sec:M4}

\noindent 
\textcolor{\tcolor}{\textbf{Interface Instructions.}
As declared in Section~\ref{sec:insight}, the interface M4 should receive task-specific implementations of the function $h(\cdot)$, which are essential for intelligibility improvement on those data-driven classifiers.
To implement it, R2 typically needs to provide external tools that can extract ICs from certain types of risks.
For example, if IC is defined as the functionality for malware analysis, then to identify function boundaries, Androguard~\cite{androguard} and BinaryNinja~\cite{binaryninja} can be utilized for App and PE binaries processed by DAMD and DR-VGG, respectively;
for vulnerability analysis with VulDeePecker, if IC is defined as tokens including APIs, variables, and operators, then a lexical analysis tool would serve for the decomposition. 
Precisely defining the IC is beyond the scope of the present paper, and it depends on the context in which R2 wants for their analysis tasks.
To be more specific, the local feature space can be abstracted into different collections, and for some collections that are confirmed to be unimportant by experts, they can also be merged together in $h(\cdot)$.
As $h(\cdot)$ that indicates the definition of IC, $k$ that depends on the complexity of IC analysis is also task-specific.  
For instance, inspecting $k$ tokens in source code~(e.g., for VulDeePecker) would be much easier than inspecting the same amount of functions in disassembled languages~(e.g., for DAMD).
To this end, the complete description of the interface can be given as Listing~\ref{lst:interface}.
}

\noindent 
\textcolor{\tcolor}{\textbf{Applicable Scope.}
Technically, \framework{} is a general framework that applies to the formalized classifier and explainer whatever the application is.
The main modules would function properly with and without the involvement of R2, i.e., the methods of \texttt{\$h\$} and \texttt{set\_\$k\$} in Listing~\ref{lst:interface} can have default implementations.
Nevertheless, if features should be abstracted into different collections for intelligibility improvement, the feature engineering methods should work on independent problem-space objects so that it will apply to the individual IC decomposed by $h(\cdot)$~(as Line~6 in Algorithm~\ref{alg:ic}).
This is the actual case for the data-driven classifiers that we focus on, e.g., with features extracted from individual tokens, opcodes, and basic blocks.
Note that for classifiers with the feature extraction $\alpha$ that is a piece-wise function, for example, defined on different sections of PE files~(e.g., \texttt{.text}, \texttt{.data}, and \texttt{.bss}), the proposed IC abstraction method would still work if the $\alpha$ over any interval is data-driven.
}

\noindent \textbf{Functionality.}
With \framework{}, an enhanced ERDS would be made up of the updated classifier and the ensemble explainer, where $f(\cdot)$ and $g(\cdot)$ are optimized towards high-quality explanations for given tasks.  
Besides the class label and the explanation, it outputs quantitative scores for end users to perceive the explanation quality of ERDS and to determine how much they can rely on the explanations during security analysis. 
Besides working jointly, different stakeholders can benefit from the decoupled architecture of \framework{} to accomplish different use cases on their own. 
For example, R1 can leverage M1 to release a highly interpretable model without the leakage of training data and the concern about downstream explanation methods; R3 can utilize M2 to refine the performance of FA algorithms while task-specific explanation desiderata~(abstraction level and maximum workload) is left for R2 to specify.

\begin{figure}[tbp]
\vspace{-.3cm}
    \centering
    \includegraphics[width=.9\linewidth]{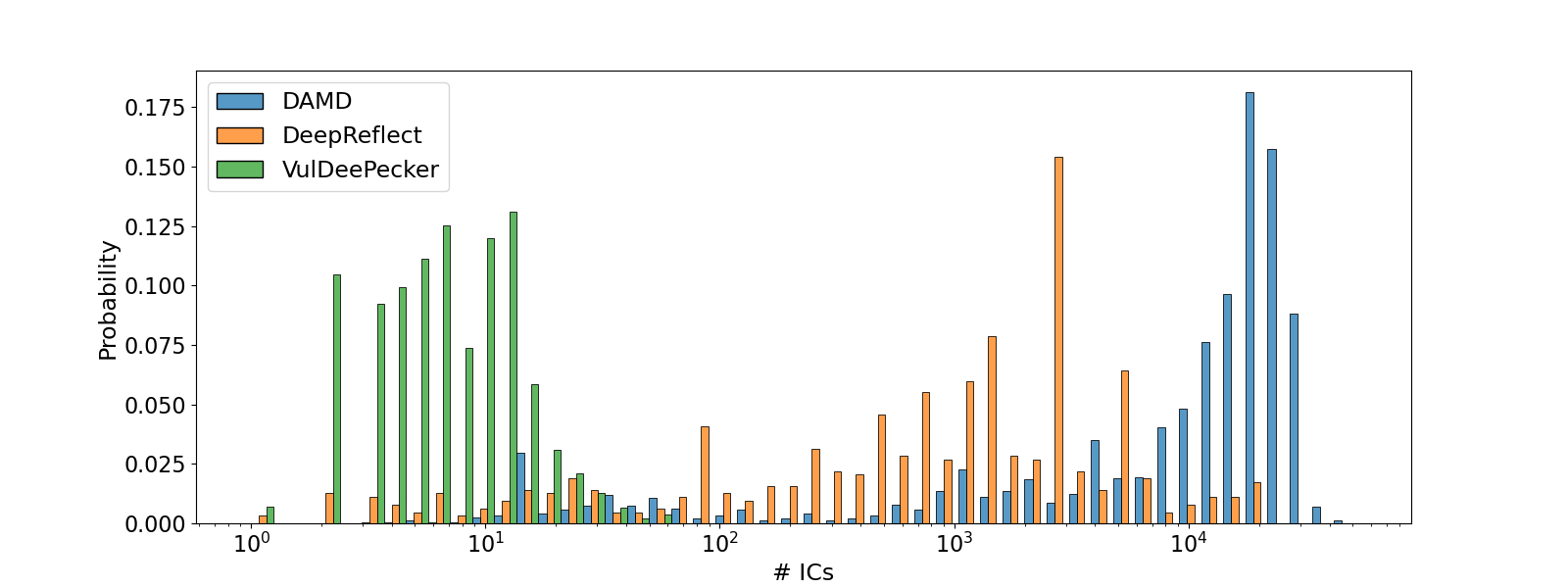}
    \caption{The distribution of IC size for each dataset.}
    \label{fig:seg_info}
\end{figure}

\lstinputlisting[caption=\textcolor{\tcolor}{Description of the IC decomposition interface}, label={lst:interface}, style=java]{interface.java}




\begin{table*}[tb]
\centering
\caption{Explanation fidelity improvement of M1 for separate explainers: MPD evaluated at a constant value of $k$, which is $25$ for the two malware detection systems and $10$ for the vulnerability detection system.}
\label{tab:fidelity}
\begin{threeparttable}
\begin{tabularx}{.98\linewidth}{*{10}C}
\toprule
\multicolumn{1}{c}{\multirow{2}{*}{\textbf{Explainer}}} & \multicolumn{3}{c}{\textbf{DAMD}} & \multicolumn{3}{c}{\textbf{DR-VGG}} & \multicolumn{3}{c}{\textbf{VulDeePecker}} \\ \cmidrule(r){2-4} \cmidrule(r){5-7} \cmidrule(r){8-10} 
\multicolumn{1}{c}{} & \textbf{w/o} & \textbf{w/} & \textbf{Improves} & \textbf{w/o} & \textbf{w/} & \textbf{Improves} & \textbf{w/o} & \textbf{w/} & \textbf{Improves} \\ \midrule
Gradients &
  0.0985 &
  0.1193 &
  \cellcolor[HTML]{EFEFEF}21.09\% &
  0.0770 &
  0.1395 &
  \cellcolor[HTML]{EFEFEF}81.03\% &
  0.3553 &
  0.4309 &
  \cellcolor[HTML]{EFEFEF}21.28\% \\
IG &
  \textbf{0.1298} &
  \textbf{0.1588} &
  \cellcolor[HTML]{EFEFEF}22.35\% &
  \textbf{0.1025} &
  \textbf{0.1865} &
  \cellcolor[HTML]{EFEFEF}82.05\% &
  0.3945 &
  0.4752 &
  \cellcolor[HTML]{EFEFEF}20.44\% \\
DeepLIFT &
  0.1176 &
  0.1320 &
  \cellcolor[HTML]{EFEFEF}12.27\% &
  0.0858 &
  0.1514 &
  \cellcolor[HTML]{EFEFEF}76.44\% &
  \textbf{0.5497} &
  \textbf{0.5719} &
  \cellcolor[HTML]{EFEFEF}4.04\% \\ 
\midrule
LIME &
  \textbf{0.1188} &
  \textbf{0.1297} &
  \cellcolor[HTML]{EFEFEF}9.16\% &
  0.0729 &
  0.1077 &
  \cellcolor[HTML]{EFEFEF}47.75\% &
  0.3446 &
  0.3596 &
  \cellcolor[HTML]{EFEFEF}4.36\% \\
LEMNA &
  0.0917 &
  0.0981 &
  \cellcolor[HTML]{EFEFEF}6.99\% &
  \textbf{0.0775} &
  \textbf{0.1111} &
  \cellcolor[HTML]{EFEFEF}43.47\% &
  0.2484 &
  0.2714 &
  \cellcolor[HTML]{EFEFEF}9.28\% \\
Shapley &
  0.1188 &
  0.1266 &
  \cellcolor[HTML]{EFEFEF}6.62\% &
  0.0626 &
  0.0916 &
  \cellcolor[HTML]{EFEFEF}46.49\% &
  \textbf{0.4856} &
  \textbf{0.5003} &
  \cellcolor[HTML]{EFEFEF}3.04\%
 \\ \bottomrule
\end{tabularx}
    \begin{tablenotes}
    \footnotesize
        \item[1] w/o: the original classifier trained without \framework{}.
        \item[2] w/: the new classifier updated with \framework{}.
    \end{tablenotes}
\end{threeparttable}
\vspace{-.3cm}
\end{table*}

\section{Systematical Evaluation}
\label{sec:eval}

\textcolor{\tcolor}{In this section, we quantitatively evaluate the performance of \framework{} in different ERDS applications.
The performance is systematically analyzed in terms of model intelligibility, model accuracy, explanation fidelity, and system efficiency.}

\subsection{Experimental Setup}

\noindent\textbf{Classifier and Dataset.}
We study the three data-driven classifiers described in Table~\ref{tab:classifiers}.
For DR-VGG, we use the ACFG+ features and the simplified VGG19 model proposed in \cite{downing2021deepreflect}.
For DAMD and VulDeePecker, we use the same implementation of model architectures as \cite{warnecke2020evaluating}.
The dataset for vulnerability detection and PE malware detection all follow their original papers, which consist of $23,307$/$36,396$ benign/malicious PE files and $29,313$/$10,444$ safe/vulnerable code gadgets, respectively.
For Android malware detection, since the original dataset is too small and too old, we make efforts to collect a new dataset, consisting of $12,807$/$4,742$ benign/malicious Android applications between the year $2017$ and $2019$. 
We split each dataset into training and testing sets with $80/20$ ratio.

\noindent\textbf{Candidate Explainers.}
We use all the explainers listed in Table~\ref{tab:explainers}.
We implement Gradients and IG in accordance to their original papers, and the number of interpolations is set to $64$ for IG. 
We make use of the SHAP toolbox~\cite{lundberg2017unified} for DeepLIFT and Shapley, where the specific class is called DeepExplainer and PartitionExplainer. 
For LIME, we use the open-source code from its authors, and for LEMNA, we override LIME's implementation by replacing the surrogate model, where the regression problem with the fussed lasso is solved with CVXPY package~\cite{diamond2016cvxpy}. 
The number of neighborhood samples for these two explainers is set to $1,000$ and the number of mixture models in LEMNA is set to $3$.

\noindent \textbf{Enhancement with \framework{}.}
For the definition and extraction of IC, we follow the example instructions in Section~\ref{sec:M4}.
To determine the ROI during data augmentation, we choose the threshold $\tau_p$ for each sample according to the top percentile $p$ of its IC attributions, 
which is $0.2\%$ for DAMD, $2.5\%$ for DR-VGG, and $10\%$ for VulDeePecker.
The different choices refer to the IC size distributions of the dataset as in Figure~\ref{fig:seg_info}:
for the two malware classifiers with large IC size, the percentile results in around $25$ functions on average~(close to the highlighted function numbers in \cite{downing2021deepreflect}), and for the vulnerability classifier, it ensures $1$ token is returned naturally on average~(for the cases where $l*r < 1$, we enforce one IC in the ROI).

\subsection{System Performance Analysis} \label{sec:evaluate-system}

\textcolor{\tcolor}{We first perform end-to-end performance analysis, and then conduct ablation experiments to demonstrate the effectiveness of the two modules, i.e., model updating and explainer ensembling.}
For each of the three applications, we compare \framework{} with the baseline system, where the classifier is the original one without updating, and the explainer uses a naive ensembling strategy where multiple explanations are summed up.
We also consider three explanation scenarios: 
(1)~Black-box where the three perturbation-based explainers that require no knowledge about $\mathcal{A}$ are used in ensembling;
(2)~Low-cost where the three gradient-based explainers with lower computational costs are used in ensembling; 
(3)~Unlimited where all the explainers are used in ensembling.
\textcolor{\tcolor}{Throughout the experiments, we use the evaluation strategies and metrics defined in the measurement module~(Section~\ref{sec:M3}) to evaluate model interpretability and explanation fidelity.}

\subsubsection{End-to-end Performance}
\textcolor{\tcolor}{We show the effectiveness of \framework{} by performing the global fidelity evaluation with different IC deduction values. 
The results are illustrated with the $k$-MPD curve in Figure~\ref{fig:sys_fidelity}.}
As in the figure, \framework{} enhances the explanation fidelity of ERDS in all three applications under different explanation scenarios.
We observe that the solid curve always has a higher average MPD score than the dashed curve in its corresponding scenario.
The number of ICs that needs to be nullified to achieve an MPD of $0.1$ is $22/35$, $25/45$, $44/\infty$ for the DAMD system with and without \framework{} in different scenarios, which means at least $37.1\%$ security analysis workload can be reduced for R2.
The enhancement with \framework{} is the most obvious on DR-VGG, where all solid curves are above the dashed curves.
We see that even black-box ensembling of \framework{} performs better than all the ensembling scenarios without \framework{}.
For VulDeePecker that works on short high-level features, the enhancement is stable in that solid curves are above their corresponding dashed curves at all values of masked IC numbers.
For an MPD score at $0.5$ where all prediction labels can be flipped, the number of saved manual workloads reaches $24.3\%$ on average.


\begin{figure*}[tb]
\myfigureshrinker
\centering
    \subfloat[DAMD]{%
        \includegraphics[width=0.28\linewidth]{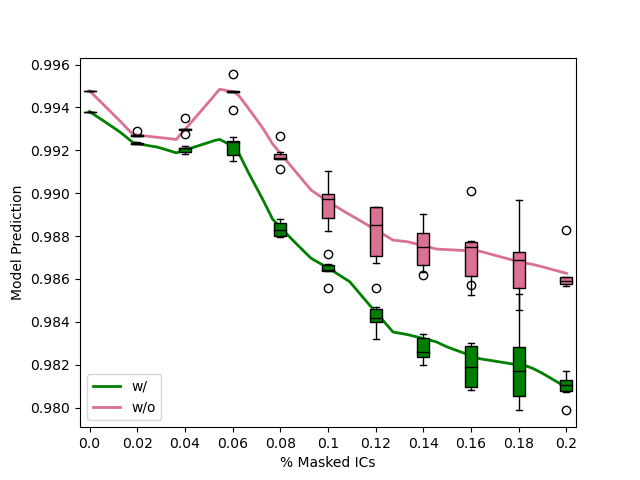}%
        \label{fig:7-a}%
        }
    \subfloat[DR-VGG]{%
        \includegraphics[width=0.28\linewidth]{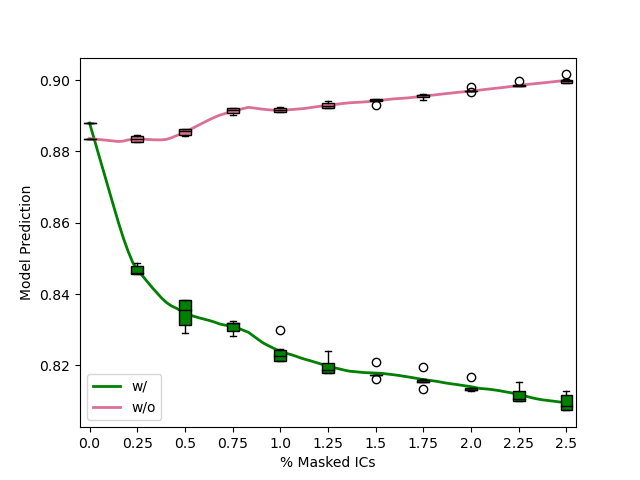}%
        \label{fig:7-b}%
        }
    \subfloat[VulDeePecker]{%
        \includegraphics[width=0.28\linewidth]{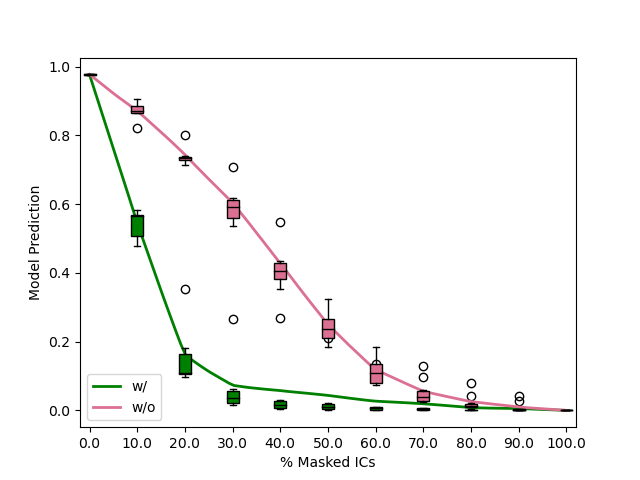} 
        \label{fig:7-c}%
        }%
    \caption{IC deduction test for classifiers with~(w/) and without~(w/o) M1 updating: average model prediction~(AMP) measured at different percentiles of masked ICs. The values on the curve are the average risk probabilities of different tests with the six candidate explainers, where their individual values are shown on the box plots. The more AMP drops, the higher the model interpretability is.}
    \label{fig:fidelity}
\vspace{.3cm}
\end{figure*}

\begin{table*}[tb]
\caption{Explanation fidelity improvement of M2. The experimental setting is the same with Table~\ref{tab:fidelity} and the classifier is the one w/ \framework{}.}
\label{tab:ensemble_fid}
\begin{threeparttable}
\begin{tabularx}{.98\linewidth}{*{10}C}
\toprule
 \multirow{2}{*}{\textbf{Scenario}} & \multicolumn{3}{c}{\textbf{DAMD}} & \multicolumn{3}{c}{\textbf{DR-VGG}} & \multicolumn{3}{c}{\textbf{VulDeePecker}} \\ \cmidrule(r){2-4} \cmidrule(r){5-7} \cmidrule(r){8-10} 
 & \textbf{w/o}   & \textbf{w/ }  & \textbf{Improves}  & \textbf{w/o}    & \textbf{w/}   & \textbf{Improves}   & \textbf{w/o}      & \textbf{w/ }    & \textbf{Improves}     \\ \midrule
Black-box &
  0.1235 &
  0.1362 &
  \cellcolor[HTML]{EFEFEF}10.30\% &
  0.1309 &
  0.1544 &
  \cellcolor[HTML]{EFEFEF}17.99\% &
  0.4068 &
  0.5169 &
  \cellcolor[HTML]{EFEFEF}22.72\% \\
Low-cost &
  0.1525 &
  0.1687 &
  \cellcolor[HTML]{EFEFEF}10.58\% &
  \textbf{0.1603} &
  0.1839 &
  \cellcolor[HTML]{EFEFEF}14.69\% &
  \textbf{0.5571} &
  0.5895 &
  \cellcolor[HTML]{EFEFEF}5.10\% \\
Unlimited &
  \textbf{0.1630} &
  \textbf{0.1784} &
  \cellcolor[HTML]{EFEFEF}9.45\% &
  0.1600 &
  \textbf{0.2084} &
  \cellcolor[HTML]{EFEFEF}30.23\% &
  0.5428 &
  \textbf{0.5916} &
  \cellcolor[HTML]{EFEFEF}9.28\% 
\\ \bottomrule
\end{tabularx}
    \begin{tablenotes}
    \footnotesize
        \item[1] w/o: the ensemble explaining without \framework{}.
        \item[2] w/: the ensemble explaining with \framework{}.
    \end{tablenotes}
\end{threeparttable}
\vspace{-.6cm}
\end{table*}


\subsubsection{Effectiveness of Model Updating}
\label{sec:result-m1}

To study the effectiveness of model updating~(the module M1), we show that model interpretability can be enhanced while there is no trade-off paid to the classification performance.
\textcolor{\tcolor}{For model interpretability, we provide the $r$-AMP curve in Figure~\ref{fig:fidelity}, and show the influence on individual explainers at a fixed workload expectation in Table~\ref{tab:fidelity}.
For classification performance, we use the traditional evaluation metrics as in Table~\ref{tab:accuracy}.}

\noindent \textbf{Model interpretability Enhancement.}
As in Figure~\ref{fig:fidelity}, there is a clear enhancement for \framework{} on all classifiers that the green curves drop more sharply than the pink ones.
We notice an abnormality that the pink curve for DR-VGG goes upward, which means the model without \framework{} can hardly be explained by a single FA method and most of its classification decisions are based on artifacts.
Table~\ref{tab:fidelity} shows how the model interpretability acts on the system explanation fidelity for downstream explanation without ensembling.
We can see that all FA methods can output higher fidelity explanations when the classifier is updated with M1.
The improvement range is $6.62\%\sim22.35\%$, $43.47\%\sim82.05\%$, $3.04\%\sim21.28\%$, respectively, on DAMD, DR-VGG, and VulDeePecker.


\begin{table}[tb]
\vspace{.45cm}
\caption{Model accuracy of classifiers before and after model updating.}
\label{tab:accuracy}
\begin{threeparttable}
\begin{tabularx}{.98\linewidth}{c*{6}C}
\toprule
\multirow{2}{*}{\textbf{Metric}} & \multicolumn{2}{c}{\textbf{DAMD}} & \multicolumn{2}{c}{\textbf{DR-VGG}} & \multicolumn{2}{c}{\textbf{VulDeePecker}} \\ \cmidrule(r){2-3} \cmidrule(r){4-5} \cmidrule(r){6-7} 
 & \textbf{w/o} & \textbf{w/} & \textbf{w/o} & \textbf{w/} & \textbf{w/o} & \textbf{w/} \\ \midrule
Accuracy & 0.983 & 0.984 & 0.888 & 0.904 & 0.918 & 0.920 \\
Precision & 0.966 & 0.972 & 0.888 & 0.895 & 0.879 & 0.876 \\
Recall & 0.971 & 0.968 & 0.889 & 0.915 & 0.797 & 0.809 \\
F1-score & 0.974 & 0.978 & 0.888 & 0.900 & 0.836 & 0.841 \\
\bottomrule
\end{tabularx}
    \begin{tablenotes}
    \footnotesize
        \item[1] w/o: the original classifier trained without \framework{}.
        \item[2] w/: the new classifier updated with \framework{}.
    \end{tablenotes}
\end{threeparttable}
\vspace{-.35cm}
\end{table}

\noindent\textbf{Impact on Classification Performance.}
As shown in Table~\ref{tab:accuracy}, the common belief that there is a trade-off between model interpretability and classification accuracy do not exist with \framework{}.
For all three classifiers, the accuracy and F1-score metric even slightly improve.
Therefore, we can infer that our explanation-guided training strategy actually helps to adjust the model decision boundary, automatically patching some classification mistakes around it.

\subsubsection{Effectiveness of Explainer Ensembling}
\label{sec:result-m2}

With the classifier fixed to the updated one, we study the effectiveness of our ensembling method in M2 by comparing it with the baseline ensembling.

\noindent \textbf{Explanation Fidelity Enhancement.}
As the MPD results in Table~\ref{tab:ensemble_fid}, though both ensembling methods are effective~(the best MPD is higher than any explanation produced by a single FA), ensembling w/ \framework{} clearly has more advancement.
Firstly, the explanation fidelity with \framework{} ensembling is better than the baseline ensembling in all scenarios, with $10.12\%$ to $17.00\%$ higher in MPD.
It is interesting to find that, while model updating makes more improvement on gradient-based methods, ensembling is more beneficial to the underprivileged perturbation-based methods.
We also observe that, while low-cost ensembling can achieve comparable~(and more often even better) results as unlimited ensembling for the baseline method, \framework{} gathers information more effectively that unlimited explaining is always better than low-cost ensembling, where the greatest advantage is $13.32\%$. 
To discuss it system-wide, \framework{} makes $9.45\% \sim 10.58\%$, $14.69\% \sim 30.23\%$, $5.10\% \sim 22.72\%$ improvement on each system.
The biggest progress is still on DR-VGG, the pseudo-image-based malware classifier with spatial information whose original model is harder to be explained. 
For another malware classifier, the improvement is relatively smaller due to the large IC size that has an order of magnitude of 4.
For VulDeePecker, although the smaller feature space is intrinsically easier for explanation, \framework{} is still helpful in that even black-box explaining can achieve an MPD greater than $0.5$.

\subsection{Analysis of IC Abstraction and Efficiency} 

\subsubsection{Effectiveness of IC Abstraction} \label{sec:eval_ic}

\textcolor{\tcolor}{
The main purpose of IC abstraction is to enhance explanation intelligibility for analysts, which we will introduce case studies and human subjective studies in Section~\ref{sec:application} and Appendix~\ref{app:case} to show the effectiveness. For quantitative evaluation in this section, we show its influence on the sub-module named domain adjustment.
Since IC is at the same level as the feature for VulDeePecker, we study the classifiers of DAMD and DR-VGG.
For ablation experiments, we use the updated classifier and observe individual FA methods.
}

\noindent\textbf{Explanation Cost Reduction.}
Firstly, the IC-based adjustment reduces much explanation cost for perturbation-based FA methods.
We randomly sample $20$ risk instances from the test dataset of DAMD and DR-VGG and analyze the average time used for explaining.
As shown in Table~\ref{tab:ic-per-time},  for Shapley, explaining the two classifiers with a large feature space can hardly be practical without IC adjustment; for LIME and LEMNA, the time consumption is largely reduced by \framework{} from $53.43\%$ to $98.58\%$.

\noindent 
\textcolor{\tcolor}{
\textbf{Explanation Fidelity Improvement.}
Secondly, for gradient-based FA methods where it is practical to calculate feature-level explanations on the whole test dataset, we introduce a naive baseline to observe the MPD change.
Specifically, the baseline method uses the standard explanation algorithm to identify a list of important features and picks up ICs that have the most selected features.
As shown in Table~\ref{tab:ic-abstraction}, the mean percentage decrease of MPD~(compared with Table~\ref{tab:fidelity}) is $55.22\%$ and $30.66\%$ for the two systems, showing the abstraction in \framework{} is effective for improving the fidelity of IC-based explanations.
To understand the difference between explanations generated by the baseline and \framework{}, we also provide the intersection size~(IS~\cite{warnecke2020evaluating}) in the table.
The average IS value of $0.48$ suggests that the two explanations are largely different, where less than half of the selected ICs are in common.
}

\subsubsection{System Efficiency Analysis}
The major overhead introduced by \framework{} is in model updating.
However, this step costs much less than retraining a new model~(intrinsically interpretable but less accurate) and can be performed offline.
In our experiments, updating respectively takes $3$, $31$, and $43$ epochs for DAMD, DR-VGG, and VulDeePecker.
Actually, \framework{} can make ERDS more efficient in two aspects.
Firstly, as discussed for IC abstraction~(Table~\ref{tab:ic-per-time}), \framework{} helps with explanation cost reduction for the black-box scenario, especially when $\mathcal{V}$ has high dimensionality.
Secondly, for the low-cost scenario, a less time-consuming FA method on the updated classifier can achieve equal/higher explanation fidelity than a more complex method on the classifier without \framework{}.  
For example, as in Appendix~\ref{app:time}, the reduced time can be estimated at $0.76$ and $3.90$ seconds per sample for DAMD and DR-VGG, respectively.


\begin{table}[t]
\caption{IC abstraction helps black-box explainers to reduce time cost.}
\label{tab:ic-per-time}
\begin{threeparttable}
\begin{tabularx}{.98\linewidth}{cCCCCCC}
\toprule
\multirow{2}{*}{} & \multicolumn{3}{c}{\textbf{DAMD}}             & \multicolumn{3}{c}{\textbf{DR-VGG}}           \\ \cmidrule(r){2-4} \cmidrule(r){5-7} 
                                    & \textbf{w/o} & \textbf{w/} & \textbf{Red.} & \textbf{w/o} & \textbf{w/} & \textbf{Red.} \\ \midrule
LIME    & 4.5E+4 & 4.7E+3 & 89.60\% & 1.3E+3 & 7.3E+1   & 94.38\% \\
LEMNA   & 5.4E+4 & 7.7E+2  & 98.58\% & 3.9E+3 & 1.8E+3 & 53.43\% \\
Shapley & N/A     & 2.1E+3 & N/A     & N/A    & 2.3E+2  & N/A     
\\ \bottomrule
\end{tabularx}
    \begin{tablenotes}
    \footnotesize
        \item[1] w/o: the explainer without IC-based adjustment.
        \item[2] w/: the explainer with IC-based adjustment.
        \item[3] N/A: run out of memory due to the large feature sizes.
    \end{tablenotes}
\end{threeparttable}
\end{table}

\begin{table}[t]
    \caption{\textcolor{\tcolor}{Gradient-based explainers without IC abstraction. To compare with adjusted explainers, Dec. stands for the percentage decrease of MPD, and IS measures the explanation similarity.}}
    \label{tab:ic-abstraction}
    \begin{tabularx}{.8\linewidth}{cCCCC}
    \toprule
     & \multicolumn{2}{c}{\textbf{DAMD}} & \multicolumn{2}{c}{\textbf{DR-VGG}} \\ \cmidrule(r){2-3} \cmidrule(r){4-5}
     & \textbf{Dec.} & \textbf{IS} & \textbf{Dec.} & \textbf{IS} \\ \midrule
    Gradients- & 56.26\% & 0.3083 & 21.16\% & 0.7841 \\
    IG- & 45.07\% & 0.4567 & 45.68\% & 0.5239 \\
    DeepLIFT- & 64.32\% & 0.3107 & 25.13\% & 0.5157    
    \\ \bottomrule
    \end{tabularx}
\vspace{-4mm}
\end{table}







\section{Task Evaluation with Ground Truth} \label{sec:application}

So far, we have validated the fidelity enhancement of \framework{} on different risk detection systems.
Since there exists little ground truth for risk explanation, we perform a case study to show how \framework{} benefits downstream tasks. 
In this section, we investigate the effectiveness of \framework{} on a ground truth dataset for malicious PE functionality localization.
We compare DR-VGG enhanced by \framework{} with two tools proposed in existing work.

\noindent \textbf{Baseline and Dataset.} The baselines include (1)~DeepReflect: a state-of-the-art tool that uses an unsupervised learning model~\cite{ronneberger2015u} and identifies malicious components with localized mean-squared-error; (2)~DR-IG: the SHAP model in \cite{downing2021deepreflect} which is FA-based that uses IG to explain the supervised VGG classifier.
The ground truth dataset has three malware samples named Rbot, Pegasus, and Carbanak, where each of them respectively has $1$, $6$, and $6$ payload files that are independently classified. The malicious functions inside each file is statically identified by security experts, which makes human-ground explanation labels. 

\begin{figure}[tb]
\vspace{-.4cm}
\centering
    \subfloat[Rbot]{%
        \includegraphics[width=0.333\linewidth]{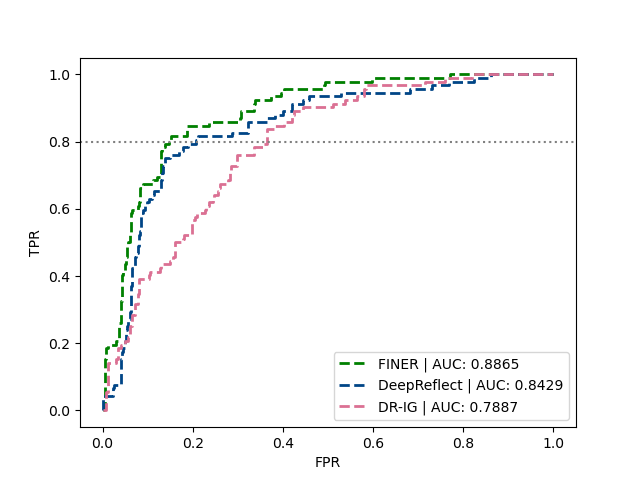}%
        \label{fig:a}%
        }
    \subfloat[Pegasus]{%
        \includegraphics[width=0.333\linewidth]{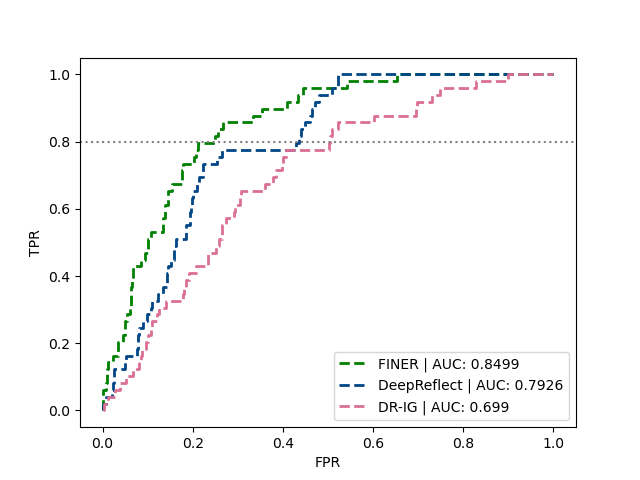}%
        \label{fig:b}%
        }
    \subfloat[Carbanak]{%
        \includegraphics[width=0.333\linewidth]{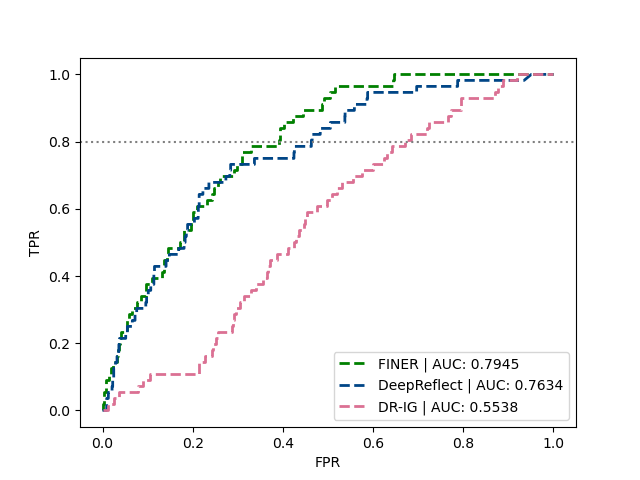} 
        \label{fig:c}%
        }%
    \caption{Global ROC curves for the three malware samples with human-annotated explanation labels.}
    \label{fig:gt}
\vspace{-1.5em}
\end{figure}

\noindent \textbf{Quantitative Results.}
We follow the evaluation practice of prior work and use the receiver operating characteristic (ROC) curve to illustrate how the attribution scores match the explanation labels.
As shown in Figure~\ref{fig:gt}, \framework{} outperforms DeepReflect and DR-IG on all malware samples.
Through the comparison with DeepReflect, we show that \framework{} can achieve state-of-the-art performance in malicious functionality detection.
By comparing \framework{} with DR-IG, we prove that our explanation enhancement strategy is still effective in human-grounded evaluation.
To take a close look, the local AUC scores for each classification unit is presented in Table~\ref{tab:gt}.
We observe that \framework{} has a rather stable performance among payloads with different sizes and maliciousness ratios.
The average increase in AUC is $17.75\%$ and $43.79\%$ when compared with DeepReflect and DR-IG, respectively. 
We also look into the two exceptional cases where DeepReflect works slightly better, and we find that \framework{} actually does not fail in the top-k recommendation contest.
For Pegasus\#idd, both methods return $3$ correct predictions out of their top $10$ results, and for Carbanak\#rdp, \framework{} actually returns the singleton malicious function quicker~(at $11$th) than DeepReflect~(at $41$th).
As DeepReflect is an unsupervised tool, users can train it from the beginning when no malware labels are available.
However, \framework{} handles existing state-of-the-art classifiers, and given the fact that binary malware detection is available from many resources~\cite{virustotal,Allix:2016:AndroZoo}, it is superior to DeepReflect for higher explanation accuracy and fewer training efforts.

\begin{figure}[t]
\myfigureshrinker
    \centering
    \subfloat[\framework{}]{%
        \includegraphics[width=\linewidth]{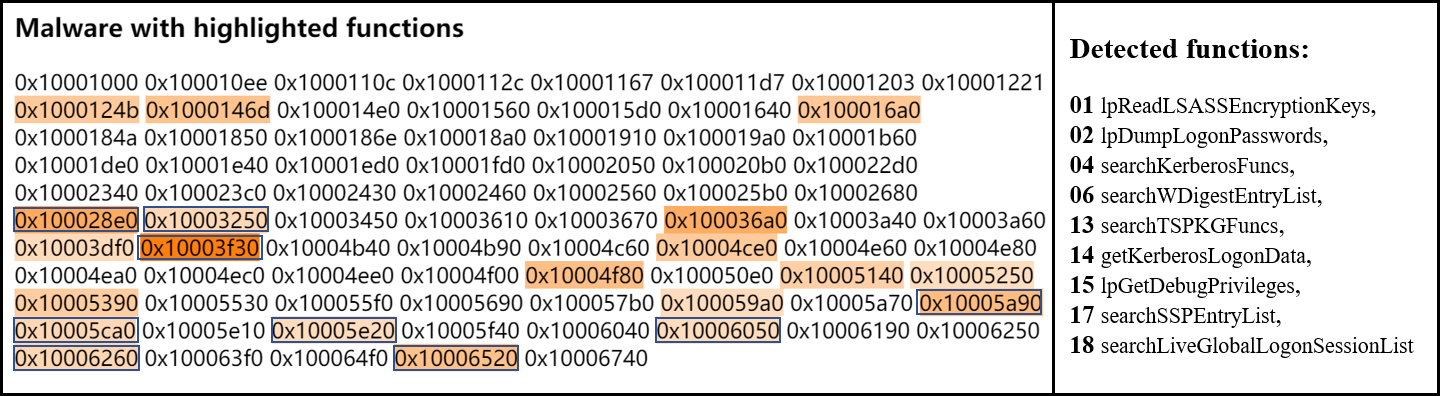}%
        \label{fig:case_a}%
        }\hfill%
    \subfloat[DeepReflect]{%
        \includegraphics[width=\linewidth]{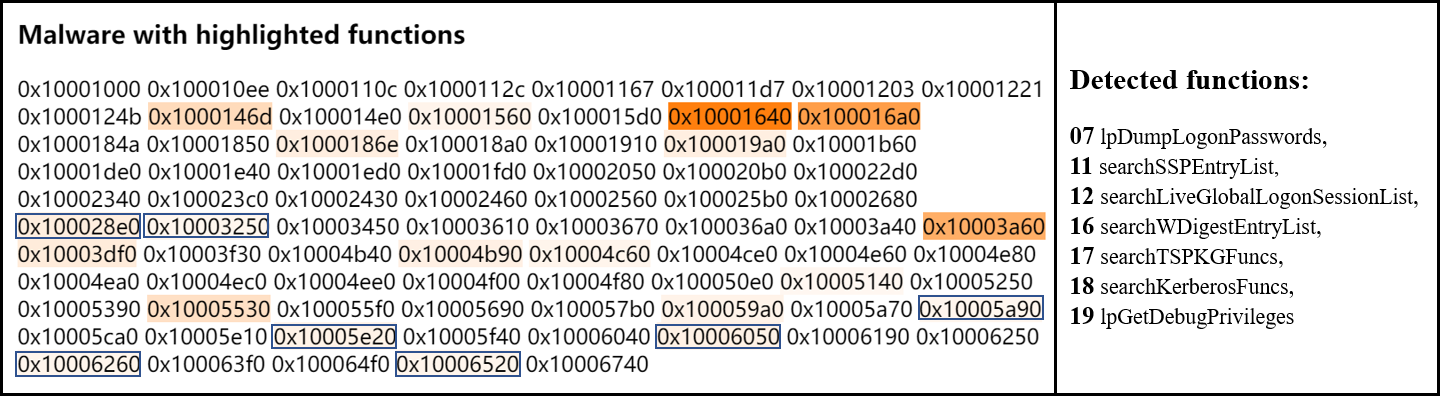}%
        \label{fig:case_b}%
        }
    \caption{Function-level malware explanations for Pegasus\#log. Important functions are highlighted on the left with their bytecode addresses, and for those match ground truth~(marked with blue boxes), the importance rankings and function names are listed on the right.}
    \label{fig:vis_pegasus_finer}
\end{figure}

\noindent 
\textcolor{\tcolor}{
\textbf{Explanation Visualization.}
To understand how \framework{} assists the malware analysis task, we show an example of visualized explanations in Figure~\ref{fig:vis_pegasus_finer}.
As illustrated in the figure, the intelligible explanation is presented as highlighted functions for analysts to inspect the corresponding bytecode addresses.
For this example, the malicious payload steals confidential logon data~(e.g., Kerberos tickets, WDigest/TsPkg/SSP passwords) through Local Security Authority Subsystem Service~(LSASS) memory dumping~\cite{lsass}.
Comparing the two explanations, we find that \framework{} is not only more accurate than DeepReflect, i.e., more malicious functions are identified, but also has much higher confidence in those correct identification, i.e., malicious functions are highlighted with darker colors.
Specifically, the two functions that implement the main logic to decrypt and dump data~(pseudo code in Listing~\ref{lst:case-study}) are exactly in \framework{}'s top two recommendations, which means analysts can waste no effort and quickly get an overview of the malicious behavior.
}

\begin{table*}[tb]
\caption{Local IC statistics and AUC scores for each malware payload of the ground truth malware samples.}
\label{tab:gt}
\begin{threeparttable}
\begin{tabularx}{\linewidth}{c*{13}C}
\toprule
\multicolumn{1}{c}{\multirow{2}{*}{}} & \multicolumn{1}{c}{\textbf{Rbot}} & \multicolumn{6}{c}{\textbf{Pegasus}} & \multicolumn{6}{c}{\textbf{Carbanak}} \\ \cmidrule(r){2-2} \cmidrule(r){3-8} \cmidrule(r){9-14}  
\multicolumn{1}{c}{} & \textbf{\#icr} & \textbf{\#log} & \textbf{\#rep} & \textbf{\#idd} & \textbf{\#net} & \textbf{\#exec} & \textbf{\#rse} & \textbf{\#auto} & \textbf{\#rdp} & \textbf{\#cmd} & \textbf{\#cve} & \textbf{\#bot} & \textbf{\#d/l}\\ \midrule
Mal./Total\tnote{*} & 92/440 & 13/81 & 15/97 & 4/98 & 10/98 & 6/69 & 1/93 & 1/66 & 1/107 & 6/804 & 2/25 & 44/999 & 2/234 \\ \midrule
DeepReflect & 0.8429 & 0.7839 & 0.7780 & \textbf{0.9016} & 0.6750 & 0.8942 & 0.8913 & 0.6226 & \textbf{0.9569} & 0.7712 & 0.8016 & 0.6087 & 0.6000 \\ 
DR-IG & 0.7887 & 0.8431 & 0.7170 & 0.6029 & 0.6522 & 0.7187 & 0.6772 & 0.5020 & 0.8208 & 0.2759 & 0.6564 & 0.5652 & 0.9692 \\ 
FINER & \textbf{0.8865} & \textbf{0.8778} & \textbf{0.8390} & 0.8761 & \textbf{0.9814} & \textbf{0.9603} & \textbf{0.9130} & \textbf{0.9231} & 0.9057 & \textbf{0.8358} & \textbf{0.9130} & \textbf{0.7568} & \textbf{0.9828}
\\ \bottomrule
\end{tabularx}
    \begin{tablenotes}
    \footnotesize
        \item[*] The number of malicious functions vs. the total number of functions inside a payload.
    \end{tablenotes}
\end{threeparttable}
\vspace{-1.5em}
\end{table*}

\section{Related Work}

Besides what has been introduced in Section~\ref{sec:erds}, we discuss other related works from three aspects.

\noindent \textbf{Intrinsic and Global XAI Methods.}
Intrinsic XAI methods are typically achieved by self-interpretable models, such as linear models, decision trees~\cite{nowozin2011decision}, and falling rule lists~\cite{wang2015falling}.
Since these methods are less complex, they usually have a lower level of accuracy, making the common belief that model interpretability is at the cost of accuracy~\cite{samtani2022explainable}. 
Global XAI methods explain the overall working mechanism of models with structures or parameters, and a typical example is the attention weights~\cite{vaswani2017attention}. 
We choose the post-hoc and instance-level FA methods as a plug-and-play toolset of ERDS. 
Note that although we update the model, the internal architecture is not changed, and our experiments~(as in Section~\ref{sec:result-m1}) validate that the interpretability-accuracy trade-off does not exist in FINER.

\noindent 
\textcolor{\tcolor}{
\textbf{Regularizing Model with Explanations.}
Recently, a few studies suggest that training models with explanation constraints help with post-hoc explaining. 
Some works~\cite{ross2017right, li2018tell} assume that explanation annotations be available and use them to supervise model gradients, but the assumption can hardly be expected to hold in risk detection scenarios.
Others~\cite{zhang2021excon, wang2022explanation, pillai2021explainable, plumb2020regularizing} leverage self-supervised learning to regularize unlabelled explanations, which can be divided into two groups:
the first focuses on a contrastive setting~\cite{schroff2015facenet} where explanations are used to generate positive and negative examples for pre-training encoder networks, thus not applicable to off-the-shelf classifiers;
the second constrains particular explainer outputs by utilizing some structural priors in data, e.g., \cite{pillai2021explainable} imposes consistency on the Grad-CAM heatmap~\cite{selvaraju2017grad} before and after image composition, which do not generalize to different data domain or downstream explainers.
Our fine-tuning method also takes inspiration from self-supervised learning. We propose explanation-guided data augmentation in the context of risk detection and novelly introduce the explanation task as classifying augmented samples, making \framework{} applicable and adaptable to ERDS.
}



\noindent 
\textcolor{\tcolor}{
\textbf{Explanation Applications in Security.}
Several works have been proposed to leverage explanations for other security purposes, such as vetting malware tags~\cite{pirch2021tagvet}, selecting concept drift samples~\cite{usenix21cade, han23anomaly}, and guiding fairness testing~\cite{fan2022explanation, wang2023towards}.
We focus on the explanation application to provide user assistance in security analysis, and these works are orthogonal to ours, which can be adopted together to develop more powerful security systems.
}

\section{Discussion} \label{sec:discussion}

\textcolor{\tcolor}{
Below, we discuss some design choices and potential future work.
}

\noindent 
\textcolor{\tcolor}{\textbf{New Explanation Methods.}
We formalize a novel explanation mechanism to tackle the challenging problems of fidelity and intelligibility, where existing FA methods are adaptably used as a toolset. 
A key observation is that much effort has been made to develop FA methods but their performance in explaining different applications remains uncertain~\cite{belaid2022we}.
As our framework is flexible to accommodate new methods, for specific security applications, greater fidelity improvement can be achieved by absorbing more dedicated explanation methods~\cite{guo2018lemna}.}

\noindent 
\textcolor{\tcolor}{\textbf{Normal Sample Explanations.}
We focus on identifying risky components from abnormal samples due to their importance in risk response. 
In practical security analysis, experts want to use their limited energy to understand and prevent critical risks~\cite{ccs21deepaid}.
Explaining normal samples would be more helpful for other targets such as model debugging and shift adaptation~\cite{cao2020benign, han23anomaly}.}

\noindent %
\textcolor{\tcolor}{
 \textbf{Supported Applications.}
In the experiments, the evaluated risk detection systems provide explanations for analyzing Android malware, PE malware, and code vulnerabilities.
We study these systems because the relevant supervised classifiers achieve state-of-the-art performance and belong to the most concerned applications that call for explainability~(Section~\ref{sec:cls_exp}).
Future work can study \framework{} on other applications such as botnet and spam detection, and also extend the framework to unsupervised anomaly detection~\cite{xnids}.}

\noindent 
\textcolor{\tcolor}{\textbf{Attacks on ERDS.}
To simplify our main design goal of handling explanation fidelity and intelligibility, we focus on the benign application scenario as explained in Section~\ref{sec:scope}. 
The robustness of ERDS against attacks is an important and complex security aspect.
A motivated adversary can perform evasion/backdoor attacks on both the classifier and the explainer~\cite{severi2021explanation, zhang2020interpretable}, which calls for different defense considerations.
Future work can investigate the performance of ERDS against these attacks and may benefit from updating the classifier with a trade-off between robustness and classification/explanation accuracy.}

\noindent 
\textcolor{\tcolor}{\textbf{Human Subject Evaluation.}
We use several human-annotated samples~(Section~\ref{sec:application}) and collect feedback from security experts~(Appendix~\ref{app:case}) to evaluate the usefulness of malware explanations.
Systematic human subject evaluation is actually challenging for ERDS. 
Firstly, human annotation requires domain knowledge and is time-consuming, and thus no large-scale annotated dataset is available.
Secondly, due to the multidisciplinary of ERDS~\cite{lopes2022xai}, there is little consensus on the participation of humans in the evaluation.
Future work can strengthen the evaluation by leveraging user prediction of model output/failure~\cite{shen2020useful} and designing post-study questionnaire~\cite{nourani2019effects} for more security applications.}

\section{Conclusion}
\label{sec:conclusion}

The black-box property of deep learning-based risk detection classifiers has been a long-standing issue.
This paper investigates the emerging field of XAI and proposes an explanation framework, named \framework{}, to promote these classifiers into explainable risk detection systems.
\framework{} considers ``what to explain'' in the context of security analysis, \textcolor{\tcolor}{``whom to explain for'}' as three main stakeholders, and \textcolor{\tcolor}{``how to explain''} in terms of fidelity and intelligibility.
Evaluation results show that \framework{} is effective to generate high-fidelity intelligible component-based explanations for different security analysis tasks.
We hope that this framework and \textcolor{\tcolor}{published code~\footnote{https://github.com/E0HYL/FINER-explain}} can inspire more research efforts on explainable security.


\bibliographystyle{ACM-Reference-Format}
\bibliography{bib}


\begin{thebibliography}{00}


\ifx \showCODEN    \undefined \def \showCODEN     #1{\unskip}     \fi
\ifx \showDOI      \undefined \def \showDOI       #1{#1}\fi
\ifx \showISBNx    \undefined \def \showISBNx     #1{\unskip}     \fi
\ifx \showISBNxiii \undefined \def \showISBNxiii  #1{\unskip}     \fi
\ifx \showISSN     \undefined \def \showISSN      #1{\unskip}     \fi
\ifx \showLCCN     \undefined \def \showLCCN      #1{\unskip}     \fi
\ifx \shownote     \undefined \def \shownote      #1{#1}          \fi
\ifx \showarticletitle \undefined \def \showarticletitle #1{#1}   \fi
\ifx \showURL      \undefined \def \showURL       {\relax}        \fi
\providecommand\bibfield[2]{#2}
\providecommand\bibinfo[2]{#2}
\providecommand\natexlab[1]{#1}
\providecommand\showeprint[2][]{arXiv:#2}

\bibitem[\protect\citeauthoryear{35}{35}{2023}]%
        {binaryninja}
\bibfield{author}{\bibinfo{person}{Vector 35}.}
  \bibinfo{year}{2023}\natexlab{}.
\newblock \bibinfo{title}{Binary Ninja}.
\newblock \bibinfo{howpublished}{\url{https://binary.ninja/}}.
  (\bibinfo{year}{2023}).
\newblock


\bibitem[\protect\citeauthoryear{Aafer, Du, and Yin}{Aafer
  et~al\mbox{.}}{2013}]%
        {aafer2013droidapiminer}
\bibfield{author}{\bibinfo{person}{Yousra Aafer}, \bibinfo{person}{Wenliang
  Du}, {and} \bibinfo{person}{Heng Yin}.} \bibinfo{year}{2013}\natexlab{}.
\newblock \showarticletitle{Droidapiminer: Mining api-level features for robust
  malware detection in android}. In \bibinfo{booktitle}{{\em International
  conference on security and privacy in communication systems}}. Springer,
  \bibinfo{pages}{86--103}.
\newblock


\bibitem[\protect\citeauthoryear{Allix, Bissyand{\'e}, Klein, and
  Le~Traon}{Allix et~al\mbox{.}}{2016}]%
        {Allix:2016:AndroZoo}
\bibfield{author}{\bibinfo{person}{Kevin Allix},
  \bibinfo{person}{Tegawend{\'e}~F. Bissyand{\'e}}, \bibinfo{person}{Jacques
  Klein}, {and} \bibinfo{person}{Yves Le~Traon}.}
  \bibinfo{year}{2016}\natexlab{}.
\newblock \showarticletitle{AndroZoo: Collecting Millions of Android Apps for
  the Research Community}. In \bibinfo{booktitle}{{\em Proceedings of the 13th
  International Conference on Mining Software Repositories}} {\em
  (\bibinfo{series}{MSR '16})}. \bibinfo{publisher}{ACM},
  \bibinfo{pages}{468--471}.
\newblock
\showISBNx{978-1-4503-4186-8}
\showDOI{%
\url{https://doi.org/10.1145/2901739.2903508}}


\bibitem[\protect\citeauthoryear{Alqaraawi, Schuessler, Wei{\ss}, Costanza, and
  Berthouze}{Alqaraawi et~al\mbox{.}}{2020}]%
        {alqaraawi2020evaluating}
\bibfield{author}{\bibinfo{person}{Ahmed Alqaraawi}, \bibinfo{person}{Martin
  Schuessler}, \bibinfo{person}{Philipp Wei{\ss}}, \bibinfo{person}{Enrico
  Costanza}, {and} \bibinfo{person}{Nadia Berthouze}.}
  \bibinfo{year}{2020}\natexlab{}.
\newblock \showarticletitle{Evaluating saliency map explanations for
  convolutional neural networks: a user study}. In \bibinfo{booktitle}{{\em
  Proceedings of the 25th International Conference on Intelligent User
  Interfaces}}. \bibinfo{pages}{275--285}.
\newblock


\bibitem[\protect\citeauthoryear{Ancona, Ceolini, {\"O}ztireli, and
  Gross}{Ancona et~al\mbox{.}}{2017}]%
        {ancona2017towards}
\bibfield{author}{\bibinfo{person}{Marco Ancona}, \bibinfo{person}{Enea
  Ceolini}, \bibinfo{person}{Cengiz {\"O}ztireli}, {and}
  \bibinfo{person}{Markus Gross}.} \bibinfo{year}{2017}\natexlab{}.
\newblock \showarticletitle{Towards better understanding of gradient-based
  attribution methods for deep neural networks}.
\newblock \bibinfo{journal}{{\em arXiv preprint arXiv:1711.06104\/}}
  (\bibinfo{year}{2017}).
\newblock


\bibitem[\protect\citeauthoryear{Anderson and Roth}{Anderson and Roth}{2018}]%
        {anderson2018ember}
\bibfield{author}{\bibinfo{person}{Hyrum~S Anderson} {and}
  \bibinfo{person}{Phil Roth}.} \bibinfo{year}{2018}\natexlab{}.
\newblock \showarticletitle{Ember: an open dataset for training static pe
  malware machine learning models}.
\newblock \bibinfo{journal}{{\em arXiv preprint arXiv:1804.04637\/}}
  (\bibinfo{year}{2018}).
\newblock


\bibitem[\protect\citeauthoryear{Arp, Quiring, Pendlebury, Warnecke, Pierazzi,
  Wressnegger, Cavallaro, and Rieck}{Arp et~al\mbox{.}}{2022}]%
        {arp2022and}
\bibfield{author}{\bibinfo{person}{Daniel Arp}, \bibinfo{person}{Erwin
  Quiring}, \bibinfo{person}{Feargus Pendlebury}, \bibinfo{person}{Alexander
  Warnecke}, \bibinfo{person}{Fabio Pierazzi}, \bibinfo{person}{Christian
  Wressnegger}, \bibinfo{person}{Lorenzo Cavallaro}, {and}
  \bibinfo{person}{Konrad Rieck}.} \bibinfo{year}{2022}\natexlab{}.
\newblock \showarticletitle{Dos and don’ts of machine learning in computer
  security}. In \bibinfo{booktitle}{{\em Proc. of the USENIX Security
  Symposium}}.
\newblock


\bibitem[\protect\citeauthoryear{Arp, Spreitzenbarth, Hubner, Gascon, Rieck,
  and Siemens}{Arp et~al\mbox{.}}{2014}]%
        {arp2014drebin}
\bibfield{author}{\bibinfo{person}{Daniel Arp}, \bibinfo{person}{Michael
  Spreitzenbarth}, \bibinfo{person}{Malte Hubner}, \bibinfo{person}{Hugo
  Gascon}, \bibinfo{person}{Konrad Rieck}, {and} \bibinfo{person}{CERT
  Siemens}.} \bibinfo{year}{2014}\natexlab{}.
\newblock \showarticletitle{Drebin: Effective and explainable detection of
  android malware in your pocket}. In \bibinfo{booktitle}{{\em Proceedings of
  the Network and Distributed Systems Security Symposium (NDSS)}},
  Vol.~\bibinfo{volume}{14}. \bibinfo{pages}{23--26}.
\newblock


\bibitem[\protect\citeauthoryear{ATT\&CK}{ATT\&CK}{2020}]%
        {lsass}
\bibfield{author}{\bibinfo{person}{MITRE ATT\&CK}.}
  \bibinfo{year}{2020}\natexlab{}.
\newblock \bibinfo{title}{OS Credential Dumping: LSASS Memory}.
\newblock
  \bibinfo{howpublished}{\url{https://attack.mitre.org/techniques/T1003/001/}}.
    (\bibinfo{year}{2020}).
\newblock


\bibitem[\protect\citeauthoryear{AV-ATLAS}{AV-ATLAS}{2023}]%
        {AV-ATLAS}
\bibfield{author}{\bibinfo{person}{AV-ATLAS}.} \bibinfo{year}{2023}\natexlab{}.
\newblock \bibinfo{title}{New malware}.
\newblock \bibinfo{howpublished}{\url{https://portal.av-atlas.org/}}.
  (\bibinfo{year}{2023}).
\newblock


\bibitem[\protect\citeauthoryear{Belaid, H{\"u}llermeier, Rabus, and
  Krestel}{Belaid et~al\mbox{.}}{2022}]%
        {belaid2022we}
\bibfield{author}{\bibinfo{person}{Mohamed~Karim Belaid}, \bibinfo{person}{Eyke
  H{\"u}llermeier}, \bibinfo{person}{Maximilian Rabus}, {and}
  \bibinfo{person}{Ralf Krestel}.} \bibinfo{year}{2022}\natexlab{}.
\newblock \showarticletitle{Do We Need Another Explainable AI Method? Toward
  Unifying Post-hoc XAI Evaluation Methods into an Interactive and
  Multi-dimensional Benchmark}.
\newblock \bibinfo{journal}{{\em arXiv preprint arXiv:2207.14160\/}}
  (\bibinfo{year}{2022}).
\newblock


\bibitem[\protect\citeauthoryear{Bhatt, Xiang, Sharma, Weller, Taly, Jia,
  Ghosh, Puri, Moura, and Eckersley}{Bhatt et~al\mbox{.}}{2020}]%
        {bhatt2020explainable}
\bibfield{author}{\bibinfo{person}{Umang Bhatt}, \bibinfo{person}{Alice Xiang},
  \bibinfo{person}{Shubham Sharma}, \bibinfo{person}{Adrian Weller},
  \bibinfo{person}{Ankur Taly}, \bibinfo{person}{Yunhan Jia},
  \bibinfo{person}{Joydeep Ghosh}, \bibinfo{person}{Ruchir Puri},
  \bibinfo{person}{Jos{\'e}~MF Moura}, {and} \bibinfo{person}{Peter
  Eckersley}.} \bibinfo{year}{2020}\natexlab{}.
\newblock \showarticletitle{Explainable machine learning in deployment}. In
  \bibinfo{booktitle}{{\em Proceedings of the 2020 conference on fairness,
  accountability, and transparency}}. \bibinfo{pages}{648--657}.
\newblock


\bibitem[\protect\citeauthoryear{Cao, Badihi, Ahmed, Xiong, and Rubin}{Cao
  et~al\mbox{.}}{2020}]%
        {cao2020benign}
\bibfield{author}{\bibinfo{person}{Michael Cao}, \bibinfo{person}{Sahar
  Badihi}, \bibinfo{person}{Khaled Ahmed}, \bibinfo{person}{Peiyu Xiong}, {and}
  \bibinfo{person}{Julia Rubin}.} \bibinfo{year}{2020}\natexlab{}.
\newblock \showarticletitle{On benign features in malware detection}. In
  \bibinfo{booktitle}{{\em Proceedings of the 35th IEEE/ACM International
  Conference on Automated Software Engineering}}. \bibinfo{pages}{1234--1238}.
\newblock


\bibitem[\protect\citeauthoryear{Cao, Sun, Bo, Wei, and Li}{Cao
  et~al\mbox{.}}{2021}]%
        {cao2021bgnn4vd}
\bibfield{author}{\bibinfo{person}{Sicong Cao}, \bibinfo{person}{Xiaobing Sun},
  \bibinfo{person}{Lili Bo}, \bibinfo{person}{Ying Wei}, {and}
  \bibinfo{person}{Bin Li}.} \bibinfo{year}{2021}\natexlab{}.
\newblock \showarticletitle{Bgnn4vd: constructing bidirectional graph
  neural-network for vulnerability detection}.
\newblock \bibinfo{journal}{{\em Information and Software Technology\/}}
  \bibinfo{volume}{136} (\bibinfo{year}{2021}), \bibinfo{pages}{106576}.
\newblock


\bibitem[\protect\citeauthoryear{Casey, Farhangi, and Vogl}{Casey
  et~al\mbox{.}}{2019}]%
        {casey2019rethinking}
\bibfield{author}{\bibinfo{person}{Bryan Casey}, \bibinfo{person}{Ashkon
  Farhangi}, {and} \bibinfo{person}{Roland Vogl}.}
  \bibinfo{year}{2019}\natexlab{}.
\newblock \showarticletitle{Rethinking Explainable Machines: The GDPR's' Right
  to Explanation'Debate and the Rise of Algorithmic Audits in Enterprise}.
\newblock \bibinfo{journal}{{\em Berkeley Tech. LJ\/}}  \bibinfo{volume}{34}
  (\bibinfo{year}{2019}), \bibinfo{pages}{143}.
\newblock


\bibitem[\protect\citeauthoryear{Chen, Kornblith, Norouzi, and Hinton}{Chen
  et~al\mbox{.}}{2020}]%
        {chen2020simple}
\bibfield{author}{\bibinfo{person}{Ting Chen}, \bibinfo{person}{Simon
  Kornblith}, \bibinfo{person}{Mohammad Norouzi}, {and}
  \bibinfo{person}{Geoffrey Hinton}.} \bibinfo{year}{2020}\natexlab{}.
\newblock \showarticletitle{A simple framework for contrastive learning of
  visual representations}. In \bibinfo{booktitle}{{\em International conference
  on machine learning}}. PMLR, \bibinfo{pages}{1597--1607}.
\newblock


\bibitem[\protect\citeauthoryear{Cheng, Ming, Fu, Peng, Chen, Zhang, and
  Marion}{Cheng et~al\mbox{.}}{2018}]%
        {cheng2018towards}
\bibfield{author}{\bibinfo{person}{Binlin Cheng}, \bibinfo{person}{Jiang Ming},
  \bibinfo{person}{Jianmin Fu}, \bibinfo{person}{Guojun Peng},
  \bibinfo{person}{Ting Chen}, \bibinfo{person}{Xiaosong Zhang}, {and}
  \bibinfo{person}{Jean-Yves Marion}.} \bibinfo{year}{2018}\natexlab{}.
\newblock \showarticletitle{Towards paving the way for large-scale windows
  malware analysis: Generic binary unpacking with orders-of-magnitude
  performance boost}. In \bibinfo{booktitle}{{\em Proceedings of the 2018 ACM
  SIGSAC Conference on Computer and Communications Security}}.
  \bibinfo{pages}{395--411}.
\newblock


\bibitem[\protect\citeauthoryear{Dabkowski and Gal}{Dabkowski and Gal}{2017}]%
        {dabkowski2017real}
\bibfield{author}{\bibinfo{person}{Piotr Dabkowski} {and}
  \bibinfo{person}{Yarin Gal}.} \bibinfo{year}{2017}\natexlab{}.
\newblock \showarticletitle{Real time image saliency for black box
  classifiers}.
\newblock \bibinfo{journal}{{\em Advances in neural information processing
  systems\/}}  \bibinfo{volume}{30} (\bibinfo{year}{2017}).
\newblock


\bibitem[\protect\citeauthoryear{Desnos}{Desnos}{2023}]%
        {androguard}
\bibfield{author}{\bibinfo{person}{Anthony Desnos}.}
  \bibinfo{year}{2023}\natexlab{}.
\newblock \bibinfo{title}{Androguard}.
\newblock
  \bibinfo{howpublished}{\url{https://github.com/androguard/androguard}}.
  (\bibinfo{year}{2023}).
\newblock


\bibitem[\protect\citeauthoryear{Diamond and Boyd}{Diamond and Boyd}{2016}]%
        {diamond2016cvxpy}
\bibfield{author}{\bibinfo{person}{Steven Diamond} {and}
  \bibinfo{person}{Stephen Boyd}.} \bibinfo{year}{2016}\natexlab{}.
\newblock \showarticletitle{CVXPY: A Python-embedded modeling language for
  convex optimization}.
\newblock \bibinfo{journal}{{\em The Journal of Machine Learning Research\/}}
  \bibinfo{volume}{17}, \bibinfo{number}{1} (\bibinfo{year}{2016}),
  \bibinfo{pages}{2909--2913}.
\newblock


\bibitem[\protect\citeauthoryear{Downing, Mirsky, Park, and Lee}{Downing
  et~al\mbox{.}}{2021}]%
        {downing2021deepreflect}
\bibfield{author}{\bibinfo{person}{Evan Downing}, \bibinfo{person}{Yisroel
  Mirsky}, \bibinfo{person}{Kyuhong Park}, {and} \bibinfo{person}{Wenke Lee}.}
  \bibinfo{year}{2021}\natexlab{}.
\newblock \showarticletitle{DeepReflect: Discovering Malicious Functionality
  through Binary Reconstruction}. In \bibinfo{booktitle}{{\em 30th USENIX
  Security Symposium (USENIX Security 21)}}.
\newblock


\bibitem[\protect\citeauthoryear{Du, Liu, and Hu}{Du et~al\mbox{.}}{2019}]%
        {du2019iml_techniques}
\bibfield{author}{\bibinfo{person}{Mengnan Du}, \bibinfo{person}{Ninghao Liu},
  {and} \bibinfo{person}{Xia Hu}.} \bibinfo{year}{2019}\natexlab{}.
\newblock \showarticletitle{Techniques for interpretable machine learning}.
\newblock \bibinfo{journal}{{\it Commun. ACM}} \bibinfo{volume}{63},
  \bibinfo{number}{1} (\bibinfo{year}{2019}), \bibinfo{pages}{68--77}.
\newblock


\bibitem[\protect\citeauthoryear{Elmasry, Akbulut, and Zaim}{Elmasry
  et~al\mbox{.}}{2020}]%
        {elmasry2020evolving}
\bibfield{author}{\bibinfo{person}{Wisam Elmasry}, \bibinfo{person}{Akhan
  Akbulut}, {and} \bibinfo{person}{Abdul~Halim Zaim}.}
  \bibinfo{year}{2020}\natexlab{}.
\newblock \showarticletitle{Evolving deep learning architectures for network
  intrusion detection using a double PSO metaheuristic}.
\newblock \bibinfo{journal}{{\em Computer Networks\/}}  \bibinfo{volume}{168}
  (\bibinfo{year}{2020}), \bibinfo{pages}{107042}.
\newblock


\bibitem[\protect\citeauthoryear{Fan, Wei, Jin, Yang, and Liu}{Fan
  et~al\mbox{.}}{2022}]%
        {fan2022explanation}
\bibfield{author}{\bibinfo{person}{Ming Fan}, \bibinfo{person}{Wenying Wei},
  \bibinfo{person}{Wuxia Jin}, \bibinfo{person}{Zijiang Yang}, {and}
  \bibinfo{person}{Ting Liu}.} \bibinfo{year}{2022}\natexlab{}.
\newblock \showarticletitle{Explanation-guided fairness testing through genetic
  algorithm}. In \bibinfo{booktitle}{{\em Proceedings of the 44th International
  Conference on Software Engineering}}. \bibinfo{pages}{871--882}.
\newblock


\bibitem[\protect\citeauthoryear{Gamage and Samarabandu}{Gamage and
  Samarabandu}{2020}]%
        {gamage2020deep}
\bibfield{author}{\bibinfo{person}{Sunanda Gamage} {and}
  \bibinfo{person}{Jagath Samarabandu}.} \bibinfo{year}{2020}\natexlab{}.
\newblock \showarticletitle{Deep learning methods in network intrusion
  detection: A survey and an objective comparison}.
\newblock \bibinfo{journal}{{\em Journal of Network and Computer
  Applications\/}}  \bibinfo{volume}{169} (\bibinfo{year}{2020}),
  \bibinfo{pages}{102767}.
\newblock


\bibitem[\protect\citeauthoryear{Grosse, Papernot, Manoharan, Backes, and
  McDaniel}{Grosse et~al\mbox{.}}{2017}]%
        {grosse2017adversarial}
\bibfield{author}{\bibinfo{person}{Kathrin Grosse}, \bibinfo{person}{Nicolas
  Papernot}, \bibinfo{person}{Praveen Manoharan}, \bibinfo{person}{Michael
  Backes}, {and} \bibinfo{person}{Patrick McDaniel}.}
  \bibinfo{year}{2017}\natexlab{}.
\newblock \showarticletitle{Adversarial examples for malware detection}. In
  \bibinfo{booktitle}{{\em Computer Security--ESORICS 2017: 22nd European
  Symposium on Research in Computer Security, Oslo, Norway, September 11-15,
  2017, Proceedings, Part II 22}}. Springer, \bibinfo{pages}{62--79}.
\newblock


\bibitem[\protect\citeauthoryear{Gunning, Stefik, Choi, Miller, Stumpf, and
  Yang}{Gunning et~al\mbox{.}}{2019}]%
        {gunning2019xai}
\bibfield{author}{\bibinfo{person}{David Gunning}, \bibinfo{person}{Mark
  Stefik}, \bibinfo{person}{Jaesik Choi}, \bibinfo{person}{Timothy Miller},
  \bibinfo{person}{Simone Stumpf}, {and} \bibinfo{person}{Guang-Zhong Yang}.}
  \bibinfo{year}{2019}\natexlab{}.
\newblock \showarticletitle{XAI—Explainable artificial intelligence}.
\newblock \bibinfo{journal}{{\em Science robotics\/}} \bibinfo{volume}{4},
  \bibinfo{number}{37} (\bibinfo{year}{2019}), \bibinfo{pages}{eaay7120}.
\newblock


\bibitem[\protect\citeauthoryear{Guo, Mu, Xu, Su, Wang, and Xing}{Guo
  et~al\mbox{.}}{2018}]%
        {guo2018lemna}
\bibfield{author}{\bibinfo{person}{Wenbo Guo}, \bibinfo{person}{Dongliang Mu},
  \bibinfo{person}{Jun Xu}, \bibinfo{person}{Purui Su}, \bibinfo{person}{Gang
  Wang}, {and} \bibinfo{person}{Xinyu Xing}.} \bibinfo{year}{2018}\natexlab{}.
\newblock \showarticletitle{Lemna: Explaining deep learning based security
  applications}. In \bibinfo{booktitle}{{\em Proceedings of the 2018 ACM SIGSAC
  Conference on Computer and Communications Security}}.
  \bibinfo{pages}{364--379}.
\newblock


\bibitem[\protect\citeauthoryear{Han, Wang, Chen, Wang, Yu, Wang, Zhang, Wang,
  Jin, Yang, et~al\mbox{.}}{Han et~al\mbox{.}}{2023}]%
        {han23anomaly}
\bibfield{author}{\bibinfo{person}{Dongqi Han}, \bibinfo{person}{Zhiliang
  Wang}, \bibinfo{person}{Wenqi Chen}, \bibinfo{person}{Kai Wang},
  \bibinfo{person}{Rui Yu}, \bibinfo{person}{Su Wang}, \bibinfo{person}{Han
  Zhang}, \bibinfo{person}{Zhihua Wang}, \bibinfo{person}{Minghui Jin},
  \bibinfo{person}{Jiahai Yang}, {et~al\mbox{.}}}
  \bibinfo{year}{2023}\natexlab{}.
\newblock \showarticletitle{Anomaly Detection in the Open World: Normality
  Shift Detection, Explanation, and Adaptation}.
\newblock  (\bibinfo{year}{2023}).
\newblock


\bibitem[\protect\citeauthoryear{Han, Wang, Chen, Zhong, Wang, Zhang, Yang,
  Shi, and Yin}{Han et~al\mbox{.}}{2021}]%
        {ccs21deepaid}
\bibfield{author}{\bibinfo{person}{Dongqi Han}, \bibinfo{person}{Zhiliang
  Wang}, \bibinfo{person}{Wenqi Chen}, \bibinfo{person}{Ying Zhong},
  \bibinfo{person}{Su Wang}, \bibinfo{person}{Han Zhang},
  \bibinfo{person}{Jiahai Yang}, \bibinfo{person}{Xingang Shi}, {and}
  \bibinfo{person}{Xia Yin}.} \bibinfo{year}{2021}\natexlab{}.
\newblock \showarticletitle{DeepAID: Interpreting and Improving Deep
  Learning-Based Anomaly Detection in Security Applications}. In
  \bibinfo{booktitle}{{\em Proceedings of the 2021 ACM SIGSAC Conference on
  Computer and Communications Security}} {\em (\bibinfo{series}{CCS '21})}.
  \bibinfo{publisher}{Association for Computing Machinery},
  \bibinfo{address}{New York, NY, USA}, \bibinfo{pages}{3197–3217}.
\newblock
\showISBNx{9781450384544}
\showDOI{%
\url{https://doi.org/10.1145/3460120.3484589}}


\bibitem[\protect\citeauthoryear{He, Liu, Wu, Yang, Ren, and Qin}{He
  et~al\mbox{.}}{2022}]%
        {he2022msdroid}
\bibfield{author}{\bibinfo{person}{Yiling He}, \bibinfo{person}{Yiping Liu},
  \bibinfo{person}{Lei Wu}, \bibinfo{person}{Ziqi Yang}, \bibinfo{person}{Kui
  Ren}, {and} \bibinfo{person}{Zhan Qin}.} \bibinfo{year}{2022}\natexlab{}.
\newblock \showarticletitle{MsDroid: Identifying Malicious Snippets for Android
  Malware Detection}.
\newblock \bibinfo{journal}{{\em IEEE Transactions on Dependable and Secure
  Computing\/}} (\bibinfo{year}{2022}).
\newblock


\bibitem[\protect\citeauthoryear{Jeyakumar, Noor, Cheng, Garcia, and
  Srivastava}{Jeyakumar et~al\mbox{.}}{2020}]%
        {jeyakumar2020can}
\bibfield{author}{\bibinfo{person}{Jeya~Vikranth Jeyakumar},
  \bibinfo{person}{Joseph Noor}, \bibinfo{person}{Yu-Hsi Cheng},
  \bibinfo{person}{Luis Garcia}, {and} \bibinfo{person}{Mani Srivastava}.}
  \bibinfo{year}{2020}\natexlab{}.
\newblock \showarticletitle{How can i explain this to you? an empirical study
  of deep neural network explanation methods}.
\newblock \bibinfo{journal}{{\em Advances in Neural Information Processing
  Systems\/}}  \bibinfo{volume}{33} (\bibinfo{year}{2020}),
  \bibinfo{pages}{4211--4222}.
\newblock


\bibitem[\protect\citeauthoryear{Kim, Wattenberg, Gilmer, Cai, Wexler, Viegas,
  et~al\mbox{.}}{Kim et~al\mbox{.}}{2018b}]%
        {kim2018interpretability}
\bibfield{author}{\bibinfo{person}{Been Kim}, \bibinfo{person}{Martin
  Wattenberg}, \bibinfo{person}{Justin Gilmer}, \bibinfo{person}{Carrie Cai},
  \bibinfo{person}{James Wexler}, \bibinfo{person}{Fernanda Viegas},
  {et~al\mbox{.}}} \bibinfo{year}{2018}\natexlab{b}.
\newblock \showarticletitle{Interpretability beyond feature attribution:
  Quantitative testing with concept activation vectors (tcav)}. In
  \bibinfo{booktitle}{{\em International conference on machine learning}}.
  PMLR, \bibinfo{pages}{2668--2677}.
\newblock


\bibitem[\protect\citeauthoryear{Kim, Kang, Rho, Sezer, and Im}{Kim
  et~al\mbox{.}}{2018a}]%
        {kim2018multimodal}
\bibfield{author}{\bibinfo{person}{TaeGuen Kim}, \bibinfo{person}{BooJoong
  Kang}, \bibinfo{person}{Mina Rho}, \bibinfo{person}{Sakir Sezer}, {and}
  \bibinfo{person}{Eul~Gyu Im}.} \bibinfo{year}{2018}\natexlab{a}.
\newblock \showarticletitle{A multimodal deep learning method for android
  malware detection using various features}.
\newblock \bibinfo{journal}{{\em IEEE Transactions on Information Forensics and
  Security\/}} \bibinfo{volume}{14}, \bibinfo{number}{3}
  (\bibinfo{year}{2018}), \bibinfo{pages}{773--788}.
\newblock


\bibitem[\protect\citeauthoryear{Kingma and Ba}{Kingma and Ba}{2014}]%
        {kingma2014adam}
\bibfield{author}{\bibinfo{person}{Diederik~P Kingma} {and}
  \bibinfo{person}{Jimmy Ba}.} \bibinfo{year}{2014}\natexlab{}.
\newblock \showarticletitle{Adam: A method for stochastic optimization}.
\newblock \bibinfo{journal}{{\em arXiv preprint arXiv:1412.6980\/}}
  (\bibinfo{year}{2014}).
\newblock


\bibitem[\protect\citeauthoryear{Kong and Yan}{Kong and Yan}{2013}]%
        {kong2013discriminant}
\bibfield{author}{\bibinfo{person}{Deguang Kong} {and} \bibinfo{person}{Guanhua
  Yan}.} \bibinfo{year}{2013}\natexlab{}.
\newblock \showarticletitle{Discriminant malware distance learning on
  structural information for automated malware classification}. In
  \bibinfo{booktitle}{{\em Proceedings of the 19th ACM SIGKDD international
  conference on Knowledge discovery and data mining}}.
  \bibinfo{pages}{1357--1365}.
\newblock


\bibitem[\protect\citeauthoryear{Laskov et~al\mbox{.}}{Laskov
  et~al\mbox{.}}{2014}]%
        {laskov2014practical}
\bibfield{author}{\bibinfo{person}{Pavel Laskov} {et~al\mbox{.}}}
  \bibinfo{year}{2014}\natexlab{}.
\newblock \showarticletitle{Practical evasion of a learning-based classifier: A
  case study}. In \bibinfo{booktitle}{{\em 2014 IEEE symposium on security and
  privacy}}. IEEE, \bibinfo{pages}{197--211}.
\newblock


\bibitem[\protect\citeauthoryear{Li, Wu, Peng, Ernst, and Fu}{Li
  et~al\mbox{.}}{2018a}]%
        {li2018tell}
\bibfield{author}{\bibinfo{person}{Kunpeng Li}, \bibinfo{person}{Ziyan Wu},
  \bibinfo{person}{Kuan-Chuan Peng}, \bibinfo{person}{Jan Ernst}, {and}
  \bibinfo{person}{Yun Fu}.} \bibinfo{year}{2018}\natexlab{a}.
\newblock \showarticletitle{Tell me where to look: Guided attention inference
  network}. In \bibinfo{booktitle}{{\em Proceedings of the IEEE conference on
  computer vision and pattern recognition}}. \bibinfo{pages}{9215--9223}.
\newblock


\bibitem[\protect\citeauthoryear{Li, Zou, Xu, Ou, Jin, Wang, Deng, and
  Zhong}{Li et~al\mbox{.}}{2018b}]%
        {li2018vuldeepecker}
\bibfield{author}{\bibinfo{person}{Zhen Li}, \bibinfo{person}{Deqing Zou},
  \bibinfo{person}{Shouhuai Xu}, \bibinfo{person}{Xinyu Ou},
  \bibinfo{person}{Hai Jin}, \bibinfo{person}{Sujuan Wang},
  \bibinfo{person}{Zhijun Deng}, {and} \bibinfo{person}{Yuyi Zhong}.}
  \bibinfo{year}{2018}\natexlab{b}.
\newblock \showarticletitle{Vuldeepecker: A deep learning-based system for
  vulnerability detection}.
\newblock \bibinfo{journal}{{\em arXiv preprint arXiv:1801.01681\/}}
  (\bibinfo{year}{2018}).
\newblock


\bibitem[\protect\citeauthoryear{Liao and Varshney}{Liao and Varshney}{2021}]%
        {liao2021human}
\bibfield{author}{\bibinfo{person}{Q~Vera Liao} {and} \bibinfo{person}{Kush~R
  Varshney}.} \bibinfo{year}{2021}\natexlab{}.
\newblock \showarticletitle{Human-centered explainable ai (xai): From
  algorithms to user experiences}.
\newblock \bibinfo{journal}{{\em arXiv preprint arXiv:2110.10790\/}}
  (\bibinfo{year}{2021}).
\newblock


\bibitem[\protect\citeauthoryear{Lin, Wen, Han, Zhang, and Xiang}{Lin
  et~al\mbox{.}}{2020}]%
        {lin2020software}
\bibfield{author}{\bibinfo{person}{Guanjun Lin}, \bibinfo{person}{Sheng Wen},
  \bibinfo{person}{Qing-Long Han}, \bibinfo{person}{Jun Zhang}, {and}
  \bibinfo{person}{Yang Xiang}.} \bibinfo{year}{2020}\natexlab{}.
\newblock \showarticletitle{Software vulnerability detection using deep neural
  networks: a survey}.
\newblock \bibinfo{journal}{{\it Proc. IEEE}} \bibinfo{volume}{108},
  \bibinfo{number}{10} (\bibinfo{year}{2020}), \bibinfo{pages}{1825--1848}.
\newblock


\bibitem[\protect\citeauthoryear{Liu, Lin, Han, Wen, Zhang, and Xiang}{Liu
  et~al\mbox{.}}{2019}]%
        {liu2019deepbalance}
\bibfield{author}{\bibinfo{person}{Shigang Liu}, \bibinfo{person}{Guanjun Lin},
  \bibinfo{person}{Qing-Long Han}, \bibinfo{person}{Sheng Wen},
  \bibinfo{person}{Jun Zhang}, {and} \bibinfo{person}{Yang Xiang}.}
  \bibinfo{year}{2019}\natexlab{}.
\newblock \showarticletitle{DeepBalance: Deep-learning and fuzzy oversampling
  for vulnerability detection}.
\newblock \bibinfo{journal}{{\em IEEE Transactions on Fuzzy Systems\/}}
  \bibinfo{volume}{28}, \bibinfo{number}{7} (\bibinfo{year}{2019}),
  \bibinfo{pages}{1329--1343}.
\newblock


\bibitem[\protect\citeauthoryear{Lopes, Silva, Braga, Oliveira, and
  Rosado}{Lopes et~al\mbox{.}}{2022}]%
        {lopes2022xai}
\bibfield{author}{\bibinfo{person}{Pedro Lopes}, \bibinfo{person}{Eduardo
  Silva}, \bibinfo{person}{Cristiana Braga}, \bibinfo{person}{Tiago Oliveira},
  {and} \bibinfo{person}{Lu{\'\i}s Rosado}.} \bibinfo{year}{2022}\natexlab{}.
\newblock \showarticletitle{XAI Systems Evaluation: A Review of Human and
  Computer-Centred Methods}.
\newblock \bibinfo{journal}{{\em Applied Sciences\/}} \bibinfo{volume}{12},
  \bibinfo{number}{19} (\bibinfo{year}{2022}), \bibinfo{pages}{9423}.
\newblock


\bibitem[\protect\citeauthoryear{Lundberg and Lee}{Lundberg and Lee}{2017}]%
        {lundberg2017unified}
\bibfield{author}{\bibinfo{person}{Scott~M Lundberg} {and}
  \bibinfo{person}{Su-In Lee}.} \bibinfo{year}{2017}\natexlab{}.
\newblock \showarticletitle{A unified approach to interpreting model
  predictions}.
\newblock \bibinfo{journal}{{\em Advances in neural information processing
  systems\/}}  \bibinfo{volume}{30} (\bibinfo{year}{2017}).
\newblock


\bibitem[\protect\citeauthoryear{Lundstrom, Huang, and Razaviyayn}{Lundstrom
  et~al\mbox{.}}{2022}]%
        {lundstrom2022rigorous}
\bibfield{author}{\bibinfo{person}{Daniel~D Lundstrom},
  \bibinfo{person}{Tianjian Huang}, {and} \bibinfo{person}{Meisam Razaviyayn}.}
  \bibinfo{year}{2022}\natexlab{}.
\newblock \showarticletitle{A rigorous study of integrated gradients method and
  extensions to internal neuron attributions}. In \bibinfo{booktitle}{{\em
  International Conference on Machine Learning}}. PMLR,
  \bibinfo{pages}{14485--14508}.
\newblock


\bibitem[\protect\citeauthoryear{Mamalakis, Barnes, and Ebert-Uphoff}{Mamalakis
  et~al\mbox{.}}{2022}]%
        {mamalakis2022carefully}
\bibfield{author}{\bibinfo{person}{Antonios Mamalakis},
  \bibinfo{person}{Elizabeth~A Barnes}, {and} \bibinfo{person}{Imme
  Ebert-Uphoff}.} \bibinfo{year}{2022}\natexlab{}.
\newblock \showarticletitle{Carefully choose the baseline: Lessons learned from
  applying XAI attribution methods for regression tasks in geoscience}.
\newblock \bibinfo{journal}{{\em Artificial Intelligence for the Earth
  Systems\/}} (\bibinfo{year}{2022}), \bibinfo{pages}{1--18}.
\newblock


\bibitem[\protect\citeauthoryear{Mane and Rao}{Mane and Rao}{2021}]%
        {xnids}
\bibfield{author}{\bibinfo{person}{Shraddha Mane} {and}
  \bibinfo{person}{Dattaraj Rao}.} \bibinfo{year}{2021}\natexlab{}.
\newblock \bibinfo{title}{Explaining Network Intrusion Detection System Using
  Explainable AI Framework}.
\newblock   (\bibinfo{year}{2021}).
\newblock
\showDOI{%
\url{https://doi.org/10.48550/ARXIV.2103.07110}}


\bibitem[\protect\citeauthoryear{Mantovani, Aonzo, Fratantonio, and
  Balzarotti}{Mantovani et~al\mbox{.}}{2022}]%
        {antovaniAFB22}
\bibfield{author}{\bibinfo{person}{Alessandro Mantovani},
  \bibinfo{person}{Simone Aonzo}, \bibinfo{person}{Yanick Fratantonio}, {and}
  \bibinfo{person}{Davide Balzarotti}.} \bibinfo{year}{2022}\natexlab{}.
\newblock \showarticletitle{RE-Mind: a First Look Inside the Mind of a Reverse
  Engineer}. In \bibinfo{booktitle}{{\em 31st {USENIX} Security Symposium,
  {USENIX} Security 2022, Boston, MA, USA, August 10-12, 2022}},
  \bibfield{editor}{\bibinfo{person}{Kevin R.~B. Butler} {and}
  \bibinfo{person}{Kurt Thomas}} (Eds.). \bibinfo{publisher}{{USENIX}
  Association}, \bibinfo{pages}{2727--2745}.
\newblock


\bibitem[\protect\citeauthoryear{marcotcr}{marcotcr}{2023}]%
        {lime}
\bibfield{author}{\bibinfo{person}{marcotcr}.} \bibinfo{year}{2023}\natexlab{}.
\newblock \bibinfo{title}{LIME}.
\newblock \bibinfo{howpublished}{\url{https://github.com/marcotcr/lime}}.
  (\bibinfo{year}{2023}).
\newblock


\bibitem[\protect\citeauthoryear{Mariconti, Onwuzurike, Andriotis,
  De~Cristofaro, Ross, and Stringhini}{Mariconti et~al\mbox{.}}{2017}]%
        {mariconti2016mamadroid}
\bibfield{author}{\bibinfo{person}{E Mariconti}, \bibinfo{person}{L
  Onwuzurike}, \bibinfo{person}{P Andriotis}, \bibinfo{person}{E
  De~Cristofaro}, \bibinfo{person}{G Ross}, {and} \bibinfo{person}{G
  Stringhini}.} \bibinfo{year}{2017}\natexlab{}.
\newblock \showarticletitle{MamaDroid: Detecting Android Malware by Building
  Markov Chains of Behavioral Models}. In \bibinfo{booktitle}{{\em Proceedings
  of the Network and Distributed Systems Security Symposium (NDSS)}}.
\newblock


\bibitem[\protect\citeauthoryear{Masum and Shahriar}{Masum and
  Shahriar}{2019}]%
        {masum2019droid}
\bibfield{author}{\bibinfo{person}{Mohammad Masum} {and}
  \bibinfo{person}{Hossain Shahriar}.} \bibinfo{year}{2019}\natexlab{}.
\newblock \showarticletitle{Droid-NNet: Deep learning neural network for
  android malware detection}. In \bibinfo{booktitle}{{\em 2019 IEEE
  International Conference on Big Data (Big Data)}}. IEEE,
  \bibinfo{pages}{5789--5793}.
\newblock


\bibitem[\protect\citeauthoryear{McLaughlin, Martinez~del Rincon, Kang, Yerima,
  Miller, Sezer, Safaei, Trickel, Zhao, Doup{\'e}, et~al\mbox{.}}{McLaughlin
  et~al\mbox{.}}{2017}]%
        {mclaughlin2017deep}
\bibfield{author}{\bibinfo{person}{Niall McLaughlin}, \bibinfo{person}{Jesus
  Martinez~del Rincon}, \bibinfo{person}{BooJoong Kang},
  \bibinfo{person}{Suleiman Yerima}, \bibinfo{person}{Paul Miller},
  \bibinfo{person}{Sakir Sezer}, \bibinfo{person}{Yeganeh Safaei},
  \bibinfo{person}{Erik Trickel}, \bibinfo{person}{Ziming Zhao},
  \bibinfo{person}{Adam Doup{\'e}}, {et~al\mbox{.}}}
  \bibinfo{year}{2017}\natexlab{}.
\newblock \showarticletitle{Deep android malware detection}. In
  \bibinfo{booktitle}{{\em Proceedings of the seventh ACM on conference on data
  and application security and privacy}}. \bibinfo{pages}{301--308}.
\newblock


\bibitem[\protect\citeauthoryear{Mikolov, Chen, Corrado, and Dean}{Mikolov
  et~al\mbox{.}}{2013}]%
        {mikolov2013efficient}
\bibfield{author}{\bibinfo{person}{Tomas Mikolov}, \bibinfo{person}{Kai Chen},
  \bibinfo{person}{Greg Corrado}, {and} \bibinfo{person}{Jeffrey Dean}.}
  \bibinfo{year}{2013}\natexlab{}.
\newblock \showarticletitle{Efficient estimation of word representations in
  vector space}.
\newblock \bibinfo{journal}{{\em arXiv preprint arXiv:1301.3781\/}}
  (\bibinfo{year}{2013}).
\newblock


\bibitem[\protect\citeauthoryear{Nourani, Kabir, Mohseni, and Ragan}{Nourani
  et~al\mbox{.}}{2019}]%
        {nourani2019effects}
\bibfield{author}{\bibinfo{person}{Mahsan Nourani}, \bibinfo{person}{Samia
  Kabir}, \bibinfo{person}{Sina Mohseni}, {and} \bibinfo{person}{Eric~D
  Ragan}.} \bibinfo{year}{2019}\natexlab{}.
\newblock \showarticletitle{The effects of meaningful and meaningless
  explanations on trust and perceived system accuracy in intelligent systems}.
  In \bibinfo{booktitle}{{\em Proceedings of the AAAI Conference on Human
  Computation and Crowdsourcing}}, Vol.~\bibinfo{volume}{7}.
  \bibinfo{pages}{97--105}.
\newblock


\bibitem[\protect\citeauthoryear{Nowozin, Rother, Bagon, Sharp, Yao, and
  Kohli}{Nowozin et~al\mbox{.}}{2011}]%
        {nowozin2011decision}
\bibfield{author}{\bibinfo{person}{Sebastian Nowozin}, \bibinfo{person}{Carsten
  Rother}, \bibinfo{person}{Shai Bagon}, \bibinfo{person}{Toby Sharp},
  \bibinfo{person}{Bangpeng Yao}, {and} \bibinfo{person}{Pushmeet Kohli}.}
  \bibinfo{year}{2011}\natexlab{}.
\newblock \showarticletitle{Decision tree fields}. In \bibinfo{booktitle}{{\em
  2011 International Conference on Computer Vision}}. IEEE,
  \bibinfo{pages}{1668--1675}.
\newblock


\bibitem[\protect\citeauthoryear{Obaidat, Sridhar, Pham, and Phung}{Obaidat
  et~al\mbox{.}}{2022}]%
        {obaidat2022jadeite}
\bibfield{author}{\bibinfo{person}{Islam Obaidat}, \bibinfo{person}{Meera
  Sridhar}, \bibinfo{person}{Khue~M Pham}, {and} \bibinfo{person}{Phu~H
  Phung}.} \bibinfo{year}{2022}\natexlab{}.
\newblock \showarticletitle{Jadeite: A novel image-behavior-based approach for
  java malware detection using deep learning}.
\newblock \bibinfo{journal}{{\em Computers \& Security\/}}
  \bibinfo{volume}{113} (\bibinfo{year}{2022}), \bibinfo{pages}{102547}.
\newblock


\bibitem[\protect\citeauthoryear{Pierazzi, Pendlebury, Cortellazzi, and
  Cavallaro}{Pierazzi et~al\mbox{.}}{2020}]%
        {pierazzi2020intriguing}
\bibfield{author}{\bibinfo{person}{Fabio Pierazzi}, \bibinfo{person}{Feargus
  Pendlebury}, \bibinfo{person}{Jacopo Cortellazzi}, {and}
  \bibinfo{person}{Lorenzo Cavallaro}.} \bibinfo{year}{2020}\natexlab{}.
\newblock \showarticletitle{Intriguing properties of adversarial ml attacks in
  the problem space}. In \bibinfo{booktitle}{{\em 2020 IEEE symposium on
  security and privacy (SP)}}. IEEE, \bibinfo{pages}{1332--1349}.
\newblock


\bibitem[\protect\citeauthoryear{Pillai and Pirsiavash}{Pillai and
  Pirsiavash}{2021}]%
        {pillai2021explainable}
\bibfield{author}{\bibinfo{person}{Vipin Pillai} {and} \bibinfo{person}{Hamed
  Pirsiavash}.} \bibinfo{year}{2021}\natexlab{}.
\newblock \showarticletitle{Explainable models with consistent
  interpretations}. In \bibinfo{booktitle}{{\em Proceedings of the AAAI
  Conference on Artificial Intelligence}}, Vol.~\bibinfo{volume}{35}.
  \bibinfo{pages}{2431--2439}.
\newblock


\bibitem[\protect\citeauthoryear{Pirch, Warnecke, Wressnegger, and Rieck}{Pirch
  et~al\mbox{.}}{2021a}]%
        {Pirch21TagVet}
\bibfield{author}{\bibinfo{person}{Lukas Pirch}, \bibinfo{person}{Alexander
  Warnecke}, \bibinfo{person}{Christian Wressnegger}, {and}
  \bibinfo{person}{Konrad Rieck}.} \bibinfo{year}{2021}\natexlab{a}.
\newblock \showarticletitle{TagVet: Vetting Malware Tags Using Explainable
  Machine Learning} {\em (\bibinfo{series}{EuroSec '21})}.
  \bibinfo{publisher}{Association for Computing Machinery},
  \bibinfo{address}{New York, NY, USA}, 7.
\newblock
\showISBNx{9781450383370}
\showDOI{%
\url{https://doi.org/10.1145/3447852.3458719}}


\bibitem[\protect\citeauthoryear{Pirch, Warnecke, Wressnegger, and Rieck}{Pirch
  et~al\mbox{.}}{2021b}]%
        {pirch2021tagvet}
\bibfield{author}{\bibinfo{person}{Lukas Pirch}, \bibinfo{person}{Alexander
  Warnecke}, \bibinfo{person}{Christian Wressnegger}, {and}
  \bibinfo{person}{Konrad Rieck}.} \bibinfo{year}{2021}\natexlab{b}.
\newblock \showarticletitle{Tagvet: Vetting malware tags using explainable
  machine learning}. In \bibinfo{booktitle}{{\em Proceedings of the 14th
  European Workshop on Systems Security}}. \bibinfo{pages}{34--40}.
\newblock


\bibitem[\protect\citeauthoryear{Plumb, Al-Shedivat, Cabrera, Perer, Xing, and
  Talwalkar}{Plumb et~al\mbox{.}}{2020}]%
        {plumb2020regularizing}
\bibfield{author}{\bibinfo{person}{Gregory Plumb}, \bibinfo{person}{Maruan
  Al-Shedivat}, \bibinfo{person}{{\'A}ngel~Alexander Cabrera},
  \bibinfo{person}{Adam Perer}, \bibinfo{person}{Eric Xing}, {and}
  \bibinfo{person}{Ameet Talwalkar}.} \bibinfo{year}{2020}\natexlab{}.
\newblock \showarticletitle{Regularizing black-box models for improved
  interpretability}.
\newblock \bibinfo{journal}{{\em Advances in Neural Information Processing
  Systems\/}}  \bibinfo{volume}{33} (\bibinfo{year}{2020}),
  \bibinfo{pages}{10526--10536}.
\newblock


\bibitem[\protect\citeauthoryear{Ribeiro, Singh, and Guestrin}{Ribeiro
  et~al\mbox{.}}{2016}]%
        {ribeiro2016should}
\bibfield{author}{\bibinfo{person}{Marco~Tulio Ribeiro},
  \bibinfo{person}{Sameer Singh}, {and} \bibinfo{person}{Carlos Guestrin}.}
  \bibinfo{year}{2016}\natexlab{}.
\newblock \showarticletitle{`Why should i trust you?' Explaining the
  predictions of any classifier}. In \bibinfo{booktitle}{{\em Proceedings of
  the 22nd ACM SIGKDD international conference on knowledge discovery and data
  mining}}. \bibinfo{pages}{1135--1144}.
\newblock


\bibitem[\protect\citeauthoryear{Ronneberger, Fischer, and Brox}{Ronneberger
  et~al\mbox{.}}{2015}]%
        {ronneberger2015u}
\bibfield{author}{\bibinfo{person}{Olaf Ronneberger}, \bibinfo{person}{Philipp
  Fischer}, {and} \bibinfo{person}{Thomas Brox}.}
  \bibinfo{year}{2015}\natexlab{}.
\newblock \showarticletitle{U-net: Convolutional networks for biomedical image
  segmentation}. In \bibinfo{booktitle}{{\em International Conference on
  Medical image computing and computer-assisted intervention}}. Springer,
  \bibinfo{pages}{234--241}.
\newblock


\bibitem[\protect\citeauthoryear{Ross, Hughes, and Doshi-Velez}{Ross
  et~al\mbox{.}}{2017}]%
        {ross2017right}
\bibfield{author}{\bibinfo{person}{Andrew~Slavin Ross},
  \bibinfo{person}{Michael~C Hughes}, {and} \bibinfo{person}{Finale
  Doshi-Velez}.} \bibinfo{year}{2017}\natexlab{}.
\newblock \showarticletitle{Right for the right reasons: training
  differentiable models by constraining their explanations}. In
  \bibinfo{booktitle}{{\em Proceedings of the 26th International Joint
  Conference on Artificial Intelligence}}. \bibinfo{pages}{2662--2670}.
\newblock


\bibitem[\protect\citeauthoryear{Roy, DeLoach, Li, Herndon, Caragea, Ou,
  Ranganath, Li, and Guevara}{Roy et~al\mbox{.}}{2015}]%
        {roy2015experimental}
\bibfield{author}{\bibinfo{person}{Sankardas Roy}, \bibinfo{person}{Jordan
  DeLoach}, \bibinfo{person}{Yuping Li}, \bibinfo{person}{Nic Herndon},
  \bibinfo{person}{Doina Caragea}, \bibinfo{person}{Xinming Ou},
  \bibinfo{person}{Venkatesh~Prasad Ranganath}, \bibinfo{person}{Hongmin Li},
  {and} \bibinfo{person}{Nicolais Guevara}.} \bibinfo{year}{2015}\natexlab{}.
\newblock \showarticletitle{Experimental study with real-world data for android
  app security analysis using machine learning}. In \bibinfo{booktitle}{{\em
  Proceedings of the 31st Annual Computer Security Applications Conference}}.
  \bibinfo{pages}{81--90}.
\newblock


\bibitem[\protect\citeauthoryear{Saeed and Omlin}{Saeed and Omlin}{2021}]%
        {saeed2021explainable}
\bibfield{author}{\bibinfo{person}{Waddah Saeed} {and}
  \bibinfo{person}{Christian Omlin}.} \bibinfo{year}{2021}\natexlab{}.
\newblock \showarticletitle{Explainable ai (xai): A systematic meta-survey of
  current challenges and future opportunities}.
\newblock \bibinfo{journal}{{\em arXiv preprint arXiv:2111.06420\/}}
  (\bibinfo{year}{2021}).
\newblock


\bibitem[\protect\citeauthoryear{Samtani, Chen, Kantarcioglu, and
  Thuraisingham}{Samtani et~al\mbox{.}}{2022}]%
        {samtani2022explainable}
\bibfield{author}{\bibinfo{person}{Sagar Samtani}, \bibinfo{person}{Hsinchun
  Chen}, \bibinfo{person}{Murat Kantarcioglu}, {and} \bibinfo{person}{Bhavani
  Thuraisingham}.} \bibinfo{year}{2022}\natexlab{}.
\newblock \showarticletitle{Explainable Artificial Intelligence for Cyber
  Threat Intelligence (XAI-CTI)}.
\newblock \bibinfo{journal}{{\em IEEE Transactions on Dependable and Secure
  Computing\/}} \bibinfo{volume}{19}, \bibinfo{number}{04}
  (\bibinfo{year}{2022}), \bibinfo{pages}{2149--2150}.
\newblock


\bibitem[\protect\citeauthoryear{Schroff, Kalenichenko, and Philbin}{Schroff
  et~al\mbox{.}}{2015}]%
        {schroff2015facenet}
\bibfield{author}{\bibinfo{person}{Florian Schroff}, \bibinfo{person}{Dmitry
  Kalenichenko}, {and} \bibinfo{person}{James Philbin}.}
  \bibinfo{year}{2015}\natexlab{}.
\newblock \showarticletitle{Facenet: A unified embedding for face recognition
  and clustering}. In \bibinfo{booktitle}{{\em Proceedings of the IEEE
  conference on computer vision and pattern recognition}}.
  \bibinfo{pages}{815--823}.
\newblock


\bibitem[\protect\citeauthoryear{Selvaraju, Cogswell, Das, Vedantam, Parikh,
  and Batra}{Selvaraju et~al\mbox{.}}{2017}]%
        {selvaraju2017grad}
\bibfield{author}{\bibinfo{person}{Ramprasaath~R Selvaraju},
  \bibinfo{person}{Michael Cogswell}, \bibinfo{person}{Abhishek Das},
  \bibinfo{person}{Ramakrishna Vedantam}, \bibinfo{person}{Devi Parikh}, {and}
  \bibinfo{person}{Dhruv Batra}.} \bibinfo{year}{2017}\natexlab{}.
\newblock \showarticletitle{Grad-cam: Visual explanations from deep networks
  via gradient-based localization}. In \bibinfo{booktitle}{{\em Proceedings of
  the IEEE international conference on computer vision}}.
  \bibinfo{pages}{618--626}.
\newblock


\bibitem[\protect\citeauthoryear{Severi, Meyer, Coull, and Oprea}{Severi
  et~al\mbox{.}}{2021}]%
        {severi2021explanation}
\bibfield{author}{\bibinfo{person}{Giorgio Severi}, \bibinfo{person}{Jim
  Meyer}, \bibinfo{person}{Scott Coull}, {and} \bibinfo{person}{Alina Oprea}.}
  \bibinfo{year}{2021}\natexlab{}.
\newblock \showarticletitle{$\{$Explanation-Guided$\}$ Backdoor Poisoning
  Attacks Against Malware Classifiers}. In \bibinfo{booktitle}{{\em 30th USENIX
  Security Symposium (USENIX Security 21)}}. \bibinfo{pages}{1487--1504}.
\newblock


\bibitem[\protect\citeauthoryear{Shen and Huang}{Shen and Huang}{2020}]%
        {shen2020useful}
\bibfield{author}{\bibinfo{person}{Hua Shen} {and} \bibinfo{person}{Ting-Hao
  Huang}.} \bibinfo{year}{2020}\natexlab{}.
\newblock \showarticletitle{How useful are the machine-generated
  interpretations to general users? A human evaluation on guessing the
  incorrectly predicted labels}. In \bibinfo{booktitle}{{\em Proceedings of the
  AAAI Conference on Human Computation and Crowdsourcing}},
  Vol.~\bibinfo{volume}{8}. \bibinfo{pages}{168--172}.
\newblock


\bibitem[\protect\citeauthoryear{Shin, Song, and Moazzezi}{Shin
  et~al\mbox{.}}{2015}]%
        {shin2015recognizing}
\bibfield{author}{\bibinfo{person}{Eui Chul~Richard Shin},
  \bibinfo{person}{Dawn Song}, {and} \bibinfo{person}{Reza Moazzezi}.}
  \bibinfo{year}{2015}\natexlab{}.
\newblock \showarticletitle{Recognizing functions in binaries with neural
  networks}. In \bibinfo{booktitle}{{\em 24th USENIX security symposium (USENIX
  Security 15)}}. \bibinfo{pages}{611--626}.
\newblock


\bibitem[\protect\citeauthoryear{Shrikumar, Greenside, and Kundaje}{Shrikumar
  et~al\mbox{.}}{2017}]%
        {shrikumar2017learning}
\bibfield{author}{\bibinfo{person}{Avanti Shrikumar}, \bibinfo{person}{Peyton
  Greenside}, {and} \bibinfo{person}{Anshul Kundaje}.}
  \bibinfo{year}{2017}\natexlab{}.
\newblock \showarticletitle{Learning important features through propagating
  activation differences}. In \bibinfo{booktitle}{{\em International conference
  on machine learning}}. PMLR, \bibinfo{pages}{3145--3153}.
\newblock


\bibitem[\protect\citeauthoryear{Simonyan, Vedaldi, and Zisserman}{Simonyan
  et~al\mbox{.}}{2014}]%
        {simonyan2014deep}
\bibfield{author}{\bibinfo{person}{Karen Simonyan}, \bibinfo{person}{Andrea
  Vedaldi}, {and} \bibinfo{person}{Andrew Zisserman}.}
  \bibinfo{year}{2014}\natexlab{}.
\newblock \showarticletitle{Deep inside convolutional networks: Visualising
  image classification models and saliency maps}. In \bibinfo{booktitle}{{\em
  In Workshop at International Conference on Learning Representations}}.
  Citeseer.
\newblock


\bibitem[\protect\citeauthoryear{Smutz and Stavrou}{Smutz and Stavrou}{2012}]%
        {smutz2012malicious}
\bibfield{author}{\bibinfo{person}{Charles Smutz} {and}
  \bibinfo{person}{Angelos Stavrou}.} \bibinfo{year}{2012}\natexlab{}.
\newblock \showarticletitle{Malicious PDF detection using metadata and
  structural features}. In \bibinfo{booktitle}{{\em Proceedings of the 28th
  annual computer security applications conference}}.
  \bibinfo{pages}{239--248}.
\newblock


\bibitem[\protect\citeauthoryear{Sundararajan, Taly, and Yan}{Sundararajan
  et~al\mbox{.}}{2017}]%
        {sundararajan2017axiomatic}
\bibfield{author}{\bibinfo{person}{Mukund Sundararajan}, \bibinfo{person}{Ankur
  Taly}, {and} \bibinfo{person}{Qiqi Yan}.} \bibinfo{year}{2017}\natexlab{}.
\newblock \showarticletitle{Axiomatic attribution for deep networks}. In
  \bibinfo{booktitle}{{\em International conference on machine learning}}.
  PMLR, \bibinfo{pages}{3319--3328}.
\newblock


\bibitem[\protect\citeauthoryear{Vaswani, Shazeer, Parmar, Uszkoreit, Jones,
  Gomez, Kaiser, and Polosukhin}{Vaswani et~al\mbox{.}}{2017}]%
        {vaswani2017attention}
\bibfield{author}{\bibinfo{person}{Ashish Vaswani}, \bibinfo{person}{Noam
  Shazeer}, \bibinfo{person}{Niki Parmar}, \bibinfo{person}{Jakob Uszkoreit},
  \bibinfo{person}{Llion Jones}, \bibinfo{person}{Aidan~N Gomez},
  \bibinfo{person}{{\L}ukasz Kaiser}, {and} \bibinfo{person}{Illia
  Polosukhin}.} \bibinfo{year}{2017}\natexlab{}.
\newblock \showarticletitle{Attention is all you need}.
\newblock \bibinfo{journal}{{\em Advances in neural information processing
  systems\/}}  \bibinfo{volume}{30} (\bibinfo{year}{2017}).
\newblock


\bibitem[\protect\citeauthoryear{VirusTotal}{VirusTotal}{2023}]%
        {virustotal}
\bibfield{author}{\bibinfo{person}{VirusTotal}.}
  \bibinfo{year}{2023}\natexlab{}.
\newblock \bibinfo{title}{VirusTotal}.
\newblock \bibinfo{howpublished}{\url{https://virustotal.com/}}.
  (\bibinfo{year}{2023}).
\newblock


\bibitem[\protect\citeauthoryear{Wang and Rudin}{Wang and Rudin}{2015}]%
        {wang2015falling}
\bibfield{author}{\bibinfo{person}{Fulton Wang} {and} \bibinfo{person}{Cynthia
  Rudin}.} \bibinfo{year}{2015}\natexlab{}.
\newblock \showarticletitle{Falling rule lists}. In \bibinfo{booktitle}{{\em
  Artificial intelligence and statistics}}. PMLR, \bibinfo{pages}{1013--1022}.
\newblock


\bibitem[\protect\citeauthoryear{Wang, Lim, Liu, and Zhao}{Wang
  et~al\mbox{.}}{2022}]%
        {wang2022explanation}
\bibfield{author}{\bibinfo{person}{Lei Wang}, \bibinfo{person}{Ee-Peng Lim},
  \bibinfo{person}{Zhiwei Liu}, {and} \bibinfo{person}{Tianxiang Zhao}.}
  \bibinfo{year}{2022}\natexlab{}.
\newblock \showarticletitle{Explanation guided contrastive learning for
  sequential recommendation}. In \bibinfo{booktitle}{{\em Proceedings of the
  31st ACM International Conference on Information \& Knowledge Management}}.
  \bibinfo{pages}{2017--2027}.
\newblock


\bibitem[\protect\citeauthoryear{Wang, Zhou, Qiu, Haque, Brown, He, Wang, Lo,
  and Zhang}{Wang et~al\mbox{.}}{2023}]%
        {wang2023towards}
\bibfield{author}{\bibinfo{person}{Zichong Wang}, \bibinfo{person}{Yang Zhou},
  \bibinfo{person}{Meikang Qiu}, \bibinfo{person}{Israat Haque},
  \bibinfo{person}{Laura Brown}, \bibinfo{person}{Yi He},
  \bibinfo{person}{Jianwu Wang}, \bibinfo{person}{David Lo}, {and}
  \bibinfo{person}{Wenbin Zhang}.} \bibinfo{year}{2023}\natexlab{}.
\newblock \showarticletitle{Towards Fair Machine Learning Software:
  Understanding and Addressing Model Bias Through Counterfactual Thinking}.
\newblock \bibinfo{journal}{{\em arXiv preprint arXiv:2302.08018\/}}
  (\bibinfo{year}{2023}).
\newblock


\bibitem[\protect\citeauthoryear{Warnecke, Arp, Wressnegger, and
  Rieck}{Warnecke et~al\mbox{.}}{2020}]%
        {warnecke2020evaluating}
\bibfield{author}{\bibinfo{person}{Alexander Warnecke}, \bibinfo{person}{Daniel
  Arp}, \bibinfo{person}{Christian Wressnegger}, {and} \bibinfo{person}{Konrad
  Rieck}.} \bibinfo{year}{2020}\natexlab{}.
\newblock \showarticletitle{Evaluating explanation methods for deep learning in
  security}. In \bibinfo{booktitle}{{\em 2020 IEEE european symposium on
  security and privacy (EuroS\&P)}}. IEEE, \bibinfo{pages}{158--174}.
\newblock


\bibitem[\protect\citeauthoryear{Wei, Li, Zhao, and Hu}{Wei
  et~al\mbox{.}}{2023}]%
        {weixnids}
\bibfield{author}{\bibinfo{person}{Feng Wei}, \bibinfo{person}{Hongda Li},
  \bibinfo{person}{Ziming Zhao}, {and} \bibinfo{person}{Hongxin Hu}.}
  \bibinfo{year}{2023}\natexlab{}.
\newblock \showarticletitle{XNIDS: Explaining Deep Learning-based Network
  Intrusion Detection Systems for Active Intrusion Responses}.
\newblock  (\bibinfo{year}{2023}).
\newblock


\bibitem[\protect\citeauthoryear{Weinberger, Dasgupta, Langford, Smola, and
  Attenberg}{Weinberger et~al\mbox{.}}{2009}]%
        {weinberger2009feature}
\bibfield{author}{\bibinfo{person}{Kilian Weinberger}, \bibinfo{person}{Anirban
  Dasgupta}, \bibinfo{person}{John Langford}, \bibinfo{person}{Alex Smola},
  {and} \bibinfo{person}{Josh Attenberg}.} \bibinfo{year}{2009}\natexlab{}.
\newblock \showarticletitle{Feature hashing for large scale multitask
  learning}. In \bibinfo{booktitle}{{\em Proceedings of the 26th annual
  international conference on machine learning}}. \bibinfo{pages}{1113--1120}.
\newblock


\bibitem[\protect\citeauthoryear{Yang, Guo, Hao, Ciptadi, Ahmadzadeh, Xing, and
  Wang}{Yang et~al\mbox{.}}{2021}]%
        {usenix21cade}
\bibfield{author}{\bibinfo{person}{Limin Yang}, \bibinfo{person}{Wenbo Guo},
  \bibinfo{person}{Qingying Hao}, \bibinfo{person}{Arridhana Ciptadi},
  \bibinfo{person}{Ali Ahmadzadeh}, \bibinfo{person}{Xinyu Xing}, {and}
  \bibinfo{person}{Gang Wang}.} \bibinfo{year}{2021}\natexlab{}.
\newblock \showarticletitle{CADE: Detecting and Explaining Concept Drift
  Samples for Security Applications}. In \bibinfo{booktitle}{{\em 30th USENIX
  Security Symposium (USENIX Security 21)}}.
\newblock


\bibitem[\protect\citeauthoryear{Yeh, Lee, Liu, and Ravikumar}{Yeh
  et~al\mbox{.}}{2022}]%
        {yeh2022threading}
\bibfield{author}{\bibinfo{person}{Chih-Kuan Yeh}, \bibinfo{person}{Kuan-Yun
  Lee}, \bibinfo{person}{Frederick Liu}, {and} \bibinfo{person}{Pradeep
  Ravikumar}.} \bibinfo{year}{2022}\natexlab{}.
\newblock \showarticletitle{Threading the Needle of On and Off-Manifold Value
  Functions for Shapley Explanations}. In \bibinfo{booktitle}{{\em
  International Conference on Artificial Intelligence and Statistics}}. PMLR,
  \bibinfo{pages}{1485--1502}.
\newblock


\bibitem[\protect\citeauthoryear{Zantedeschi, Nicolae, and Rawat}{Zantedeschi
  et~al\mbox{.}}{2017}]%
        {zantedeschi2017efficient}
\bibfield{author}{\bibinfo{person}{Valentina Zantedeschi},
  \bibinfo{person}{Maria-Irina Nicolae}, {and} \bibinfo{person}{Ambrish
  Rawat}.} \bibinfo{year}{2017}\natexlab{}.
\newblock \showarticletitle{Efficient defenses against adversarial attacks}. In
  \bibinfo{booktitle}{{\em Proceedings of the 10th ACM Workshop on Artificial
  Intelligence and Security}}. \bibinfo{pages}{39--49}.
\newblock


\bibitem[\protect\citeauthoryear{Zhang, Wang, Shen, Ji, Luo, and Wang}{Zhang
  et~al\mbox{.}}{2020a}]%
        {zhang2020interpretable}
\bibfield{author}{\bibinfo{person}{Xinyang Zhang}, \bibinfo{person}{Ningfei
  Wang}, \bibinfo{person}{Hua Shen}, \bibinfo{person}{Shouling Ji},
  \bibinfo{person}{Xiapu Luo}, {and} \bibinfo{person}{Ting Wang}.}
  \bibinfo{year}{2020}\natexlab{a}.
\newblock \showarticletitle{Interpretable deep learning under fire}. In
  \bibinfo{booktitle}{{\em 29th $\{$USENIX$\}$ Security Symposium
  ($\{$USENIX$\}$ Security 20)}}.
\newblock


\bibitem[\protect\citeauthoryear{Zhang, Zhang, Zhong, Ding, Cao, Zhang, Zhang,
  and Yang}{Zhang et~al\mbox{.}}{2020b}]%
        {zhang2020enhancing}
\bibfield{author}{\bibinfo{person}{Xiaohan Zhang}, \bibinfo{person}{Yuan
  Zhang}, \bibinfo{person}{Ming Zhong}, \bibinfo{person}{Daizong Ding},
  \bibinfo{person}{Yinzhi Cao}, \bibinfo{person}{Yukun Zhang},
  \bibinfo{person}{Mi Zhang}, {and} \bibinfo{person}{Min Yang}.}
  \bibinfo{year}{2020}\natexlab{b}.
\newblock \showarticletitle{Enhancing state-of-the-art classifiers with API
  semantics to detect evolved android malware}. In \bibinfo{booktitle}{{\em
  Proceedings of the 2020 ACM SIGSAC Conference on Computer and Communications
  Security}}. \bibinfo{pages}{757--770}.
\newblock


\bibitem[\protect\citeauthoryear{Zhang, Jang, Trabelsi, Li, Sanner, Jeong, and
  Shim}{Zhang et~al\mbox{.}}{2021}]%
        {zhang2021excon}
\bibfield{author}{\bibinfo{person}{Zhibo Zhang}, \bibinfo{person}{Jongseong
  Jang}, \bibinfo{person}{Chiheb Trabelsi}, \bibinfo{person}{Ruiwen Li},
  \bibinfo{person}{Scott Sanner}, \bibinfo{person}{Yeonjeong Jeong}, {and}
  \bibinfo{person}{Dongsub Shim}.} \bibinfo{year}{2021}\natexlab{}.
\newblock \showarticletitle{ExCon: Explanation-driven supervised contrastive
  learning for image classification}.
\newblock \bibinfo{journal}{{\em arXiv preprint arXiv:2111.14271\/}}
  (\bibinfo{year}{2021}).
\newblock


\bibitem[\protect\citeauthoryear{Zhou, Khosla, Lapedriza, Oliva, and
  Torralba}{Zhou et~al\mbox{.}}{2016}]%
        {zhou2016learning}
\bibfield{author}{\bibinfo{person}{Bolei Zhou}, \bibinfo{person}{Aditya
  Khosla}, \bibinfo{person}{Agata Lapedriza}, \bibinfo{person}{Aude Oliva},
  {and} \bibinfo{person}{Antonio Torralba}.} \bibinfo{year}{2016}\natexlab{}.
\newblock \showarticletitle{Learning deep features for discriminative
  localization}. In \bibinfo{booktitle}{{\em Proceedings of the IEEE conference
  on computer vision and pattern recognition}}. \bibinfo{pages}{2921--2929}.
\newblock


\bibitem[\protect\citeauthoryear{Zhou and Jiang}{Zhou and Jiang}{2012}]%
        {zhou2012dissecting}
\bibfield{author}{\bibinfo{person}{Yajin Zhou} {and} \bibinfo{person}{Xuxian
  Jiang}.} \bibinfo{year}{2012}\natexlab{}.
\newblock \showarticletitle{Dissecting android malware: Characterization and
  evolution}. In \bibinfo{booktitle}{{\em 2012 IEEE symposium on security and
  privacy}}. IEEE, \bibinfo{pages}{95--109}.
\newblock


\end{thebibliography}

\appendix

\section{Notations} \label{app:notation}

\textcolor{\tcolor}{
To improve the readability, we list the important notations defined in this paper as in Table~\ref{tab:notation}.
}

\begin{table}[htbp]
    \caption{\textcolor{\tcolor}{Important acronyms and symbols.}}
    \label{tab:notation}
    \begin{tabularx}{\linewidth}{Cc}
    \toprule
    \textbf{Notation}      & \textbf{Description}     \\ \midrule
    FA            & Feature attribution               \\
    ERDS          & Explainable risk detection system \\
    MPD          & Model prediction drop~(for fidelity evaluation) \\ \midrule
    $\mathcal{Z}$; $\mathcal{F}$; $\mathcal{V}$  & Problem/Feature/Vector space \\
    $\mathcal{C}$ & Class label set (binary)          \\
    $x$; $e$           & Sample/Explanation~(subscript: different spaces) \\
    $\alpha$; $\varphi$      & Feature extraction / Vector representation function       \\
    $f$; $g$           & DNN/FA algorithms    \\
    $\omega$      & Explanation presentation function \\
    \bottomrule
    \end{tabularx}
\end{table}

\section{Feature and Vector Space} \label{app:exp_domain}

\begin{figure}[htb]
    \centering
    \includegraphics[width=.92\linewidth]{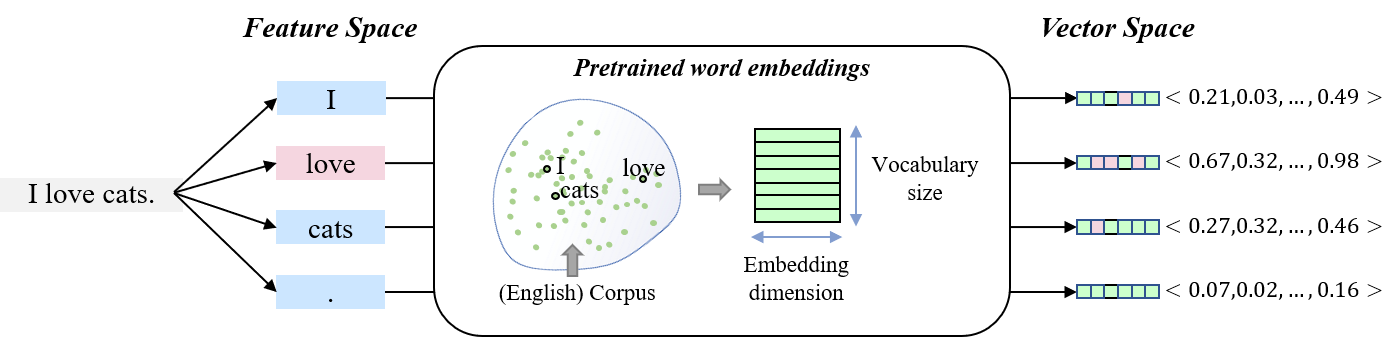}
    \caption{\textcolor{\tcolor}{The separation of $\mathcal{F}$ and $\mathcal{V}$ in text domain, where features are vectorized with a look-up table. The pink color indicates explanations in different spaces. For the pretrained embedding vector, each dimension does not have a special meaning.}}
    \label{fig:word2vec}
\end{figure}

The definition of the feature space $\mathcal{F}$ and the vector space $\mathcal{V}$ might be a little confusing since the mapping between them is typically implicit in other domains. 
For example, $\mathcal{F}$ and $\mathcal{V}$ can be naturally the same~(both being pixel values) for image classification, and translating words in $\mathcal{F}$ to a vector space $\mathcal{V}$ commonly adopts word embedding techniques~(e.g., Word2Vec~\cite{mikolov2013efficient}, as illustrated in Figure~\ref{fig:word2vec}) for text classification.
As a matter of fact, it would be acceptable to omit the difference between them in other researches~\cite{pierazzi2020intriguing}.
However, as for providing explanations, $\mathcal{F}$ decides what semantics can be conveyed to human while $\mathcal{V}$ influences how FA algorithms work.
Therefore, we differentiate the two spaces to emphasize their different impacts on explanations and to better illustrate our design. 


\section{Efficiency Analysis} \label{app:time}

As in Table~\ref{tab:time}, we provide the time used for individual FA methods on the three risk detection classifiers.
Gradient-based methods is faster than perturbation-based methods by $1243.5$, $567.7$, $1.8$ seconds on average for DAMD, DR-VGG, and VulDeePecker, where the average IC number is at a magnitude of $4$, $3$, and $1$, respectively. This proves the discussion in Section~\ref{sec:cls_exp}. 

\begin{figure}[htbp]
\myfigureshrinker
    \centering
    \subfloat[DAMD]{%
        \includegraphics[width=.4\linewidth]{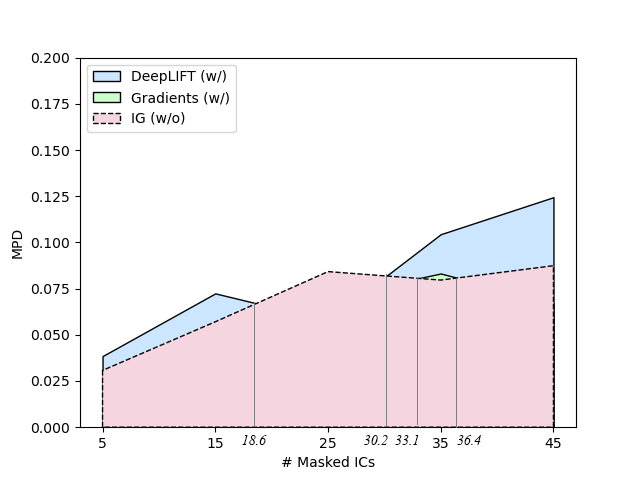}%
        \label{fig:damdq3}%
        }
    \subfloat[DR-VGG]{%
        \includegraphics[width=.4\linewidth]{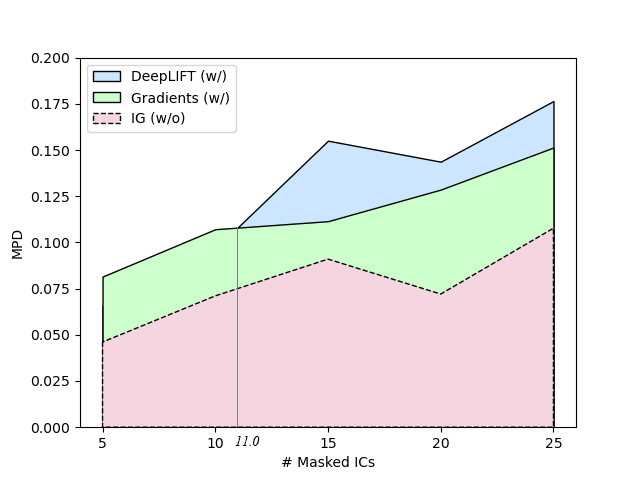}%
        \label{fig:drq3}%
        }
    \caption{Reduced explanation costs for white-box explaining.}
    \label{fig:white-save}
\end{figure}

\begin{table}[htbp]
\caption{Efficiency study for different explainers.}
\label{tab:time}
\begin{tabularx}{.98\linewidth}{CCCCC}
\toprule
\textbf{Explainer} & \textbf{DAMD} & \textbf{DR-VGG} & \textbf{VulDeeP.} & \textbf{Avg.} \\ \midrule
Gradients & 2.5E-2 & 1.0E-1 & 1.6E-3 & 4.3E-2 \\
IG & 2.4E+0 & 4.0E+0 & 5.3E-1 & 2.3E+0 \\
DeepLIFT & 1.6E+0 & 1.7E+0 & 5.8E-2 & 1.1E+0 \\ \midrule
LIME & 4.6E+2 & 5.4E+1 & 3.4E+0 & 1.7E+2 \\
LEMNA & 1.8E+3 & 1.4E+3 & 4.3E+1 & 1.1E+3 \\
Shapley & 1.5E+3 & 2.1E+2 & 8.2E+0 & 5.6E+2 
\\ \bottomrule
\end{tabularx}
\end{table}
For the low-cost scenario, a less time-consuming FA method on the updated classifier can achieve equal/higher explanation fidelity than more complex method on the classifier without \framework{}.
We explain it  with Figure~\ref{fig:white-save}, where the chance for Gradients/DeepLIFT to replace IG is $8.25\%$/$71.00\%$ for DAMD and $30\%/100\%$ for DR-VGG.
As the these methods are faster than IG by $2.38/0.80$ and $3.90/2.30$ seconds per sample, the saved time can then be estimated at $0.76$ and $3.90$ seconds for the two systems, respectively.

\section{Human Subject Evaluation} \label{app:case}

\textcolor{\tcolor}{
Following the paradigm in related works~\cite{downing2021deepreflect}, we collect the feedback of five experienced malware analysts to evaluate the usefulness of the explanation.
Firstly, we ask them about how they feel about the IC-level~(function) explanation and the feature-level~(opcode) explanation for DAMD, and all of them think highlighted functions would be more useful.
Secondly, we provide malware samples in the dataset of DR-VGG for them to analyze and evaluate the performance of different tools in Section~\ref{sec:application}.
To ensure the consistency in detecting function boundaries, we release the database files~(.bndb) of BinaryNinja for them to read~\footnote{https://github.com/XdlSec/FINER-questionnaire-gadgets}.
The sampled malwares have $71$ functions on average and would take analysts around $26$ minutes per sample to analyze.
After reading the disassembled language, we ask them to choose from two explanations~(generated by \framework{} and DeepReflect), and \framework{} has been their first choice in all cases. 
When asking them to score the selected explanation with five grades, the average score reaches $4$, and for some cases, the analysts claim that more than $70\%$ manual efforts can be reduced with the explanation.
}

\noindent 
\textcolor{\tcolor}{\textbf{Other Insights.}
Experts all believe that labeling suspicious functions is helpful for malware analysis, but at the same time, they suggest that connecting the functions with data and control flows would be more beneficial.
To solve this, efforts should be made beyond the explanation algorithms. 
For example, involving more domain knowledge for the feature extraction $\alpha$ and making more delicate design for the explanation presentation $\omega$.
}

\lstinputlisting[caption=\textcolor{\tcolor}{Pseudo code of the key function in Pegasus\#log}, label={lst:case-study}, style=mystyle]{LogonPasswords.cpp} 

\section{The MPD Value}

\textcolor{\tcolor}{
\textbf{Higher Workload Cost.}
The absolute value of MPD seems low for the two malware classifiers in Table~\ref{tab:fidelity} because $k$ is relatively small compared with the IC size~(see Figure~\ref{fig:seg_info}). We make such illustration since we are more concerned about the enhancement at a smaller $k$, which indicates less downstream workload, e.g., the number of functions that should be inspected by analysts. Actually, the two systems with \framework{} can also reach the 0.5 level MPD if the number of removed ICs increases, i.e., $k=1000$ for DAMD and $k=350$ for DR-VGG as in Table~\ref{tab:higherk}.
}

\begin{table}[t]
\caption{\textcolor{\tcolor}{MPD at higher $k$ for the two malware classifiers.}}
\label{tab:higherk}
\begin{threeparttable}
\begin{tabularx}{\linewidth}{cCCCCCCC}
\toprule
\textbf{DAMD} & \textbf{500}    & \textbf{1000}   & \textbf{1500}   & \textbf{2000}   & \textbf{2500}   & \textbf{3000}   \\ \midrule
w/o           & 0.3756          & 0.4809          & 0.5596          & 0.5471          & 0.5936          & 0.5863          \\
w/            & 0.4295          & 0.5118          & 0.6019          & 0.6108          & 0.6256          & 0.6532          \\
\rowcolor[HTML]{EFEFEF} 
Imp. & 14.35\% & 6.43\% & 7.56\% & 11.64\% & 5.39\% & 11.41\% \\ \midrule
\textbf{DR-VGG} & \textbf{100}    & \textbf{150}    & \textbf{200}    & \textbf{250}    & \textbf{300}    & \textbf{350}    \\ \midrule
w/o             & 0.1542          & 0.1949          & 0.2551          & 0.2516          & 0.2781          & 0.2656          \\
w/              & 0.3047          & 0.3609          & 0.3702          & 0.4045          & 0.4460          & 0.5113           \\
\rowcolor[HTML]{EFEFEF} 
Imp. & 97.60\% & 85.17\% & 45.12\% & 60.77\% & 60.37\% & 92.51\%
\\ \bottomrule
\end{tabularx}
    \begin{tablenotes}
    \footnotesize
        \item[1] w/o: the original classifier.
        \item[2] w/: the updated classifier.
    \end{tablenotes}
\end{threeparttable}
\end{table}

\noindent 
\textcolor{\tcolor}{\textbf{Feature-level Masking.}
Since the target of ERDS is IC-based explanation, we use IC masking to evaluate model interpretability.
To justify our measurement, we also take the baseline approach to mask the most important features.
Specifically, for each sample, we first count the number of features that are contained in its most important ICs, next mask the same amount of features according to the importance in $e_v$~(if practical for calculation, see Section~\ref{sec:eval_ic}), and finally calculate the MPD. 
As shown in Table~\ref{tab:top_feature_mask}, the updated classifiers still outperform the original classifiers, and the MPD score is improved by $21.39\%$ and $92.83\%$ on average for the two systems.
Additionally, comparing with the IC masking results~(in Table~\ref{tab:fidelity}), the mean percentage decrease of $53.20\%$ and $125.42\%$ shows the effectiveness of IC abstraction during fidelity evaluation. 
}

\begin{table}[tbhp]
\caption{\textcolor{\tcolor}{The MPD of feature-level masking. For each sample, the number of masked top features is the same as that are contained in top-$25$ ICs, and the Dec. columns indicate the percentage decrease from IC masking.}} 
\label{tab:top_feature_mask}
\begin{threeparttable}
\begin{tabularx}{\linewidth}{CCCCC}
\toprule
\textbf{DAMD} & \textbf{w/o} & \textbf{Dec.} & \textbf{w/} & \textbf{Dec.} \\ \midrule
Gradients & 0.0419 & \cellcolor[HTML]{EFEFEF}57.42\% & 0.0514 & \cellcolor[HTML]{EFEFEF}56.89\%  \\
IG & 0.0783 & \cellcolor[HTML]{EFEFEF}39.68\% & 0.0836 & \cellcolor[HTML]{EFEFEF}47.37\%  \\
DeepLIFT & 0.0439 & \cellcolor[HTML]{EFEFEF}62.68\% & 0.0592 & \cellcolor[HTML]{EFEFEF}55.19\%  \\ \midrule
\textbf{DR-VGG} & \textbf{w/o} & \textbf{Dec.} & \textbf{w/} & \textbf{Dec.} \\ \midrule
Gradients & -0.0594 & \cellcolor[HTML]{EFEFEF}177.10\% & -0.0008 & \cellcolor[HTML]{EFEFEF}100.55\% \\
IG  & -0.0333 & \cellcolor[HTML]{EFEFEF}132.48\% & -0.0038 & \cellcolor[HTML]{EFEFEF}102.05\% \\
DeepLIFT & -0.0330 & \cellcolor[HTML]{EFEFEF}138.43\% & -0.0029 & \cellcolor[HTML]{EFEFEF}101.91\% \\
\bottomrule
\end{tabularx}
    \begin{tablenotes}
    \footnotesize
        \item[1] w/o: the original classifier.
        \item[2] w/: the updated classifier.
    \end{tablenotes}
\end{threeparttable}
\end{table}

\section{IC decomposition Interface} \label{app:interface}

\textcolor{\tcolor}{The interface M2, as described in Listing~\ref{lst:interface}, has two methods for users to provide $h:\mathcal{Z} \rightarrow \mathcal{D}$ and $k \in \mathbb{N}^+$ and two methods for the framework to generate the IC indicator and query the workload expectation. 
The generation of the indicator $\mathbf{I}$ is illustrated in Algorithm~\ref{alg:ic} and we explain it in this section.}
Since IC is unique among samples and do not have fixed size in $\mathcal{V}$, for each sample, we apply $\varphi \circ \alpha$ on individual IC objects to get the counterpart and fill in the matrix $\mathbf{I}$.
As for those objects that have no counterpart in $\mathcal{V}$, they are not considered as explainable and thus are filtered out in the valid IC array $\mathcal{I}$.


\begin{algorithm}[t]
\DontPrintSemicolon
  \SetKwFunction{FMain}{get\_ic\_indicator}
  \SetKwProg{Pn}{Function}{:}{\KwRet}
  \Pn{\FMain{$x$, $h, \alpha$, $\varphi$}}{
    Get vector-space representation $x_v = \varphi \circ \alpha (x)$ \;
    Init $\mathbf{I}$ to $[0]_{m \times n}$ \;  
    Init valid IC array $\mathcal{I}$ to $\emptyset$, current IC index $i$ to $0$ \; 
    \For{each domain-space IC $d \in h(x)$}{ 
        Get vector-space counterpart $d_v=\varphi \circ \alpha (d)$ \;
        Get regions $\mathcal{r}$ where $Ind(x_v, d_v)=1$ \;
        \If{$\mathcal{r}$ is not empty}{
            Assign $i$ to elements in $\mathbf{I}[\mathcal{r}]$ \;
            Assign $d$ to $\mathcal{I}[i]$; Add $1$ to $i$
        }
    }
    \KwRet $\mathbf{I}$, $\mathcal{I}$\;
  }
\caption{IC Indicator Generation}
\label{alg:ic}
\end{algorithm}

\section{Pattern-driven Classifiers} \label{app:pattern}

\textcolor{\tcolor}{
We focus on explaining data-driven risk detection classifiers to achieve high semantic capacity of explanations and to solve the extreme difficulty in explaining them~(Section~\ref{sec:scope}).
In this section, we further explain the idea by discussing the explainability of those pattern-driven classifiers.}

\begin{figure}[tbhp]
\myfigureshrinker
    \centering
    \subfloat[Sample Explanations]{%
        \includegraphics[width=.585\linewidth]{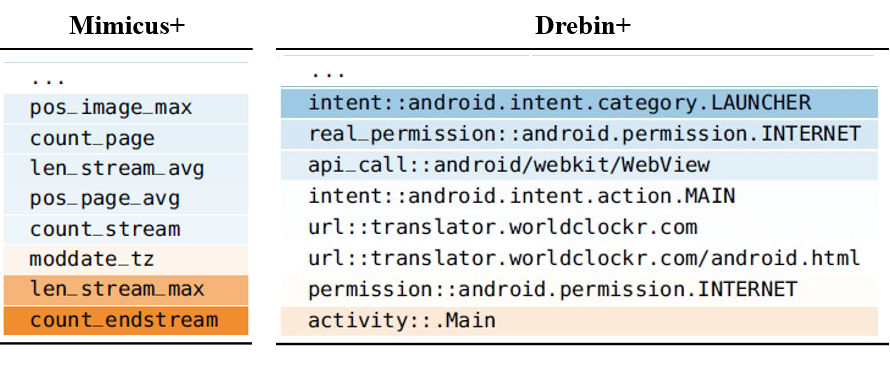}%
        \label{fig:mimicus-drebin-exp}%
        }
    \subfloat[MPD Curves]{%
        \includegraphics[width=.32\linewidth]{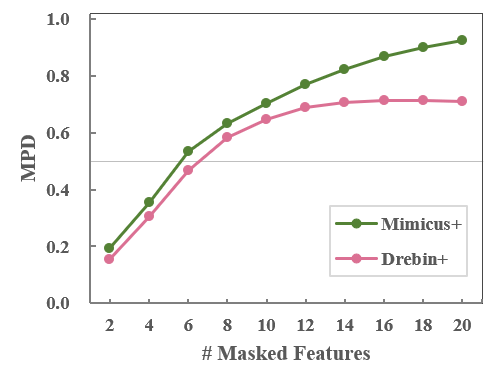}%
        \label{fig:mimicus-drebin-mpd}%
        }
    \caption{Feature-level explanation results for Mimicus+ and Drebin+ from prior works. The fidelity is shown with the MPD curves of IG~(right), and the intelligibility is shown with example outputs~(left).}
    \label{fig:drebin-exp}
\end{figure}

\textcolor{\tcolor}{
As mentioned in Table~\ref{tab:classifiers}, the two classifiers named Mimicus+ and Drebin+ are pattern-driven because their feature engineering methods adopt hand-crafted patterns that are human readable~\cite{laskov2014practical, grosse2017adversarial}. 
In other words, when designing the feature space~(before ERDS takes effect), a certain level of abstraction has already been reached by security experts.
For example, Drebin+ extracts $8$ sets of features from the manifest file and disassembled code, which represent the existence of components that are believed to serve for malicious behaviors, such as the intent \texttt{BOOT\_COMPLETED}~(to trigger malicious activity after rebooting), the permission \texttt{SEND\_SMS}~(to send sensitive data with messages), and the API call \texttt{Runtime.exec}~(to execute external commands).
As a result, explaining these classifiers would only reflect patterns that are recognized by human, i.e., \textit{low semantic capacity}.
On the other hand, existing FA methods can be regarded as a better fit for them in terms of
(1)~\textit{high intelligibility}: feature-level explanations that can be linked to known patterns are more friendly for security experts~(R2) to understand;
(2)~\textit{high fidelity}: post-hoc approximations on the lightweight model~(MLP) that works on the simple vector space~(binary vectors of Tabular data) is easier for explainers to make. 
The conclusion can also be validated with prior works~\cite{warnecke2020evaluating} that investigate the two systems, where the MPD value reaches $0.5$ at a small deduction rate~(Figure~\ref{fig:mimicus-drebin-mpd}) and analysts take several useful insights from the generated explanations~(Figure~\ref{fig:mimicus-drebin-exp}).
}

\textcolor{\tcolor}{
As a matter of fact, the heuristic feature engineering methods are common before the wide adoption of deep learning in risk detection.
The features of the two aforementioned systems are originally developed by Mimicus~\cite{smutz2012malicious} and Drebin~\cite{arp2014drebin}, which are proposed in the year of 2012 and 2014. In the original papers, the adopted classification models are simple and intrinsically interpretable, i.e., Random Forest and linear SVM.
Similar examples include Ember~\cite{anderson2018ember}, a PE malware detector that extracts $8$ groups of features to form vectors of $2,351$ dimensions and leverages the LightGBM model. 
For Ember, features are more general patterns that include the hashing of string lists and the histogram of raw bytes.
Although involving more properties, the semantic capacity becomes lower due to the black-box and lossy feature compression, i.e., hashing trick~\cite{weinberger2009feature} and histogram summary.
From another perspective, these examples actually have problem-space objects that are independently handled, which are different sections of the file structures. Our framework can handle the scenario to abstract different sections, but it would be meaningless since the generated explanations would be too coarse-grained for security analysis.
Anyhow, for these feature engineering methods, the adoption of lightweight deep learning models has been proven to generally improve the classification accuracy and the model robustness~\cite{severi2021explanation, masum2019droid, zhang2020enhancing}.
Benefiting from FA methods, the improvement can be achieved without loss of interpretability.
}

\end{document}